\documentclass[preprint,3p,12pt]{elsarticle}

\usepackage[acronym]{glossaries}
\usepackage{amsmath}

\usepackage{amssymb}
\usepackage{bm}
\usepackage{color}
\usepackage{subfig}
\usepackage{hyperref}
\hypersetup{colorlinks=true,linkcolor=blue,citecolor=blue,urlcolor=blue}

\newcommand{\beq}{\begin{equation}}
\newcommand{\eeq}{\end{equation}}
\newcommand{\beqar}{\begin{eqnarray}}
\newcommand{\eeqar}{\end{eqnarray}}
\newcommand{\bea}{\begin{eqnarray}}
\newcommand{\eea}{\end{eqnarray}}
\newcommand{\bcen}{\begin{center}}
\newcommand{\ecen}{\end{center}}

\newcommand{\bra}[1]{\left< #1 \right|}
\newcommand{\ket}[1]{\left| #1 \right>}
\newcommand{\ketbra}[2]{\left| #1 \right> \left< #2 \right|}

\newcommand{\tr}{\mathrm{tr}}
\newcommand{\ave}[1]{\left< #1 \right>}

\newcommand{\Iop}{\hat{I}}

\makenoidxglossaries

\newacronym{qhe}{QHE}{Quantum Heat Engines}
\newacronym{qe}{QE}{Quantum Engines}
\newacronym{qed}{QED}{Quantum Electrodynamics}
\newacronym{fts}{FTs}{Fluctuation Theorems}
\newacronym{nv}{NV}{Nitrogen Vacancy}
\newacronym{nmr}{NMR}{Nuclear Magnetic Resonance}
\newacronym{tpm}{TPM}{Two-Point-Measurement}
\newacronym{tur}{TUR}{Thermodynamics Uncertainty Relations}
\newacronym{tls}{TLS}{Two-Level System}
\newacronym{ho}{HO}{Harmonic Oscillator}
\newacronym{qr}{QR}{Quantum Refrigerator}
\newacronym{lgks}{LGKS}{Lindblad-Gorini-Kossakowski-Sudarshan}
\newacronym{md}{MD}{Maxwell's Demon}
\newacronym{MBL}{MBL}{Many-body Localization}
\newacronym{STA}{STA}{Shortcut to Adiabaticity}
\newacronym{QD}{QD}{Quantum Dot}
\newacronym{CB}{CB}{Coulomb Blockade}
\newacronym{BCS}{BCS}{Bardeen–Cooper–Schrieffer}
\newacronym{DOS}{DOS}{Density of States}
\newacronym{RC}{RC}{Reaction Coordinates}
\newacronym{lmg}{LMG}{Lipkin Meshkov Glick}
\newacronym{kz}{KZ}{Kibble-Zurek}
\newacronym{tcl}{TCL}{Time Convolution-Less}
\newacronym{negf}{NEGF}{Nonequilibrium Green Function}
\newacronym{qpc}{QPC}{Quantum Point Contact}
\newacronym{chsh}{CHSH}{Clauser-Horne-Shimony-Holt}
\newacronym{lhs}{LHS}{Local Hidden State}
\newacronym{cps}{CPS}{Cooper Pair Splitter}
\newacronym{mps}{MPS}{Matrix Product States}
\newacronym{tn}{TN}{Tensor Networks}
\newacronym{dft}{DFT}{Density Functional Theory}
\newacronym{lda}{LDA}{Local Density Approximation}
\newacronym{rl}{RL}{Reinforcement Learning}
\newacronym{sshe}{SSHE}{Steady-State Heat Engine}
\newacronym{eh}{EH}{Electron Hole}
\newacronym{ec}{EC}{Elastic Cotunneling}
\newacronym{2deg}{2DEG}{Two-Dimensional Electron Gas}
\newacronym{lwi}{LWI}{Lasing Without Inversion}
\newacronym{eit}{EIT}{Electromagnetically Induced Transparency}
\newacronym{povm}{POVM}{Positive Operator-Valued Measure}
\newacronym{qar}{QAR}{Quantum Absorption Refrigerator}
\newacronym{dp}{DP}{Differential Programming}
\newacronym{sln}{SLN}{Stochastic Liouville-von Neumann}
\newacronym{cop}{COP}{Coefficient of Performance}
\journal{Physics Reports}

\begin{document}

\begin{frontmatter}



\title{Quantum Engines and Refrigerators}


\author{Loris Maria Cangemi \fnref{fn}}
\ead{lorismaria.cangemi@gmail.com}

\author{Chitrak Bhadra \fnref{fn}}
\ead{chitrak.iitb@gmail.com}

\author{Amikam Levy \corref{cor}}
\ead{amikam.levy@biu.ac.il}
\ead[https://levyresearch.com/]{levyresearch.com}
\cortext[cor]{Corresponding author}
\fntext[fn]{Both authors contributed equally to this work}


\begin{abstract}
Engines are systems and devices that convert one form of energy into another, typically into a more useful form that can perform work. 
In the classical setup, physical, chemical, and biological engines largely involve the conversion of heat into work. This energy conversion is at the core of thermodynamic laws and principles and is codified in textbook material.
In the quantum regime, however, the principles of energy conversion become ambiguous, since quantum phenomena come into play. As with classical thermodynamics, fundamental principles can be explored through engines and refrigerators, but, in the quantum case, these devices are miniaturized and their operations involve uniquely quantum effects.
Our work provides a broad overview of this active field of quantum engines and refrigerators, reviewing the latest theoretical proposals and experimental realizations. 
We cover myriad aspects of these devices, starting with the basic concepts of quantum analogs to the classical thermodynamic cycle and continuing with different quantum features of energy conversion that span many branches of quantum mechanics. These features include quantum fluctuations that become dominant in the microscale, non-thermal resources that fuel the engines, and the possibility of scaling up the working medium's size, to account for collective phenomena in many-body heat engines. Furthermore, we review studies of quantum engines operating in the strong system-bath coupling regime and those that include non-Markovian phenomena. Recent advances in thermoelectric devices and quantum information perspectives, including quantum measurement and feedback in quantum engines, are also presented.  

\end{abstract}

\begin{graphicalabstract}
\includegraphics{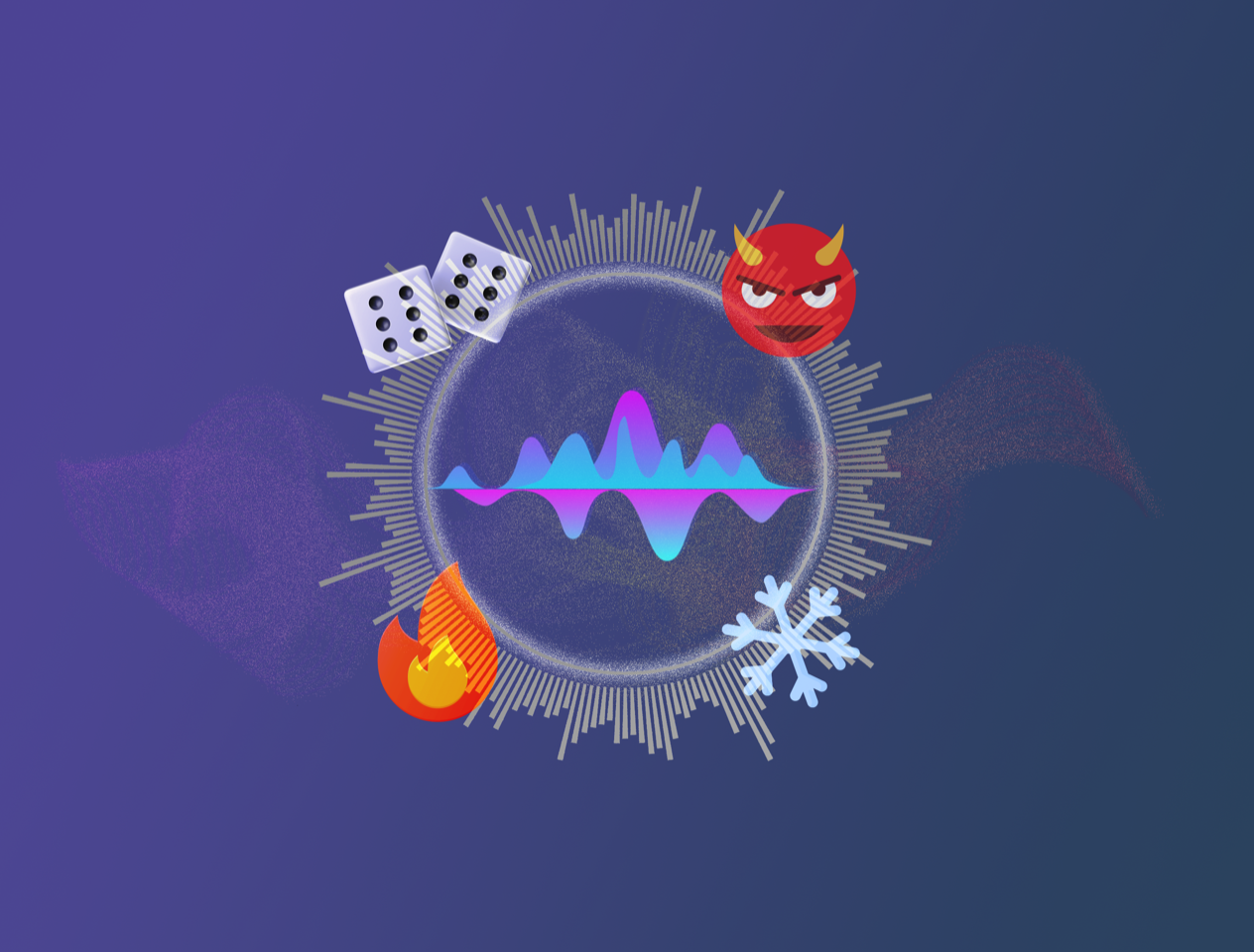}
\end{graphicalabstract}

\begin{highlights}
\item Quantum Engines provide a unique platform for studying the interplay of quantum phenomena and nonequilibrium thermodynamics.
\item The role of quantum fluctuations, non-Markovianity, and strong coupling in energy conversion can be investigated through Quantum Engines.
\item Many-body systems and non-thermal baths are building blocks of Quantum Engines.
\item Recent developments in thermoelectric devices open new experimental possibilities.
\item Quantum correlation measurements and feedback can serve as resources for work extraction and cooling. 
\end{highlights}


\end{frontmatter}


\tableofcontents

\clearpage

\printnoidxglossary[type=\acronymtype,nonumberlist]

\clearpage

\section{Introduction}
The transformation of energy from one form into another is a fundamental concept that is studied across many fields of research, ranging from physics to chemistry and biology. Exploring the principles and mechanisms of energy transformation is essential to understanding the behavior of both natural and human-made systems. Such explorations are crucial, moreover, for designing and improving technologies and finding alternative energy sources. Thermodynamics, the field that investigates the conversion of work and heat on the macro scale, was initiated by the theoretical work of Carnot on heat engines in 1824. In this work~\cite{carnot1824reflections}, ``Reflection on Motive Power,''\footnote{Translation to English by R. H. Thurston from 1890 can be found online.} Carnot paved the way toward the basic understanding that work can be produced using two thermal heat baths at different temperatures, the common theoretical setting of heat engines even today. Furthermore, the first step in formulating the second law of thermodynamics was achieved thanks to his study on the maximal efficiency of heat engines, which set an upper bound on the efficiency of heat conversion into work in such cyclic operating devices.

Engines and refrigerators have provided a platform for exploring the basic principles of energy conversion and thermodynamics for nearly 200 years.  It isn't surprising, therefore, that not long after its invention, the maser (i.e., microwave amplification via stimulated emission of radiation) was analyzed as an engine operation.   
A pioneering work done by Scovil and Schulz-DuBois~\cite{Scovil:1959} in 1959 is traditionally considered the first study of quantum heat engines. Yet it provides only a quasistatic picture; its main quantum feature is the discreet energy levels of the working medium, as will be discussed in the next section. 
It was only much later that a dynamical picture of quantum heat engines within an open quantum system formalism was established by Alicki~\cite{alicki79} and Kosloff~\cite{kosloff84}.
Over the next several decades, the study of quantum engines and refrigerators\footnote{Extensive work on Laser cooling took place during those years, but its interpretation in terms of the quantum thermodynamic framework was still missing.} still took place within a relatively small community. 
In the last decade, however, the field has seen tremendous growth, thanks in large part to experimental advances in nanoscale devices, superconducting circuits, circuit \acrshort{qed}s, and quantum many-body devices. 
Both theoretically and experimentally, this study of quantum engines and refrigerators has expanded into many areas of quantum mechanics, including quantum information, many-body systems, nonequilibrium quantum dynamics, quantum fluctuations, quantum control, thermoelectric devices, and more.

After giving a short account of the most striking contributions to the field, the majority of which were made in the last fifteen years, we focus on theoretical proposals and experimental findings put forward over the last six years, a period covering the emergence of quantum engines in several different branches of quantum mechanics, as specified in the table of contents.  
A previous review~\cite{levy14a} from 2014 focused solely on continuous thermal devices, and more recent ones~\cite{myers2022quantum,arrachea2022energy}, published during the period in which we composed the present work, offer a detailed list of promising platforms for experimental studies of quantum heat engines (\acrshort{qhe}s) and explore energy dynamics, heat production and heat-to-work conversion with qubits. In each of the main sections, we also provide references to reviews salient to the specific topic considered. To keep our discussion self-contained, we had to choose the research papers under our consideration quite selectively, and, thus, covering all relevant studies was not possible.
	
The work is organized as follows: in Sec.~\ref{sec:basic}, we summarize basic concepts and models of driven and autonomous QHEs. 
We discuss both stroke engines that, analogously to their classical counterparts, undergo thermodynamic cycles, and continuous thermal machines that can provide thermodynamic insight into Lasers and Masers.
In these models, the working medium is a quantum object that is driven out-of-equilibrium, by means of control fields and exchanges of work and heat with a number of thermal reservoirs.

In Sec.~\ref{sec:fluctuations}, 
we discuss the significance of fluctuations in the operations of QHEs. At the quantum scale, fluctuation may dominate the behavior of the engine. We therefore introduce different protocols for characterizing the fluctuations in the work, heat and efficiency of the devices, and examine their possible pitfalls.  Furthermore, we stress the connection between thermodynamic uncertainty relations and bounds on the efficiency of QHEs.  

In Sec.~\ref{sec:neqbath}, we consider several proposals for prototypical non-thermal QHEs.
This class of engines is energized by means of non-thermal reservoirs, i.e., physical systems whose quantum states cannot be described with a thermal Gibbs state.
Engines belonging to this class harness additional resources with respect to their thermal counterparts, e.g., coherent and squeezed baths.
Due to their special features that provide additional sources of work, the efficiency of these engines has been shown to exceed, perhaps not surprisingly, conventional heat engine bounds.
These violations are only apparent, and they can be avoided via a suitable theory presented here, which takes into account additional thermodynamic resources linked to the non-equilibrium nature of the bath. 
Moreover, this section also reviews the effect of correlations between the QHE's different baths.

In Sec.~\ref{sec:manybody}, we focus on recent  theoretical proposals of QHEs, whose working mediums consist of driven quantum many-body systems.
This branch of research is motivated by the search for genuine quantum effects that could provide an advantage in the performance of QHEs, with respect to their classical analogs.
We take into account several features of many-body systems that could lead to various advantages in the performance of heat engines.
For instance, entanglement and other forms of quantum correlations have been proven to boost the output power and efficiency of QHEs, with respect to an ensemble of noninteracting single-particle engines.
Moreover, criticality due to interactions among the constituents of the working medium can modify the scaling of the power output with the engine size, such that the engine could achieve finite output power with ideal efficiency.
Critical QHEs working in the presence of non-equilibrium phase transitions, ruled by the Kibble-Zurek \acrshort{kz} mechanism, have also been studied, as have statistically indistinguishable bosonic particle-based QHEs. We will review the main theoretical proposals related to many-body QHEs, underlining the different definitions of quantum advantage, as well as experimental platforms suitable for implementing such engines.

Sec.~\ref{sec:nonMarkov} provides an overview of recent theoretical approaches and proposals that address the impact of the working medium's non-Markovian dynamics on the performance of QHEs.
Non-Markovianity in an open quantum system entails a wide set of physical effects; these typically arise in the limit of strong system-bath interactions and low-temperature regimes, where system-bath correlation decay times are of the order of the system's characteristic timescales.
A non-Markovian evolution can be described in terms of the occurrence of memory effects of the open system dynamics.
Such effects can lead, in turn, to environment backaction and information backflows, which are absent in the Markovian setting.
As a result, from the theoretical perspective, these dynamical effects cannot be caught within the framework of a quantum Markovian semigroup.
This phenomenon holds especially true for open quantum systems subject to driving fields, as nonequilibrium dynamics are dominated by the interplay between the driving fields and dissipation.
We will thus focus on recent proposals of QHEs undergoing arbitrary finite-time thermodynamic cycles, where the coupling with thermal baths can also be suitably modulated in time. Furthermore, we will present special proposals of QHEs whose working medium is strongly coupled to heat baths, describing their features as compared to more conventional settings.      
 
In Sec.~\ref{sec:thermoel}, we provide a short introduction to the recent theoretical and experimental literature on nanoscale thermoelectric devices.
This kind of QHE belongs to the class of autonomous or continuous engines.
Unlike driven QHEs, such devices convert steady-state heat currents into charge currents against a voltage bias, achieving heat to work conversion at the nanoscale.
In addition to heat-to-work conversion, steady-state machines can be employed as heat rectificators, refrigerators, and on-chip coolers, working at sub-Kelvin temperatures and employing superconducting circuit elements.
In the first part of the section, we outline the main theoretical approaches to modeling these devices, referring to the excellent reviews available in the literature. Subsequently, we review novel theoretical proposals and experimental realizations, including the concepts of hybrid thermal machines and non-local and non-linear superconducting thermoelectric devices.  
  
In Sec.~\ref{sec:information}, we provide an overview of recent theoretical and experimental advances that investigate the fundamental relationship between quantum thermal devices and information theory. The cornerstone of this research in the classical setting, Maxwell's demon, has received great interest in recent years, prompted by experimental advances in the control of nonequilibrium processes. After summarizing the main steps taken during the two decades of research on this topic, we then focus on recent studies that shed light on the consequences of genuine quantum features in the measurement-feedback process, which have been described using the framework of resource theories. We also review novel theoretical proposals of quantum devices that harness quantum measurement as an additional resource for achieving thermodynamic tasks, such as heat-to-work conversion and refrigeration. Finally, we provide a brief account of autonomous quantum heat engines that allow for markedly nonclassical behavior, such as entanglement generation.

\section{Basic concepts}\label{sec:basic}
\subsection{Quantum thermodynamic processes}
Classically, a thermodynamic process is defined as a change from an initial equilibrium state to a final one, during which thermodynamic state variables such as pressure, volume, and temperature may vary. 
In the quantum realm, the system undergoing the thermodynamic process is on the scale of a single atom, and among the relevant variables are the eigenenergies and the energy-level occupations of the system. 
%

%

The closest quantum counterpart to a quasistatic process, where the system is changing slowly such that it remains in thermodynamic equilibrium throughout the process, is the quantum adiabatic process\footnote{ The quantum adiabatic process should not be confused with an adiabatic process in thermodynamics which implies that no heat is exchanged with the environment.}. In this process, for closed quantum systems, the control Hamiltonian changes gradually, ensuring the system's state aligns with the instantaneous eigenenergies while maintaining a constant population. In the context of open quantum systems interacting with a thermal environment —the analog of the isothermal process— the system connects to a heat reservoir, and its energy levels adjust slowly to maintain thermal equilibrium. If the thermalization time is significantly shorter than the internal dynamics of the system, it stays in a Gibbs state at a fixed temperature, aligned with the instantaneous Hamiltonian, experiencing both work and heat transformations.

Conversely, the quantum isochoric process entails modifications only in the population distribution across energy levels, with the eigenenergies staying the same. Here, the energy interaction with the thermal bath is recognized as heat, resulting in changes in entropy.

These processes are the building blocks that constitute quantum thermodynamic cycles, including the Carnot and Otto cycles discussed below. In most cases, these basic processes demonstrate a performance very similar to their classical counterparts. 
Much of their quantum behavior, however, becomes conspicuous when finite-time processes, non-equilibrium baths, and not fully thermalizing processes are considered, all of which will be the subject of detailed discussion in this review.  

\subsection{Quantum engines and refrigeration cycles }\label{subsec:Quantengrefr}
\subsubsection{Carnot cycle}\label{subsec:Carnotcycle}
\begin{figure}[h]
    \centering
    \includegraphics[scale=1.5]{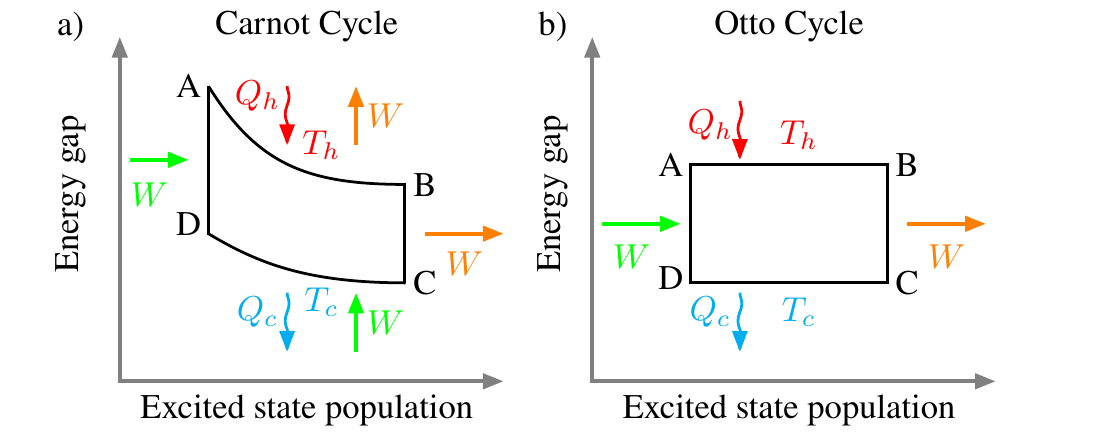}
    \caption{Schematic diagrams of the quantum Carnot cycle a) and the quantum Otto cycle b) for a TLS. Arrows pointing towards the branches indicate work performed on the TLS and heat flowing into the system from the bath. Arrows pointing outwards indicate work performed by the TLS. a) $\rm{A}\rightarrow \rm{B}$ and $\rm{C}\rightarrow \rm{D}$ are the isothermal strokes, and  $\rm{B}\rightarrow \rm{C}$ and $\rm{D}\rightarrow \rm{A}$  are the adiabatic ones. b) $\rm{A}\rightarrow \rm{B}$ and $\rm{C}\rightarrow \rm{D}$ are the isochoric strokes, and  $\rm{B}\rightarrow \rm{C}$ and $\rm{D}\rightarrow \rm{A}$  are the adiabatic ones. The plotted cycles demonstrate the operation of heat engines when the cycle goes clockwise, via reversal of the direction in which a refrigeration process can be realized. 
    }
    \label{fig1}
\end{figure}

To understand the basic operation of an ideal quantum Carnot cycle, here we consider the cycle of a heat engine with a two-level-system (\acrshort{tls}) working medium, which operates quasistatically between two heat baths at temperatures $T_{ h} > T_{ c}$.
The cyclic operation of the engine consists of two isothermal strokes and two adiabatic ones, as shown in Fig.~\ref{fig1}a.
In the \textit{first (isothermal) stroke} ($\rm{A}\rightarrow \rm{B}$), the TLS is coupled to a hot bath with temperature $T_{h}$, and the energy gap of the TLS is reduced from $\varepsilon_{A}$ to $\varepsilon_{B}$, while the temperature remains constant. This condition is permissible given the energy level changes on a time scale much larger than the system's relaxation time. 
 In this case, the TLS remains in equilibrium with temperature $T_{h}$ at all times, such that the population ratio of the exited and ground state is given by the Boltzmann factor $P_1/P_0=\exp\left[-\beta_{ h}  \varepsilon(t)\right]$, with  $\varepsilon(t)$ the instantaneous energy gap, and the inverse temperature  $\beta_{ h}=(k_{\rm B} T_{h})^{-1}$ where $k_{\rm B}$ is the Boltzmann constant.         
During this process heat flows into the TLS and work is done by the TLS as both the population and energy gap change simultaneously.

In the subsequent \textit{second (adiabatic) stroke} ($\rm{B}\rightarrow \rm{C}$), the TLS is decoupled from the heat bath and follows a unitary evolution according to some protocol that further reduces the gap from $\varepsilon_{B}$ to $\varepsilon_{C}$. 
In this case, the entropy of the system is preserved. When the system is driven quantum adiabatically, the population remains the same, implying that no excitation or heat is being generated, and work is further performed by the system. 
For the TLS, an effective temperature at the end of the stroke can be expressed as $k_{\rm B} T_{\rm eff}= \varepsilon_{C}\ln^{-1}(P_1/P_0)$. This point is crucial as, in the next stroke, we wish to couple the TLS to a cold bath with temperature $T_{ c}=T_{\rm eff}$, guaranteeing thermodynamic reversibility.   
We note in passing that, for a multilevel system, an effective temperature is achieved if all energy levels are shifted uniformly, such that the ratio between energy gaps at the beginning and end of the stroke is constant and equal to~$T_{h}/T_{ c}$~\cite{bender2000quantum,nori07}.  

In the \textit{third (isothermal) stroke} ($\rm{C}\rightarrow \rm{D}$), the TLS is kept in thermal equilibrium as the energy gap is enlarged to $\varepsilon_{D}$. During this process, heat is removed from the system into a cold bath with temperature $T_{c}$, while work is performed on the TLS. The cycle is completed once in the \textit{fourth (adiabtic) stroke} ($\rm{D}\rightarrow \rm{A}$) the system is brought back to its initial condition at point $(\rm{A})$ via a unitary evolution. In this stroke, work is performed on the system. 

The Carnot theorem of maximal efficiency stems from the fact that at the beginning of strokes ($\rm{A}\rightarrow \rm{B}$) and ($\rm{C}\rightarrow \rm{D}$), the system's temperature and that of the baths are equal. Heat is exchanged with the baths only when there is no temperature difference, leading to an efficient conversion of heat to work. The efficiency is calculated simply via
\beq
\label{eq:carnot_efficiency}
\eta_{\rm{C}}=\frac{|W|}{Q_{h}}=\frac{Q_{h}+Q_{c}}{Q_{h}}=1+\frac{T_{c}\Delta S_{c}}{T_{h}\Delta S_{h}}=1-\frac{T_{c}}{T_{h}}.
\eeq
Here $W$ is the total average work gain over a cycle, which is equal to the total energy exchanged with the baths.  The heat exchange with the hot and cold bath ${Q}_{h}=T_h \Delta S_h$ and $Q_{c}=T_c \Delta S_c$  respectively. Since the cycle is thermodynamically reversible  and the entropy is invariant in the adiabatic strokes, the change of entropy $\Delta S_{c(h)}$ in the two isothermal strokes is equal in magnitude but with the opposite sign. We defined $\Delta S_h=S(\rho_B)-S(\rho_A)$ and $\Delta S_c=S(\rho_D)-S(\rho_C)$, with $S(\rho_X)=-\tr[{\rho_X \ln(\rho_X)}]$ the von-Neumann entropy at point $X$.
The work in the isothermal stroke~($\rm{A}\rightarrow \rm{B}$) is then directly calculated from $W_{BA}=\Delta U_{BA}-T_h \Delta S_h$, which amounts to the free energy. Here,  $\Delta U_{BA}=\ave{H}_B-\ave{H}_A$ is the average change in energy. Similarly, we can calculate the work in the isothermal stroke~($\rm{D}\rightarrow \rm{C}$). In the second and last strokes, heat is not exchanged and the work is given by the energy difference at the end and beginning of the stroke. The total work over a cycle can be either calculated directly by summing all the contributions of the different strokes or from the first law of thermodynamics and the fact that over a cycle the change in energy is zero we have $W=-(Q_h+Q_c)=-(T_h-T_c) \Delta S_h$. 
For the TLS considered above, the net average work over a cycle is then given by
\beq
W = \eta_{\rm C} \left[ k_{\rm B} T_{h}\ln \left(\frac{Z(B)}{Z(A)}\right) + \varepsilon_B\tanh(\beta_h \varepsilon_B)-\varepsilon_A\tanh(\beta_h \varepsilon_A)   \right],
\eeq
with partition function  $Z(X)=\exp(\beta_h \varepsilon_X)+\exp(-\beta_h \varepsilon_X)$.
The analysis of the quantum Carnot cycle follows reasoning similar to that of its classical counterpart, and the cycle will exhibit the same efficiency. Yet other characteristics of the quantum engine (QE) performance, such as the work gained from the engine, will depend on the explicit details of the working medium, e.g., whether it is a TLS, harmonic oscillator, or particle in a box~\cite{geva1992classical,bender2000quantum,nori07}. 

The Carnot cycle describes an ideal thermodynamic reversible cycle. In a realistic cycle, however, i.e., finite-time thermodynamics, a dynamical picture~\cite{kosloff13} of the engine evolution and its dissipation should be adopted. When a weak coupling between the system and the bath is assumed, it is the custom to apply the Lindblad-Gorini-Kossakowski-Sudarshan (\acrshort{lgks}) Master equation approach to open quantum systems~\cite{lindblad1976generators,gorini276,alicki87}.
A dynamical treatment gives rise to new questions and emphasizes the role of quantum phenomena in the operation of the devices~\cite{levy2018book}. The finite-time operations in the  Carnot-type-cycles~\cite{geva1992quantum,kosloff2002discrete,ccakmak2020quantum,dann2020quantum} lead to processes of dephasing, loss of correlations, and quantum friction~\cite{feldmann2003quantum}, which become essential to optimizing the power output of the QHE, the cooling rate of the Quantum refrigerator (\acrshort{qr}), and the efficiency of these devices. Optimization of Carnot cycle performance was studied in different setups, including a single particle in a potential well~\cite{abe2012role}. 

\subsubsection{Otto cycle}\label{subsubsec:Ottocycle}
The quantum analog of the  Otto cycle has become the canonical model for exploring quantum engines and refrigerators, mainly due to its relatively simple analysis. The separation of heat and work strokes in the cycle makes it easier to define these magnitudes in the quantum regime, especially when considering finite-time cycles.

In the Otto cycle, the two isothermal strokes are replaced by isochoric analog strokes, as shown in Fig.~\ref{fig1}b.
In these strokes, only heat is exchanged between the working medium and the bath. 
In stroke ($\rm{A}\rightarrow \rm{B}$), the energy gap of the TLS, $\Delta_h$ remains invariant, while the exited state occupation is increasing from $n_c$ to $n_h$, where $n_x=(1+e^{\beta_x\Delta_x})^{-1}$ for $x=h,c$, and the TLS is thermalized at the bath temperature $T_h$. This thermalization stage is associated with heat absorbed by the TLS and equals $Q_h=\Delta_h(n_h-n_c)$. 
The stroke ($\rm{C}\rightarrow \rm{D}$) describes the opposite population transition from $n_h$ to $n_c$. However, since this transition accrues in a smaller energy gap $\Delta_c < \Delta_h$, less heat is transferred to the cold bath, $Q_c=\Delta_c(n_c-n_h)$.
Here the sign of the heat determines the direction of heat flow. Positive heat implies that heat is absorbed by the working medium, whereas negative heat implies that it is released.    
The work output over a closed cycle of the quantum Otto engine can be easily calculated according to the first law of thermodynamics
\beq
W=-(Q_h+Q_c)=-(\Delta_h-\Delta_c)(n_h-n_c).
\eeq
Negative work implies that energy is extracted from the working medium.
The efficiency of the Otto engine,
\beq
\eta_O=\frac{|W|}{Q_h}=1-\frac{\Delta_c}{\Delta_h}\leq 1-\frac{T_c}{T_h} = \eta_{\rm C}, 
\eeq
is bounded by the Carnot efficiency, as can be immediately observed from the condition for the cycle to operate as an engine,  $n_c < n_h \Rightarrow (1+e^{\beta_c\Delta_c})^{-1}<(1+e^{\beta_h\Delta_h})^{-1}$  which boils down to the condition on the compression ratio and the temperature ratio $1<\Delta_h/\Delta_c < T_h/T_c$.

\begin{figure}%
    \centering
    \includegraphics[width=7.5cm]{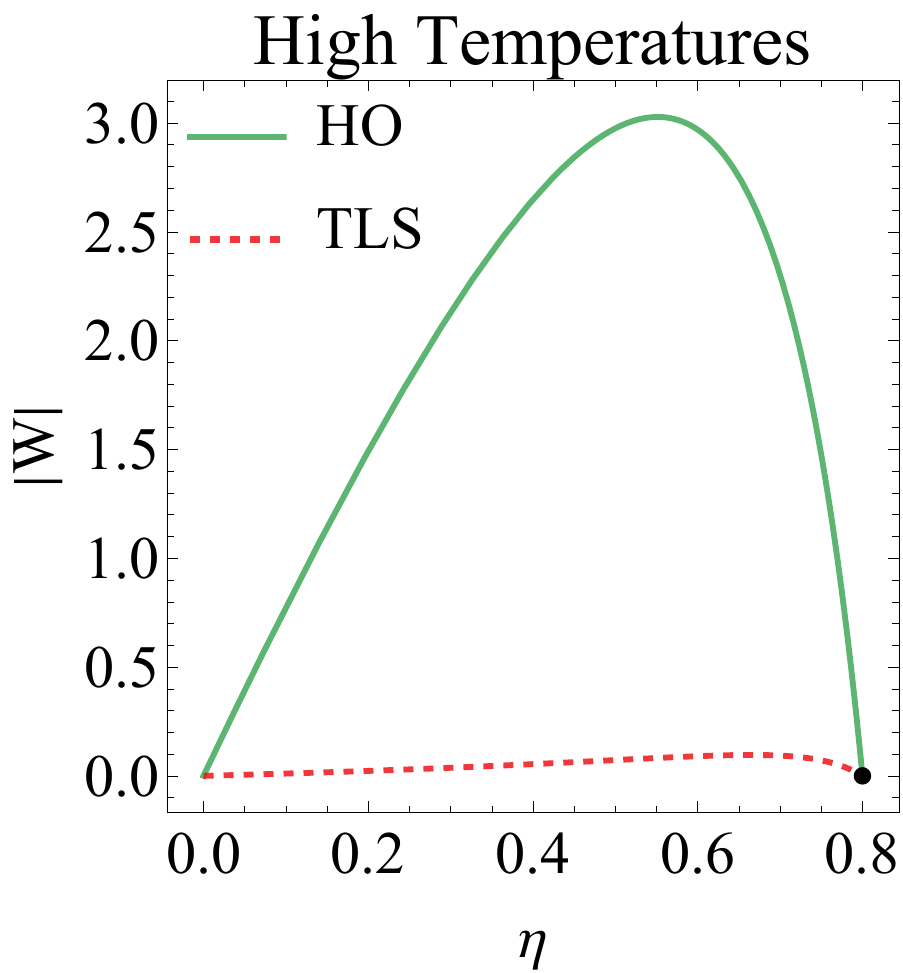} %
    \qquad
    \includegraphics[width=7.5cm]{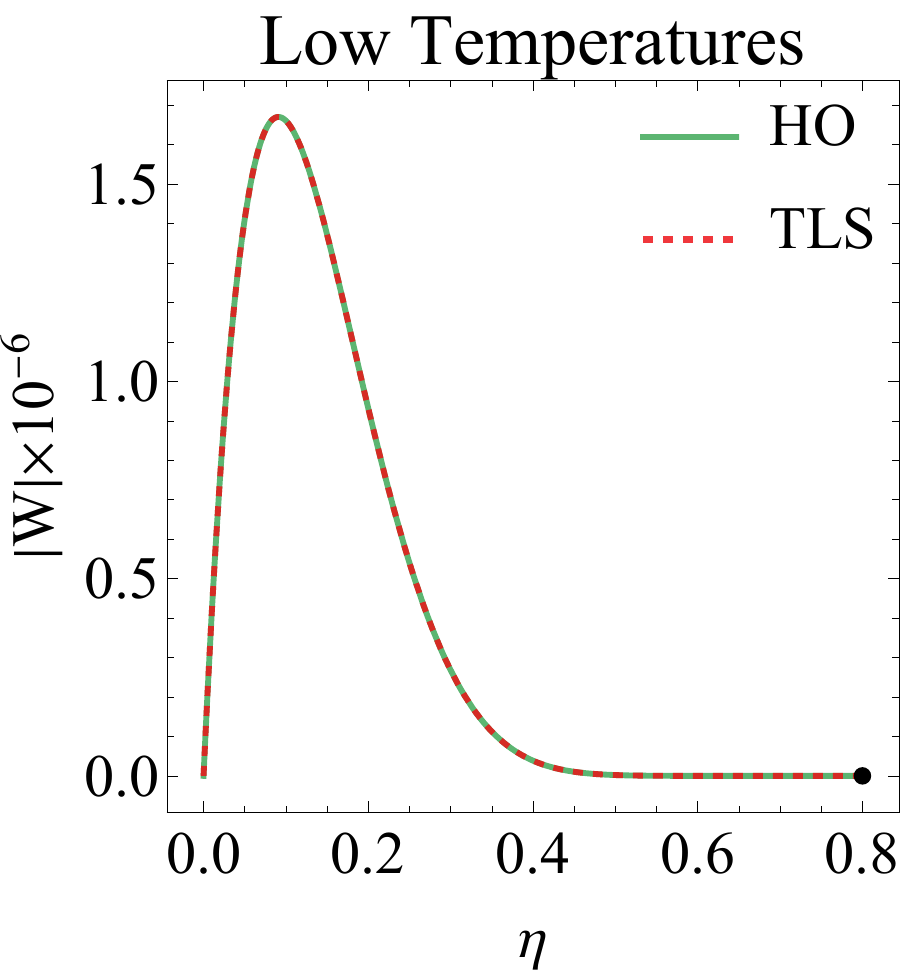} %
    \caption{ The Work gained from the Otto engine over one cycle vs. the engine's efficiency at high- and low- temperatures in arbitrary units. The results for the harmonic oscillator (HO) are plotted in the solid green lines and those of the two-level-system (TLS) in dashed red lines, the black point marks the Carnot efficiency. At this point, the amount of work extracted over a cycle vanishes. Parameters: $\omega_c = 1$ and we vary $\omega_h=\Delta_h$ in the interval $\omega_h =\Delta_h \in \left[ \omega_c , 5  \omega_c \right]$. At high temperature $T_h=10\omega_c$, $T_c=2\omega_c$ and at low temperatures $T_h=0.1\omega_c$, $T_c=0.02\omega_c$. }%
    \label{fig:WvsEta}%
\end{figure}

When the working medium is considered to be a single Quantum Harmonic Oscillator (\acrshort{ho}), the efficiency of the Otto engine will be similar to that of the TLS working medium; however, a difference is observed in the work output
\beq
\eta_O^{\rm HO}=1-\frac{\omega_c}{\omega_h}, \; \; \; \; \; W^{\rm HO}=-(\omega_h-\omega_c)(n_h^{\rm HO}-n_c^{\rm HO}).
\eeq
Here $\omega_h$ and $\omega_c$ are the HO frequency at the end of strokes ($\rm{D}\rightarrow \rm{A}$) and ($\rm{B}\rightarrow \rm{C}$) respectively. Additionally, in this case, the ``compression'' ratio, i.e., the ratio between the energy gaps, determines the efficiency. The work, on the other hand, is governed by the thermal occupation, $n_x^{\rm HO}=(e^{\beta_x\omega_x}-1)^{-1}$, which exceeds that of the TLS for a given energy gap. 
This is observed in Fig.~\ref{fig:WvsEta} where we compare the work and efficiency of the HO and that of the TLS at high and low temperatures. At low temperatures, the behavior of the TLS is similar to that of the HO as in this limit $n_x^{\rm HO} \approx n_x^{\rm TLS}$, meaning only the first excited state plays a role in the work extraction process. On the other hand, in the high-temperature limit $n_x^{\rm HO} \gg n_x^{\rm TLS}$, thus more levels of the HO participate in the energy transformation.
This result is also manifested in other types of engines, including those with finite-time operation, and becomes significant when optimizing the efficiency at maximum power~\cite{levy14a}.

For a comparative study of the Carnot and Otto engines operating in the quasistatic limit, see Ref.\cite{nori07}. We note in passing that the cycles described in Figs.~\ref{fig1} can be reversed and operate as a refrigerator, in which heat is absorbed from the cold bath and is transferred to the hot one. This reversal occurs when the cycle is completed anticlockwise.

Much as in the Carnot QHE, in the Otto engine, quantum phenomena emerge when finite-time processes are considered and a dynamical picture is  adopted~\cite{rezek2006irreversible,gelbwaser13,kosloff17entropy,dann2020finite}.  In addition, a dynamical picture is pivotal to address many relevant issues regarding the optimization of the engine performance \cite{allahverdyanwork2008,allahverdyanopti2010}. As it will be clear in the following sections, optimization is not an easy task. Apart from requiring a high level of control of the dynamics of the working medium \cite{gelbwaser2015thermodynamics,koch2022quantum,zhang2022dynamical}, it involves the optimization of many engine's features at the same time, e.g., the output power, the efficiency, and output power fluctuations (see Sec.~\ref{sec:fluctuations}). Quite recently, one aspect of this multifaceted problem has been addressed by designing optimal cycles for quantum engines, where the engine strokes can be designed in the slow and in the fast driving regime using analytical \cite{Abiusooptimal2020,Cavinamaximum21,terrenalonso2022}, and machine learning approaches \cite{erdman2022_machine_learning,erdman2023pareto}. 

\subsection{Continuous and autonomous thermal machines}
\label{subsec:Enginesintro}
Thermodynamic analysis of heat engines is usually based on a cyclic thermodynamic process, in which the system returns to its initial state at the end of the last stroke.
However, other engine types that are not stroke-based can provide a thermodynamic description for many processes and devices in biology, chemistry, and physics. 
These are sometimes referred to as continuous engines,~i.e.~, a working medium simultaneously coupled to all heat baths and work sources. Ref.~\cite{levy14a} provides an extended review of the basic operation of continuous quantum engines and refrigerators.

\subsubsection{The tree-level maser  quantum heat engine: Static and dynamical pictures}
\label{subsec:maser}
\begin{figure}[h]
    \centering
    \includegraphics[scale=1.5]{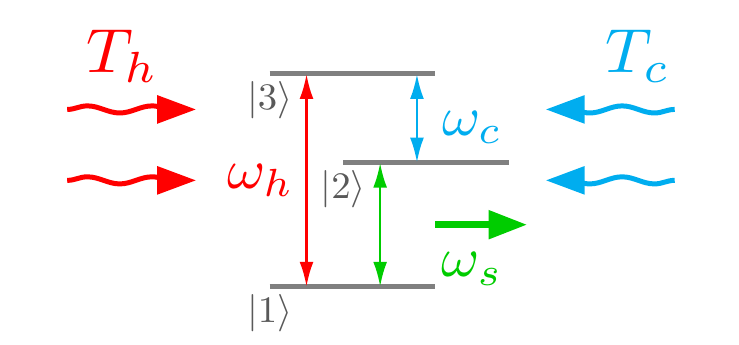}
    \caption{Schematic diagram of a three-level Maser QHE. Hot (cold) reservoirs are modeled with thermal light sources at temperatures $T_{h} (T_{c})$.~Filter waveguides allow for the absorption of hot (cold) photons of frequency $\omega_{h}$($\omega_{c}$).~The population inversion takes place between levels 1 and 2. Work is performed by stimulated emission of radiation of frequency $\omega_{s}$. 
    }
	\label{fig:three_level}
\end{figure}

The earliest thermal QHE proposal, dating to 1959~\cite{Scovil:1959}, can be considered a continuous operating engine and is based on the maser (i.e., microwave amplification by stimulated emission
of radiation) mechanism (see Fig.~\ref{fig:three_level}).
A three-level quantum system acts as a working fluid, while two sources of thermal light set at different temperatures $T_{ h}$ and $T_{c}$ play the role of the heat baths.
The working fluid interacts with the baths by means of filter waveguides that only let photons of frequencies $\omega_{h}=\omega_{3}-\omega_{1}$ and $ \omega_{c}=\omega_{3}-\omega_{2}$ pass.
As a consequence, the working fluid absorbs energy $\hbar\omega_{ h}$ from the hot bath and releases part of it, namely $\hbar\omega_{c}$, to the cold bath.
Work can be produced via the stimulated emission of photons of frequency $\omega_{s}=\omega_{h}-\omega_{c}$, if, as required by the maser operating conditions, a population inversion between levels $\ket{2}$ and $\ket{1}$ is achieved.
If the engine efficiency is assumed to be $\eta_{\rm M}=\omega_{s}/\omega_{h}$ and $p_{i}$ denotes the equilibrium population of the $i$-level, it follows from the population-inversion condition $p_{2}/p_{1}> 1$ that $\eta_{\rm M} <\eta_{\rm C} = 1-T_{c}/T_{h}$,~i.e., the efficiency of the maser QHE is Carnot bounded~\cite{Scovil:1959,geusic67}. 
 This can be observed using the Boltzmann factor to write the population ratio $p_2/p_1=(p_2/p_3)(p_3/p_1)=\exp{(\hbar\omega_c/k_B T_c)} \exp{(-\hbar\omega_h/k_B T_h)}$ and then express it using the efficiencies $p_2/p_1=\exp{[(\hbar \omega_s/k_B T_c)(\eta_{\rm C}/ \eta_{\rm M}-1)]}$.
We note that the maser efficiency resembles that of the Otto cycle discussed in the previous section.

In the original paper~\cite{Scovil:1959}, no dynamical picture was associated with the engine analysis. 
Here we provide a simple dynamical picture \cite{kosloff13,levy212} that will recover the effect on the maser efficiency but will also provide further insight into and information on the energy flows between the different baths. Our description is based on the LGKS master equation for the reduced three-level system interacting with three baths.
As described above, the hot and cold baths with temperature $T_h$ and $T_c$ couple the levels $\ket{1}\leftrightarrow \ket{3}$ and $\ket{2}\leftrightarrow \ket{3}$ respectively. 
The work bath is assumed to have an infinite temperature, and thus no entropy flow is associated with the energy flow from the system to the bath~\cite{geusic67}. 
The LGKS master equation inherently mimics the waveguide frequency filtering, by assuming a weak system-bath coupling limit that emphasizes the resonant frequency.  
The master equation for the three-level working medium reads,

\beq
\frac{d\rho}{dt}=-i[H,\rho]+\mathcal{D}_h(\rho)+\mathcal{D}_c(\rho)+\mathcal{D}_s(\rho),
\eeq
with the Hamiltonian $H=\sum_{i=1}^3\hbar\omega_i \ketbra{i}{i}$ and the thermelizing disspators,
\beq
\mathcal{D}_x (\rho) = \gamma_x \left(A_x \rho A_x^{\dagger}\ -\frac{1}{2}\{A^{\dagger}_x A_x,\rho\} \right)+\gamma_x e^{-\beta_x\hbar\omega_x}\left(A_x^{\dagger}\rho A_x -\frac{1}{2}\{A_x A_x^{\dagger},\rho\} \right)
\label{eq:D}
\eeq
for $x=h,c,w$, the hot, cold and work bath.
Here $\gamma_x$ are relaxation rates that in the weak system-bath coupling limit is given by the Fourier transform of the $x$-bath correlation functions~\cite{breuer}, and $\beta_x=(k_{\rm B} T_x)^{-1}$ is the inverse temperature of the $x$-bath, which implies $\beta_w=0$ for the work bath.  
 The Boltzmann factor multiplying the second term on the right-hand side of Eq.~(\ref{eq:D}) guarantees detail balance is obeyed.
The system operators $A_x$ can be derived in the weak system-bath coupling limit and are identified as $A_h=\ketbra{1}{3}$, $A_c=\ketbra{2}{3}$, and $A_w=\ketbra{1}{2}$.  

The energy flow, l $J_x$, between the system and  the $x$-baths can be evaluated according to $J_x=\ave{\mathcal{D}_x^{\dagger}(H)}\equiv \tr{[\rho \mathcal{D}^{\dagger}(H)]}$~\cite{spohn78,levy212},  where $\mathcal{D}^{\dagger}$ is the adjoint of $\mathcal{D}$ operating on arbitrary system operator $O$, such that 
\begin{equation}
 \mathcal{D}^{\dagger}_x (O) = \gamma_x \left(A_x^{\dagger} O A_x\ -\frac{1}{2}\{A^{\dagger}_x A_x,O \} \right)+\gamma_x e^{-\beta_x\hbar\omega_x}\left(A_x O A_x^{\dagger} - \frac{1}{2}\{A_x A_x^{\dagger},O\} \right).  
 \label{eq:Ddag}
\end{equation}

\begin{figure}%
    \centering
    \includegraphics[width=8cm]{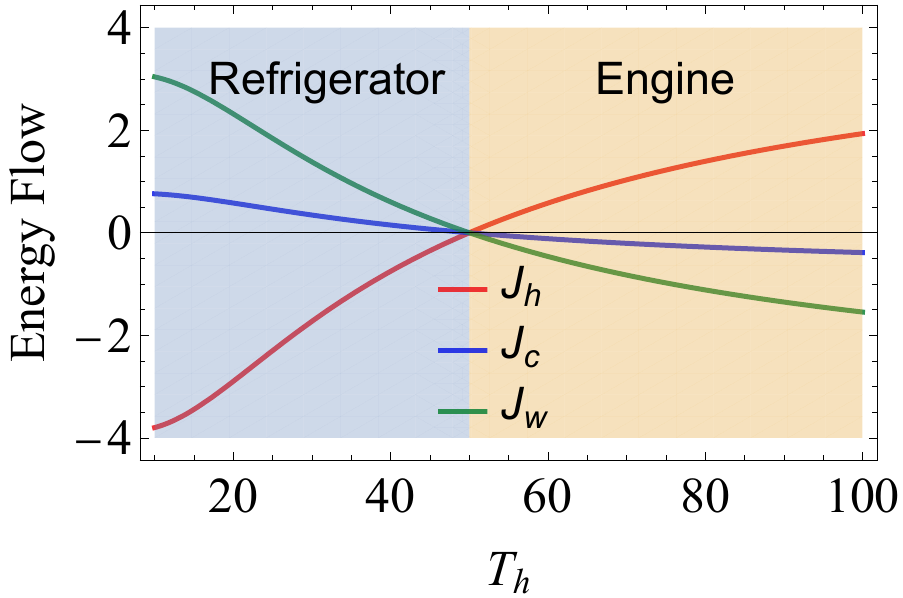} %
    \caption{ Steady state energy flow between the system and the reservoirs as a function of the temperature of the hot bath $T_h$, provided by Eq.~(\ref{eq:heatflow}). We observe two regimes of operation determined by the corporation ratio and the bath temperature ratio. When $\omega_c/\omega_h < T_c/T_h$ the machine operates as a heat engine, whereas, in the opposite region, it operates as a refrigerator. Parameters: $\gamma_h=\gamma_c=\gamma_w=\gamma, \omega_h=50\gamma, \omega_c=10\gamma,$ and $ T_c=10\gamma $.     }%
    \label{fig:energyflow}%
\end{figure}

Inserting the Hamiltonian to Eq.~(\ref{eq:Ddag}) for the different $x$ we arrive at
\begin{eqnarray}
\label{eq:heatflow}
J_h &=& \hbar \omega_h \gamma_h \left( e^{-\beta_h \hbar\omega_h}p_1 - p_3 \right),\nonumber\\
J_c &=& \hbar \omega_c \gamma_c \left( e^{-\beta_c \hbar\omega_c}p_2 - p_3 \right),\nonumber\\
J_w &\equiv& P = \hbar (\omega_c -\omega_h)\gamma_w \left( p_2 - p_1 \right),
\end{eqnarray}
where $P<0$ is the power supplied by the engine (see Fig.~\ref{fig:energyflow}).

At a steady state, the equation of motion for the population $p_i$ reduces to
\begin{eqnarray}
   \frac{dp_1}{dt} &=& \gamma_h p_3 +\gamma_w p_2-(\gamma_w+ e^{-\beta_h \hbar \omega_h}\gamma_h)p_1=0, \nonumber \\
   \frac{dp_2}{dt} &=& \gamma_c p_3 +\gamma_w p_1-(\gamma_w + e^{-\beta_c \hbar \omega_c}\gamma_c)p_2=0, \nonumber \\
   \frac{dp_3}{dt} &=& e^{-\beta_h \hbar \omega_h}\gamma_hp_1+e^{-\beta_c \hbar \omega_c}\gamma_cp_2 -(\gamma_h+\gamma_c) p_3  =0, \nonumber \\
\end{eqnarray}
which is a simple set of linear equations that can be solved analytically provided the constraint $p_1+p_2+p_3=1$. In doing so, it follows immediately that the efficiency is $\eta_{\rm M}=-P/J_h=1-\omega_c/\omega_h$.

 A more recent proposal of the autonomous maser heat engine involves a working medium made of a two-level system and an off-resonant microwave mode of a given frequency \cite{ghosh2018two}. Contrary to~\cite{Scovil:1959}, this autonomous engine does not require population inversion.

Exploring the thermodynamics of continuously operating quantum devices such as engines and refrigerators is very useful, as many quantum processes in physics, biology, and chemistry fall into this category of mechanisms. Engines and refrigerators provide a common language and set of tools and bounds on the energetic, entropic, and efficiency of these quantum processes.
The continuous models can be extended far beyond the simple three-level QHE described above, to include phenomena induced by many-body working media, time-dependent external driving, multi-terminal machines, and more,
all of which will be reviewed in the following chapters.

\subsubsection{Quantum absorption refrigerators}\label{subsubsec:absopt}

A further example of autonomous, multi-terminal devices, quantum absorption refrigerators (\acrshort{qar}) have gained widespread attention over the last decade~\cite{mitchison2019quantum}.
Starting from early theoretical proposals~\cite{palao2001quantum,linden10,levy12,levy212,mari2012,levy312}, a great amount of theoretical work has been done to elucidate the  working mechanism of this engine~\cite{palao13,Nimmrichter2017quantumclassical}, as well as to investigate the possibility of quantum advantage in its performance~\cite{correa2014quantum,levy14a} due to coherence and entanglement.
Since the majority of these developments were extensively reviewed in a recent work~\cite{mitchison2019quantum}, in what follows we limit ourselves to a brief overview of this kind of engine's working principles, focusing on more recent theoretical and experimental proposals. 

\begin{figure}[h]
		\centering
		\includegraphics[scale=0.40]{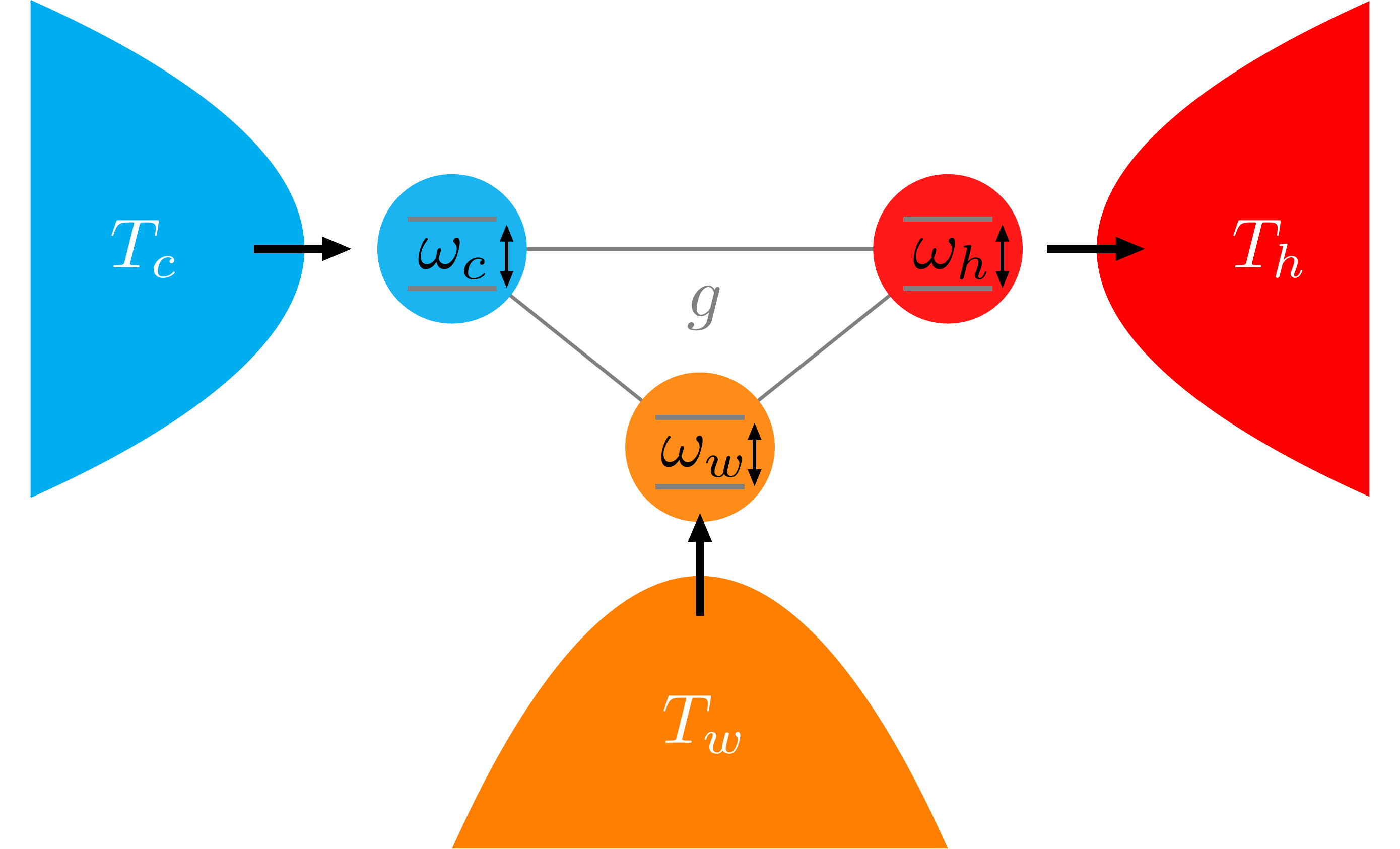}
		\caption{Sketch of a quantum absorption refrigerator's basic setup, with a three-body working medium. In the typical setting,  quantum harmonic oscillators as well as  two-level systems can be used to model the working medium. Reproduced from \cite{mitchison2019quantum}.}
		\label{fig:thermoel10} 
\end{figure}

A quantum absorption refrigerator works with three thermal reservoirs. It achieves cooling by removing heat from the cold ($c$) reservoir and dumping part of it into the hot ($h$) one.
While conventional refrigerator schemes utilize work provided from the outside -- by driving either the working medium or its coupling to the reservoir (see Sec.~\ref{sec:nonMarkov}) -- here the energy required to achieve cooling is provided by a third thermal reservoir, i.e., the work reservoir ($w$).

In its simplest setting, the working medium is composed of a three-level system, similar to a maser heat engine discussed in the previous section, operated in reverse mode. 
However, more realistic schemes employ three qubits, modeled as usual by means of two-level systems, each coupled to a different reservoir.
In addition, different settings have also been proposed that are based on three quantum harmonic oscillators~\cite{levy14a}, as well as two quantum harmonic oscillators~\cite{levy12} coupled via optomechanical interactions \cite{naseem2018thermodynamic,naseem2020two,bhandari2021minimal}.
In what follows, we focus on the three-qubit setting as depicted in Fig.~\ref{fig:thermoel10}.
The qubits are labeled as $h,c,w$, depending on the reservoir to which they are coupled.
Moreover, they interact with each other via a trilinear coupling Hamiltonian, such that the whole Hamiltonian of the working medium takes the form 
\beq\label{eq:thermoel21}
	H_{\text{wm}}=H_{0}+H_{\text{int}}= \sum_{i=h,c,w}\omega_{i}\sigma^{+}_{i}\sigma^{-}_{i} + g(\sigma^{-}_{c}\sigma^{+}_{h}\sigma^{-}_{w} + \sigma^{+}_{c}\sigma^{-}_{h}\sigma^{+}_{w}).
\eeq
Although more recent schemes involve three qubits interacting with two-body interactions~\cite{hewgill2020three}, below we focus on the conventional, trilinear interaction setting. Here, as in the maser engine, the bare qubit frequencies are chosen such that $\omega_{w}=\omega_{h}-\omega_{c}$.
With the ground and excited states of each qubit denoted with $\{\ket{0}\ket{1}\}$  respectively, the interaction Hamiltonian of the form~\eqref{eq:thermoel21} couples only two states belonging to the three qubits' Hilbert space, namely $\{\ket{0}_{c}\ket{1}_{h}\ket{0}_{w},\ket{1}_{c}\ket{0}_{h}\ket{1}_{w}\}$, which are degenerate with respect to $H_{0}$. 
It follows that excitation in the cold qubit can be coherently destroyed and created in the hot qubit, provided that the work qubit is in its excited state. Thus, it is possible to drive the cold qubit to its ground state by employing the energy coming from the work qubit. 

As each qubit is weakly coupled to its bath, cooling of the cold bath can, in principle, be achieved. However, whether the present setup can work as a cooler depends on the detailed study of its dynamics, when it is considered as an open quantum system~\cite{levy14a}.
A rather intuitive argument for finding out the temperature regime under which cooling is achieved can be found in Eq.~\eqref{eq:thermoel21}, a swap interaction between the cold qubit and a virtual two-level system defined from states $\{\ket{-}_{v}=\ket{0}_{h}\ket{1}_{w},\ket{+}_{v}=\ket{1}_{h}\ket{0}_{w}\}$ of energies $\omega^{-}_{v}=\omega_{h}-\omega_{c},$ and $\omega^{+}_{v}=\omega_{h}$.
In the absence of interaction among qubits, if the hot and work qubits are found in thermal states with inverse temperatures $\beta_{h},\beta_{w}$, the populations of the virtual states fulfill $p_{+v}/p_{-v}=\exp(-\beta_{v}\omega_c)$, where $\beta_{v}$ is the effective inverse temperature of the thermal state that reads $\beta_v=(1/\omega_c)(\beta_h \omega_h -\beta_w \omega_w)$. As a consequence, if $T_{c} > 1/\beta_{v}$, an excess of population in the excited state of the cold qubit is present with respect to the virtual one.~Once the swap interaction is switched on, the excited state of the cold qubit will start to deplete, i.e., the transition $\ket{1}_{c}\ket{-}_{v}\rightarrow \ket{0}_{c}\ket{+}_{v}$ increasingly takes place.
In the limit of weak qubit-bath coupling, thermal excitations are progressively removed from the cold bath and transferred to the hot bath. A nonequilibrium stationary state thus builds up, in which a net heat current (the cooling power) flows between the cold and the hot bath.   

Several theoretical proposals have been put forward that model quantum absorption refrigerators in superconducting circuits~\cite{hofer2016autonomous}, quantum dots~\cite{Venturelli3,erdman2018absorption,manikandan2020autonomous}, and cavity QED systems~\cite{Mitchinson:Qthermo2018,mitchison2016realising}. In~\cite{maslennikov2019quantum}, a quantum absorption refrigerator was experimentally realized via a trapped ion setup. An rf-driven Paul trap is used to confine the $^{+}\text{Yb}$ ions along a given spatial direction. 
 Here three ions are subject to a confining three-dimensional quadratic potential, they thus undergo small oscillations around their equilibrium positions. The normal modes of oscillation as well as the ion positions can be controlled by means of the trap electrodes.  However, the ions also exchange Coulomb interactions, so that the modes couple to each others~\cite{marquet2003phonon}. The trap employs two radial modes and one axial mode of frequencies $(\omega_x,\omega_y,\omega_z)$ of the order of MHz. Three ionic modes of frequencies $\omega_{h}=\sqrt{29/5}\omega_{z}$, $\omega_{w}=\sqrt{\omega^2_{x}-\omega^2_{z}}$, $\omega_{c}=\sqrt{\omega^2_{x}-(12/5)\omega^2_{z}}$ are thus selected to simulate the energies of the three-body Hamiltonian. The radial trap frequencies are tuned in order to fulfill $\omega_{h}=\omega_{w} + \omega_{c}$, a necessary condition for refrigeration. With the aid of Raman laser beams, optical dipole forces and spin-motion coupling are applied to the ions, so that they are cooled down and forced to follow controlled motional states. The three qubits in the Hamiltonian \eqref{eq:thermoel21} are thus replaced by ionic normal modes of oscillation ($\omega_{c},\omega_{h},\omega_{w}$).
Thus, the platform simulates the quantum dynamics of a trilinear Hamiltonian of the form $H=\xi( a^{\dagger}_{c}a_{h}a^{\dagger}_{w} +a_{c}a^{\dagger}_{h}a_{w})$, where  $a_i$ are bosonic ladder operators. The coupling strength, $\xi$, is related to the trap parameters and the equilibrium distance between ions aligned in the trap. 

More recently, quantum refrigeration effects were achieved employing novel experimental approaches based on indefinite causal orders \cite{rubino2017experimental,Nieindefinite2022}.

\subsection{Quantum effects and signatures}
The study of quantum processes and devices in terms of engine operation provides a means of relating quantum phenomena to energetic and entropic currents and the efficiency of different energy and quantum resource conversions (see also the book chapter~\cite{levy2018book} and references therein for a short review on the manifestation of quantum phenomena in quantum thermal devices).
The operation of an engine in the quantum regime is affected by various quantum features. The discrepancy between the classical and quantum operation of engines can already be traced to the quantization of the energy levels and to the uncertainty principle~\cite{gelbwaser2017single},  also in the relativistic regime~\cite{chattopadhyay2019relativistic}.
Other quantum effects such as coherence and quantum correlations have far-reaching consequences on engine operation and will be discussed at length in this review.  For instance, quantum coherence can affect the performance of an engine run over many cycles~\cite{watanabe2017}.

Beyond understanding the relationship between thermodynamics in the quantum regime and quantum phenomena, one of the greatest challenges in this field is identifying when these phenomena impact the process.
The customary manner of approaching this problem is revealing quantum signatures that witness the presence of quantum effects.
In the context of quantum engines, these signatures can be expressed via classical bounds on measurable thermodynamic quantities, such as heat and work, i.e., violation of these bounds is possible only in the presence of quantum effects. Such a bound on the power output of QHEs was introduced in~\cite{uzdin15} and was experimentally demonstrated in a nitrogen-vacancy (NV) center in a diamond setup~\cite{klatzow2019experimental}.
Other quantum thermodynamic signatures have been revealed and demonstrated in heat exchange processes~\cite{levy2020quasiprobability} and work protocols~\cite{hernandez2022experimental}.


\section {Fluctuations and thermodynamic uncertainty relations in  quantum heat engines
}\label{sec:fluctuations}

\subsection{The significance of thermodynamic fluctuations in small thermal machines }

The variation of heat and work in macroscopic heat engines becomes  insignificant from one cycle to another when such engines operate in the limit cycle. Moreover, in the thermodynamic limit of large systems, fluctuations vanish and the system can be fully characterized  by thermodynamic averages (excluding  critical points and phase transitions).  
As the size of the  thermal machines is scaled down to mesoscopic and quantum length scales, due to the finite nature of the working substance (single, few, or countably finite n-particle systems),  thermodynamic quantities like heat and work become truly stochastic with a distribution characterized by the average, the variance and higher order moments. Thus, evaluating the fluctuations of these thermodynamic quantities becomes essential, as they affect the different efficiency and output parameters that characterize a heat engine.

Fluctuation-dissipation relations for equilibrium  \cite{kubo1966, RevModPhys.81.1665,marconi2008fluctuation} and nonequilibrium steady-state \cite{agarwal1972fluctuation,bochkov1981nonlinear,seifert2010fluctuation} problems in classical and quantum \cite{konopik2019quantum,levy2021response} systems have been studied for quite some time. 
Over the last few decades, such fluctuation relations  have been extended to systems evolving far-from-equilibrium and are known as Fluctuation Theorems (\acrshort{fts}).
Broadly speaking, FTs relate the probability distributions of the forward and reversed nonequilibrium processes of some fluctuating quantity. 
Classically, these theorems are relevant to biomolecules, molecular motors, colloidal particles, etc. \cite{seifert2012stochastic}. 
In the quantum regime, they are applicable to, and experimentally observed in, a wide variety of quantum devices, such as trapped ions \cite{an2015experimental}, superconducting qubits \cite{pekola2015towards}, quantum dots \cite{RevModPhys.81.1665,campisi2011colloquium}, \acrshort{nmr} setups \cite{batalhao2014experimental}, and \acrshort{nv} centers in diamond \cite{hernandez2020experimental,gomez2021experimental}.
 
Fluctuation theorems reveal fundamental aspects of entropy production, irreversible work, heat, and matter transport for both open and closed systems, and thus bear important consequences for scaled-down heat engines.
The predominant relations include  Crooks FT~\cite{crooks1999entropy}, which generalizes the Jarzynski equality~\cite{jarzynski1997nonequilibrium} and relates work fluctuations of nonequilibrium processes to the difference in free energy between the initial and final states at equilibrium; FT of heat exchange between thermal states  \cite{jarzynski2004classical}; and steady-state FT of entropy production~\cite{evans1993probability}. Many of the classical FTs have been extended and derived for quantum systems~\cite{RevModPhys.81.1665,campisi2011colloquium,PRXQuantum.1.010309}.
Specifically, such relations characterize the functioning of heat engines~\cite{sinitsyn2011} and have important implications for thermal devices functioning in the quantum domain~\cite{PhysRevE.103.032130,campisi2014fluctuation,campisi2015nonequilibrium}. 

\subsection {Fluctuations of quantum heat engines: A Two-Point-Measurement-based study}
\label{sec:fluc_QHE}
Unlike classical macroscopic heat engines, QHEs are crucially affected by the tussle between non-zero quantum and thermal fluctuations, and, hence, evaluating them is critical to understanding thermal devices at the mesoscopic and quantum scale. 
To evaluate thermodynamic quantities like work and heat, the standard phenomenological apparatus in Quantum Thermodynamics is the Two-Point-Measurement (\acrshort{tpm}) scheme~\cite{PhysRevE.75.050102}.
In this scheme, two projective measurements of an observable of interest are performed on the system, one at the beginning and the other at the end of the dynamical process. This scheme can provide, for example, the energy difference statistics of the process. Yet one should keep in mind that the first measurement collapses the initial state on an eigenstate of the measured observable, and may thus interfere with the statistics.

The device-efficiency fluctuations can be evaluated by calculating the joint probability distribution for heat and work on the different strokes \cite{PhysRevResearch.2.032062}. Most of the studies have focused on the quantum Otto cycle \cite{saryal2021bounds,jiao2021quantum}, in order to avoid ambiguity in distinguishing between heat and work.  
With the Otto engine, for example, the work distribution of a certain work stroke is given by  
\begin{equation}
P(W) = \sum _ {n,m} \delta \left[W - (E_{m}^{\tau}-E_{n}^{0})\right] P_{n \rightarrow m}^{\tau} P_{n}^{0} (\beta),
\end{equation}
where $E_m^0$ and $E_m^{\tau}$ are the energy eigenvalues at the beginning and at the end of a stroke of duration $\tau$ respectively. $P_{n \rightarrow m}^{\tau}=|\langle n|U(\tau)|m \rangle|^{2}$ are the transition probabilities, and  $P_{n}^0(\beta)=\exp{(-\beta E_n^0)}/Z^0(\beta)$ are the initial thermal occupations. 
At the next stroke, the conditional probability for a heat outcome $Q$ given the work $W$ is
\begin{equation}
P(Q|W) = \sum _ {k,l} \delta \left[Q - (E_{l}^{\tau}-E_{k}^{\tau}) \right] P_{k \rightarrow l}^{\tau'} P_{k}^{\tau},
\end{equation}
with $p_k^{\tau}=\delta_{km}$ the occupation after the projective measurement at time $\tau$ on the energy eigenstate $\ket{m}$, and $P_{k \rightarrow l}^{\tau'}=p_{k}^{\tau}(\beta')$ the thermal occupation at the end of the stroke. In this case, the energy levels are those projected at the beginning of the stroke at time $\tau$, and the inverse temperature $\beta'$ is that of the thermalizing stroke.  

In a similar manner, one can construct the probability of a certain work/heat stroke conditioned on previous measurements outcomes and apply the chain rule to determine any joint probability. 
For example, an evaluation of the engine-efficiency fluctuations $\eta=\frac{W_{\rm in}+W_{\rm out}}{Q_h}$ can be carried out  using the joint probability $P(W_{\rm in}, Q_{h},W_{\rm out} )~=~P(W_{\rm out}|Q_{h},W_{\rm in})P(Q_{h}|W_{\rm in})P(W_{\rm in})$, for the work done on the system $W_{\rm in}$, with the system  $W_{\rm out}$ and the heat entering the system $Q_h$.
The efficiency distribution then reads
\begin{equation}\label{eq:fluctuations5}
P(\eta) = \int dW_{\rm in}dQ_{h}dW_{\rm out} P(W_{\rm in}, Q_{h},W_{\rm out} ) \delta \left( \eta - \frac {W_{\rm in} + W_{\rm out}}{Q_{h}}\right).
\end{equation}

Experimental studies have recovered the above distributions in several setups of QHEs. In~\cite{denzler2021nonequilibrium}, the joint distribution of heat and work and the efficiency distribution were reported in an NMR-based quantum Otto engine. On the other hand, in~\cite{solfanelli2021experimental}, the joint probability for the energy exchanged in a two-stroke engine was measured on an IBM quantum computer platform, verifying the heat-engine fluctuation relation.

Power fluctuations in miniaturized quantum engines have been studied in the same framework, with an emphasis on exploring the finiteness of the working medium's Hilbert space and the strict degeneracy of energy levels~\cite{denzler2021power}. In the quasistatic limit, all the moments of work distribution $\langle W^{n} \rangle~=~\int dW P(W) W^{n}$ can be calculated via a joint p.d.f,~$P(W)$, for the full cycle of a Quantum Carnot Engine, derived on similar lines as Eq.~\eqref{eq:fluctuations5}.
 This allows the study of the relative magnitude of the work fluctuations as compared with its average value. As a measure of the latter, the coefficient of variation $\sigma_{w}/|W|$ can be employed, where $\sigma_{w}$ is the standard deviation related to the work distribution and $W$ is the average work. This ratio can be optimized with respect of the number of levels and the degeneracies of the quantum working medium~\cite{denzler2021power}.
Strikingly, the analysis points to the supremacy in performance of a quantum machine over its nondegenerate counterparts, via optimization of the thermodynamic parameters with respect to energy spacings or level numbers.

\paragraph{Work and heat fluctuations: beyond the TPM paradigm} 
Work and heat are  thermodynamic quantities that are defined over two times. As such, the fluctuations of these quantities are described by a joint probability for the energy at the beginning and end of a given process. The TPM scheme discussed above provides such a probability distribution, but with the clear drawback of destroying all coherence and correlations in the measured energy eigenbasis of the initial state.      
To circumvent this issue, different paths have been taken to study interference and correlation effects in the work and heat distribution. These include probability distributions that can turn negative, as in the  full counting statistics method~\cite{PhysRevA.96.052115,solinas2016probing,xu2018effects}, consistent histories~\cite{miller2016time}, and the Margenau-Hill quasiprobability distribution~\cite{PhysRevE.90.032137,PRXQuantum.1.010309}. In addition to turning negative,  the Kirkwood-Dirac quasiprobability  can also have imaginary parts~\cite{lostaglio2022kirkwood}.
Methods in which the work distribution is not linear in the density operator, such as the Bohmian framework~\cite{sampaio2018quantum} and Bayesian networks~\cite{micadei2020quantum,micadei2021experimental}, offer another possible direction~\cite{gherardini2021end}. These approaches provide tools for studying energy and efficiency fluctuations of QHEs beyond the TPM scheme limitations and are motivated by a recent experiment recovering the Margenau-Hill quasiprobability distribution for the work in a quantum process~\cite{hernandez2022experimental}. Applying these ideas to stroke engines that encode multiple measurement points would require devising proper measurement schemes~\cite{halpern2018quasiprobability,alonso2019out}.

\subsection{Thermodynamic Uncertainty Relations in Quantum Heat Engines}

Thermodynamic Uncertainty Relations (\acrshort{tur}) have been an area of intense study over the past few years, in the context of stochastic thermodynamics and nonequilibrium processes in classical and quantum systems~\cite{seifert2012stochastic,PhysRevLett.114.158101,PhysRevLett.123.090604,Menczel_2021}. These relations help in providing bounds in the form of mathematical inequalities for averages, fluctuations, and higher moments of important thermodynamic quantities, like currents and entropy production for out-of-equilibrium systems. Within the ambit of heat engines, classical and quantum, TURs provide insights into the efficiency and power output measures, especially away from the quasi-static limit. 
TUR were first proposed in the classical setting to model biomolecular processes~\cite{PhysRevLett.114.158101}, e.g., molecular motors and chemical reactions~\cite{pietzonka2016universal}, as prototypical nonequilibrium dissipative systems. These tradeoff relations can be expressed in the following compact form:

\begin{equation}
\label{eq:tur}
    \frac{{\rm Var} (j) }{\langle j \rangle^{2}}   \geq \frac{2 k_{B}}{\sigma},
\end{equation}
where $\langle j \rangle $ is the average current, ${\rm Var} (j) \equiv \langle j^{2} \rangle - \langle j \rangle^{2} $ is the variance, and $\sigma$ is the entropy production, all of which are evaluated in the steady state limit. 
Physically, these relations express a trade-off between \textit{precision}, characterized by the signal-to-noise ratio of currents at a steady state, and the \textit{amount of entropy} produced in the process. 
Eq.~\eqref{eq:tur} reveals that, even away from equilibrium, dissipation continues to regulate small fluctuations~\cite{horowitz1,PhysRevB.98.155438}.
 Eq.~\eqref{eq:tur} can be derived for a wide class of steady-state engines modeled by means of stochastic thermodynamics theory \cite{Seifert_2012,seifert2019stochastic}. Here, the energy levels can make up a discrete set in the same spirit of stochastic Markovian networks \cite{pietzonka2018}, or they can be described by continuous degrees of freedom undergoing overdamped Langevin dynamics \cite{brandner2015thermodynamics}.Quite recently, generalizations of Eq.~\eqref{eq:tur} to periodic \cite{koyuk2019operationally}  and finite-time drivings \cite{koyuk2020thermodynamic} have also been achieved.

A lot of work has focused on the quantum implications of such TURs in heat transport, continuous thermal devices, and, most recently, QHEs with discrete strokes~\cite{PhysRevE.100.042101,PhysRevE.103.L060103, PhysRevLett.127.190603,PhysRevE.102.042138}. In the case of the QHEs, the effect of TURs are best analyzed when higher cumulants of thermodynamic quantities are evaluated. 
 A joint probability distribution for heat and work, based on the outcomes discussed in section~\ref{sec:fluc_QHE} and analyzed across an entire engine cycle, may serve as a framework for computing the fluctuations.
Defining $W$ as the total work over a cycle and $Q_{in}$ the input heat, it was shown in~\cite{PhysRevE.103.L060103} that in the linear response, the ratio of the two fluctuations $\eta ^{(2)} = \frac{{\rm Var} (W)}{{\rm Var}(Q_{in})} $ is bounded both from above and below,
 \beq
\eta^2 \leq \eta^{(2)} \leq \eta_{\rm C}^2, 
 \eeq
where $\eta^2$ is just the square of the efficiency and $\eta_{\rm C}$ is the Carnot efficiency. This result is derived from the Onsager reciprocity relation, and reveals the relation of $\eta^{(2)}$ with the TUR.
For a continuous engine  with multiple thermodynamic forces (affinities such as temperature and chemical potential differences) it was shown~\cite{PhysRevE.103.L060103}  that the relative uncertainty in the output power is always higher than the relative uncertainty in the heat current absorbed from the hot bath,  
\beq
{\rm Var}(\dot{Q}_{in})/\dot{Q}_{in}^2 \leq {\rm Var}(\dot W)/\dot{W}^2. 
\eeq
Similar results were obtained for refrigerator and heat pump operation of the thermal machine, and in certain cases it was shown to hold beyond the linear response regime.
Generalization of these results to higher cumulants
and for ensemble of quantum thermal machines was introduced in~\cite{gerry2022}.

In~\cite{pietzonka2018}, a tighter bound than the Carnot limit, $\eta \leq \eta_{\rm C} \:= 1- (T_{C}/T_{H})$, was found in classical Steady-State Heat Engines, as a consequence of TURs that rigorously identify the influence of the time-averaged fluctuations, $\Delta P_{W}$, of the generated power, $P_{W}$, during the engine's working cycle,

\begin{equation}\label{eq:turbound}
    \eta \leq \frac{\eta_{\rm C}}{1+ 2 T_{C} (P_{W}/\Delta P_{W})}.
\end{equation}
Such universal trade-off relations in the quantum regime remain an open problem, and a recent attempt explores the question in periodically driven heat engines~\cite{miller2019work,miller2021}. 
 In the quantum mechanical case, quantum fluctuations due to the effect of quantum friction, i.e., the non-commutativity of operators in the driving Hamiltonian, can play a major role in the performance of the QHE. Indeed, such corrections to the classical TUR bound have been found. This modifies Eq.~\eqref{eq:turbound} to 
\begin{equation}\label{eq:turcoher}
\eta\leq \frac{\eta_{\text C}}{1 + 2 T_{C}P_{W} G(P_{W},P^{QS}_{W},\Delta P_{W},\Delta I_{W})}.  
\end{equation}
Here, $P^{QS}_{W}$ is the quasi-static output power, i.e., the power generated if the state of the working medium follows the instantaneous equilibrium state at each point in time. Moreover, $\Delta I_{W}$ is a measure of the degree of quantum friction along the protocol, computed employing the Wigner-Yanase skew information \cite{WignerYanase1963}. $G$ is an analytic function of the relevant quantities that provides the correction.   Eq.~\eqref{eq:turcoher} holds valid in the limit of slow-driving and Markovian dynamics. It is proved that Eq.~\eqref{eq:turcoher} can either fall below or exceed the classical bound.
In \cite{tajimasuper2021}, the effect of quantum coherence among degenerate eigenstates of the working medium was proved to improve the classical bound to the ratio $\ave{J}^2/\sigma$. This \emph{quantum lubrication} effect can also be produced by means of collective system-bath interactions acting on degenerate subspaces of the working medium.

\section{Non-thermal quantum engines}\label{sec:neqbath}

In the quantum domain, due to the inherently quantum nature of the working medium and the baths, counterintuitive physical effects can set in, leading to apparent violations of thermodynamic laws. These effects pose many challenges to a consistent formulation of thermodynamics in the quantum regime~\cite{vinjanampathy2016_QT_review} and to the assessment of the performance of QHEs. Recently, non-thermal QHEs, i.e., engines working with baths whose quantum state differs from the Gibbs form, have been the subject of much debate. We will more appropriately refer to them as Quantum Engines (\acrshort{qe}s), omitting "heat." 

Non-thermal effects in QEs can be achieved in many ways, e.g., by employing non-linear drivings of the working medium when it couples to thermal baths \cite{ghosh2017catalysis}, or by using Kerr interactions to achieve nonlinear coherent engines \cite{opatrny2023nonlinear}. On the other hand, a non-thermal bath is typically prepared in a non-equilibrium quantum state: this setup can exhibit quantum coherence and correlations (classical and quantum), shared among constituent parts as well as with the working medium.

Moreover, a bath can be prepared in a purely non-equilibrium state, such as a squeezed state in the quantum optics setting~\cite{rossnagel14}, or go through a population inversion procedure that makes it behave like a negative effective temperature bath~\cite{de2019efficiency}. Indeed, several instances of non-thermal QEs have been theoretically devised~\cite{Leggio:PRA2015,Cherubim:2019,Carollo:PRL_2020}, and experimental proofs-of-principle have been proposed~\cite{de2019efficiency,Klaers:expsqueeze,vonLinden:PRL_2019}.
These systems' apparently dramatic implications for thermodynamic laws have been addressed in several theoretical works, which in some cases developed generalizations aimed at describing the energetics of the systems.
Below we provide an overview of several non-thermal QEs, deferring our discussion of non-Gibbsian states arising from strong correlations between the working medium and the bath to Sec.~\ref{sec:nonMarkov}.              
	 
\subsection{The phaseonium and quantum coherence induced by dissipation}\label{subsec:coherent}

Quantum coherent baths are a prominent type of non-thermal bath.
A QE operating on this type of bath shows additional thermodynamic features that are impossible to achieve by means of conventional heat engines. Non-thermal QEs were first introduced as quantum optical devices. 
The idea builds on previous breakthroughs in quantum optics, namely
micromasers, Electromagnetic Induced Transparency (\acrshort{eit}), and Lasing Without Inversion (\acrshort{lwi})~\cite{scully_zubairy_1997}. Here, a three-level atom interacts with a pair of resonant electromagnetic fields. If the atom is prepared in a state of coherent superposition of two of its lower eigenstates then, under given conditions, destructive interference between different transition processes can take place, causing the absence of photon absorption. As a consequence, 
the detailed balance between photon absorption and emission processes can break. 
In this limit, the thermodynamic features of QHEs like those described in Fig. 2 must be modified, leading to phenomena like
work extraction from a single reservoir and the breaking of the Carnot limit in two-reservoir
engines.   
    \begin{figure}[h]
    \centering
    \includegraphics[scale=1.1]{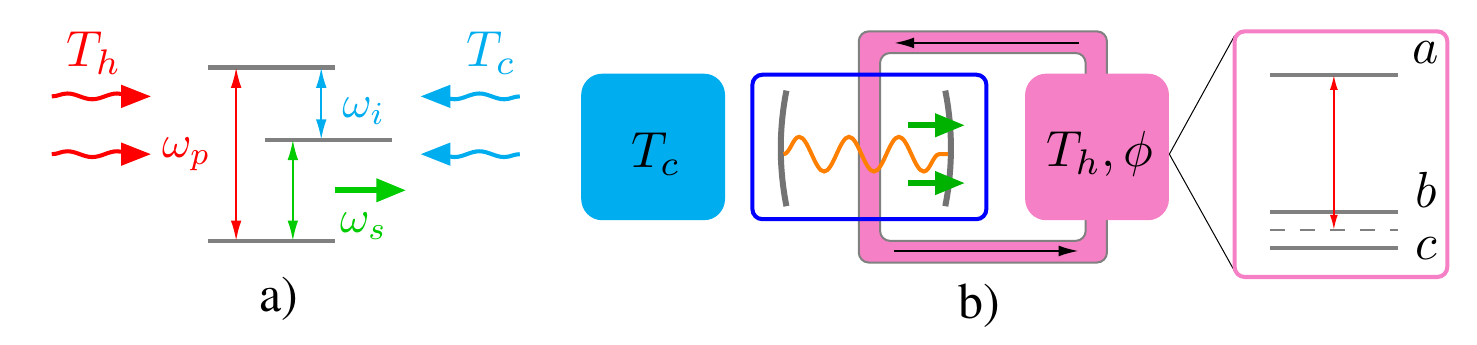}
    \caption{Schematic diagram of early prototypes of QEs. $a)$ A three-level Maser QE. Hot (cold) reservoirs are modeled with thermal light sources at temperatures $T_{h} (T_{c})$.~Filter waveguides allow the absorption of hot (cold) photons of frequency $\omega_{p}$($\omega_{i}$).~The population inversion takes place between levels 1 and 2. Work is performed by spontaneous emission of radiation of frequency $\omega_{s}$. $b)$ The Phaseonium QE.  A radiation mode inside a cavity performs work on the mirrors by means of radiation pressure, playing the role of the engine's working medium.
    The hot reservoir is an ensemble of three-level $(a,b,c)$ atoms, which are in contact with a thermal bath at temperature $T_{h}$.~The two lower levels are quasi-degenerate, such that the state of the atom can show quantum coherence, dependent on the atomic phase difference $\phi$.
    The atoms are injected into the cavity and thermalize the radiation field at temperature $T_{\phi}$, controlled by $\phi$.~The state of the reservoir is thus equivalent to a thermal state with effective temperature $T_{\phi}$ that, due to quantum coherence, can be higher than $T_{h}$.~A conventional entropy sink at temperature $T_{c}$ is also used.~Reproduced from \cite{scully03}. 
    }
	\label{neqbathfig1}
	\end{figure}


{ In their important contribution to the study of such engines, Scully et al.~\cite{scully03} proposed
a modified version of the QHE that works with a quantum coherent bath, i.e., the \emph{phaseonium} heat engine. The term ''phaseonium'' stands for phase coherent atomic ensemble \cite{SCULLYphaseonium91}. It typically denotes an ensemble of atoms, each having a ground state energy doublet and prepared in a coherent superpositions of its eigenstates.}
Unlike in conventional QHEs, the working medium is a mode of radiation inside a cavity, and the work is accomplished through radiation pressure on the cavity mirrors.
An ensemble of three-level atoms replaces the thermal light and acts as the hot bath.~Each three-level atom has a level structure as in Fig.~\ref{neqbathfig1} b, where coherence is allowed between the two lower, quasi-degenerate levels.~An entropy sink is also present, working as the cold bath.~
 Here, it is further assumed that the initial density matrix of each atom of the bath reads $\rho(0)=\rho_{aa}\ket{a}\bra{a} +\rho_{bb}\ket{b}\bra{b} +\rho_{cc}\ket{c}\bra{c} +\rho_{bc}\ket{b}\bra{c}+\rho_{cb}\ket{c}\bra{b}$, with $\rho_{bc}=|\rho_{bc}|e^{i\phi}$ and $\phi$ the coherent phase. Moreover, the populations of each energy level $\rho_{ii}=P_{i}$ for ${i=a,b,c}$ follow the Boltzmann weights with temperature $T_{\text h}$. As the atoms are injected into the cavity, the radiation inside the cavity is brought to a quantum state that depends on the phase difference $\phi$ of the two lower atomic levels.~Indeed, the mean number of photons inside the cavity, $\bar{n}_{\phi}$, obeys the following rate equation
\beq\label{eq:nonthermal1}
\dot{\overline{n}}_{\phi} =\alpha (2 P_{a}(\overline{n}_{\phi} + 1) -(P_{b}+P_{c})(1 +\varepsilon \cos \phi)\overline{n}_{\phi}),
\eeq
where $\alpha$ is a rate factor depending on the properties of the atoms and $\varepsilon=|\rho_{bc}|/2(P_{b} + P_{c})$.

It is thus clear that however small, quantum coherence in the bath affects the steady-state number of photons in the cavity.~From a microscopic treatment \cite{Scully:AIP2002}, it follows that the steady-state number of photons reads
\beq \label{eq:nonthermal2}
\overline{n}_{\phi}\simeq n_{\rm{th}}(1-n_{\rm{th}} \varepsilon\cos\phi ), 
\eeq
where $n_{\rm{th}}=((P_b + P_c)/2 P_a -1)^{-1}$ is the photon population in the absence of coherence 
and the levels spacing fulfill the limit $E_{a}-E_{b}\simeq E_{a}-E_{c}\ll k_{\text B} T_{\text h}$.~It is worth stressing that the quantum state of radiation is thermal and can be characterized by a parameter of temperature
     \beq\label{eq:nonthermal3}
	      T_{\phi}= T_{\rm{h}}(1-n_{\rm{th}} \varepsilon\cos\phi ), 
     \eeq  		
where $\varepsilon$ is a parameter linked to quantum coherence that, in the high-temperature limit, reads $\varepsilon=3 |\rho_{bc}| $. 
It follows that the proper control of phase difference $\phi$ can lead to a thermal state of radiation with effective temperature $T_{\phi}>T_{h}$.
     
A remarkable consequence is that work can be extracted from the engine even in the case of equal temperature, i.e., $T_{c}=T_{h}$, which is conceptually analogous to extracting work from a single reservoir. However, this task does not violate thermodynamic laws, as additional energy is required to prepare the initial coherences in the bath state. Thus, quantum coherence acts as an additional quantum resource \cite{baumgratzcoher2014,Streltsov:RevModPhys_2017}, similar to a special kind of \emph{fuel} for the QE \cite{Gershon16}.
Indeed, it permits energy extraction from the high-temperature bath more efficiently than traditional engines allow.

\begin{figure}%
    \centering
    \includegraphics[width=8cm]{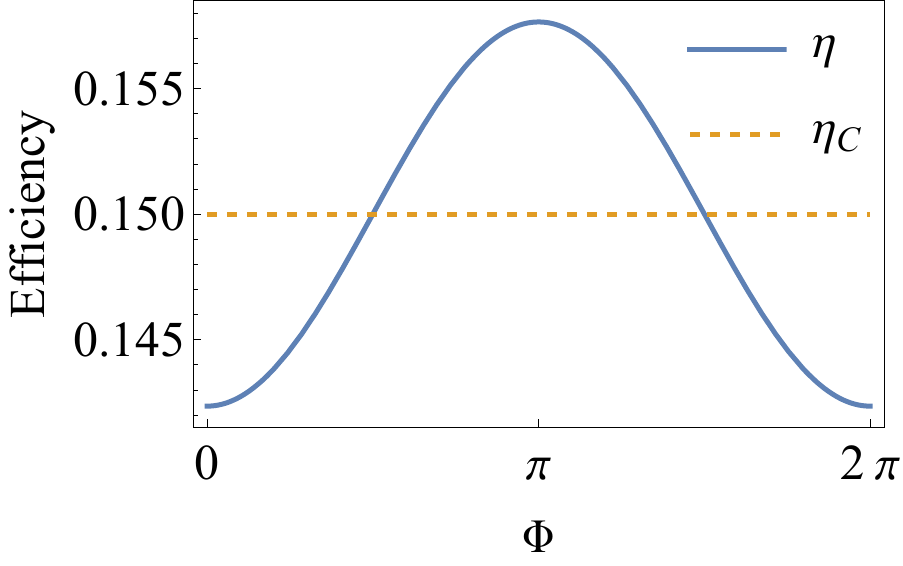} %
    \caption{ Efficiency as a function of the phase  $\phi$ at the high temperature limit, Eq.~(\ref{eq:nonthermal4}). $\eta$ is the efficiency of the phaseonium QE, and $\eta_{\rm C}$ is the Carnot efficiency. The phaseonium QE exceeds the Carnot efficiency for certain phases. Here  $T_c/T_h=0.85$, $\bar{n}=10^3$, and $|\rho_{bc}|=3\cdot 10^{-6}$. }%
    \label{fig:effphos}%
\end{figure}

Moreover, this setup devises a thermodynamic cycle, composed of an isothermal expansion at $T=T_{\phi}$, followed by an adiabatic expansion, where the atoms are removed from the cavity.
Subsequently, the radiation cavity is coupled with the entropy sink, undergoing an isothermal compression at $T=T_{c}$ followed by an adiabatic compression that brings it back to its initial state. With small cavity volume variations, using Eq.~\eqref{eq:nonthermal3} the engine efficiency can be written as     
\beq\label{eq:nonthermal4}
\eta \simeq 1-\frac{T_{c}}{T_{h}}(1 + n_{\rm{th}} \varepsilon\cos\phi )=\eta_{\rm{C}} -3\frac{T_{c}}{T_{h}}n_{\rm{th}}|\rho_{bc}|\cos\phi.
\eeq
In the presence of quantum coherences, by modulating the phase difference of the two low-lying atomic levels, the phaseonium efficiency can be tuned to exceed the Carnot bound (see Fig.~\ref{fig:effphos}). %
Yet, we emphasize that when the engine isn't powered solely by heat baths the  Carnot bound is no longer a statement of the second law of thermodynamics as other resources are involved in the generation of work. Ultimately, the correct evaluation of the consumed resources and their costs \cite{brandao2013resource,chitambarresource19} is the key to explaining the apparent violations of thermodynamics laws occurring in these engines, similarly to what happens with information engines (see Sec.8).
A recent study~\cite{guff2019power} revisiting this model explored the power-efficiency trade-off when considering finite-time processes.   

 Quantum coherence can also enhance the performance of more conventional QEs working between two thermal baths, such as a laser or photocell QE \cite{scully2010quantum,scully11}
\begin{figure}[h]
\centering
\includegraphics[scale=1.50]{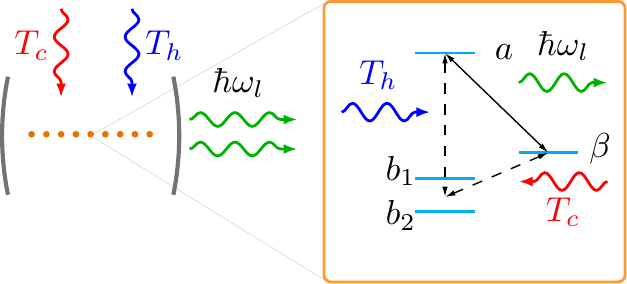}
\caption{
Schematic diagram of the photocell QE. An ensemble of three-level atoms (orange dots) immersed in a cavity acts as the working medium. Hot (cold) reservoirs are again modeled with thermal light sources at temperatures $T_{h} (T_{c})$.~Lasing transition is achieved between levels $a,\beta$, after pumping from $b_{i}$ to $a$ ($\beta$ to $b_{i}$) by means of hot (cold) photons.
Due to the degeneracy of the lower levels ($b_{1},b_{2}$), different atomic paths interfere, so that quantum coherence builds up as induced by the baths.~No additional resources are required. Reproduced from \cite{scully11}.
}
\label{neqbathfig2}
\end{figure}
The latter device, depicted in Fig.~\ref{neqbathfig2}, is more similar to conventional maser QHEs described in Sec.~\ref{subsec:maser} above.
However, here the two baths are thermal light sources at different temperatures $T_{h},T_{c}$.~The working medium is again a three-level quantum system.
The baths provide hot and cold photons to optically pump the system and achieve lasing conditions.
Unlike in conventional QHEs, the lower level is replaced by a couple of nearly degenerate levels.
This change has a crucial consequence, as stimulated emission processes from upper levels can interfere, producing coherences between the two lower states that modify the steady state of the working medium.~Moreover, the bath-induced coherences enhance the photon absorption from the reservoirs, as they lead to a faster depletion of the lower levels in favor of the upper ones.
As a result, a net improvement in the power output of the engine can be obtained.
The power can be written as
\beq\label{eq:nonthermal5}
P_{l}= A(\overline{n}_{h} -\overline{n}_{c})\hbar\omega_{l}, 
\eeq
where $\omega_{l}$ is the laser frequency and $\overline{n}_{h/c}$ is the average number of hot/cold photons.~In the presence of bath-induced coherences, it has been shown that the prefactor $A$ doubles when convenient model parameters are chosen.
If the photocell QHE is considered, better current-voltage characteristics can be obtained.
This proposal's distinctive trait is that quantum coherence is not provided by any form of external fuel but is rather induced by the noise linked to the same two baths that sustain the laser operating condition.
As a consequence, in contrast to the phaseonium engine, no additional costs to produce quantum coherence are required.
 More recent works analyzed the impact of a small amount of quantum coherence in the reservoirs on single-qubit engines by employing the framework of collisional models \cite{Hammam_2022} (see also Sec. \ref{subsec:corrbath}).
Other models and experiments \cite{hardal2015superradiant,kloc2021superradiant,kim2022photonic} have studied the enhancement of the power output due to a coherence reservoir. In these studies, the engine operates between superradiant  (see  Sec. \ref{sec:manybody} ) and thermal reservoirs.

\subsection{Squeezing as a thermodynamic resource}\label{subsec:squeezed}

Non-thermal QEs can also be powered by squeezing~\cite{scully_zubairy_1997}.~Physically, squeezing is a technique applied to a quantized field of radiation, in order to reduce the quantum fluctuations of a given quadrature of the field at the expense of the conjugate one.
A non-thermal squeezed QE is composed of a working medium, e.g., a driven harmonic oscillator, which exchanges energy between two reservoirs at inverse temperatures $\beta_{1}$ and $\beta_{2}$, such that  $\beta_{1}>\beta_{2}$.
In contrast to conventional heat engine setups, one of the two reservoirs, let us say the hot one, is prepared into a squeezed thermal state~\cite{MarianPRA93}. In the case of reservoirs modeled with noninteracting bosons, the non-equilibrium squeezed thermal state reads
\beq\label{eq:nonthermal6}
\rho_{s}= P\frac{e^{-\beta H_{B}}}{Z_{B}}P^{\dagger}, 
\eeq         
where $H_{B}=\sum_{k}\hbar \omega_{k} a^{\dagger}_{k}a_{k}$, $P=\prod_{k} P_{k}$, and $P_{k}=\exp[1/2(\xi^{*}_k a^2_k-\xi_k a^{\dagger 2}_k)]$ are the single-mode squeezing operator corresponding to the squeezing parameter $\xi_{k}=|\xi_{k}| e^{i\theta_k}$.~The squeezed thermal bath is thus properly characterized by its inverse temperature $\beta$ and the squeezing parameter $\xi_{k}$ of each mode. In a squeezed thermal bath, the fluctuations of the position and momentum of each of the bath's modes can exponentially decrease (or increase) depending on $\xi_{k}$, while their product does not vary \cite{scully_zubairy_1997}. 
 Moreover, since the state of the bath is not invariant as in the case of a thermal bath, the correlation functions of the bath are not homogeneous in time, which implies that the spectral tensor of squeezed bath explicitly depends on time \cite{breuer}.
These non-equilibrium features of the bath have severe consequences for the working medium's thermalization properties.
Indeed, in the weak-coupling, Markovian limit, and in the absence of driving fields, the stationary state of the working medium is
\beq\label{eq:nonthermal7}
\rho_{\rm{st}}= P\frac{e^{-\beta H_{S}}}{Z_{S}}P^{\dagger}, 
\eeq
and it follows that the variances in the position and momentum of the working medium are squeezed as well. 
    
\begin{figure}[h]
    	\centering
        \includegraphics[scale=1.50]{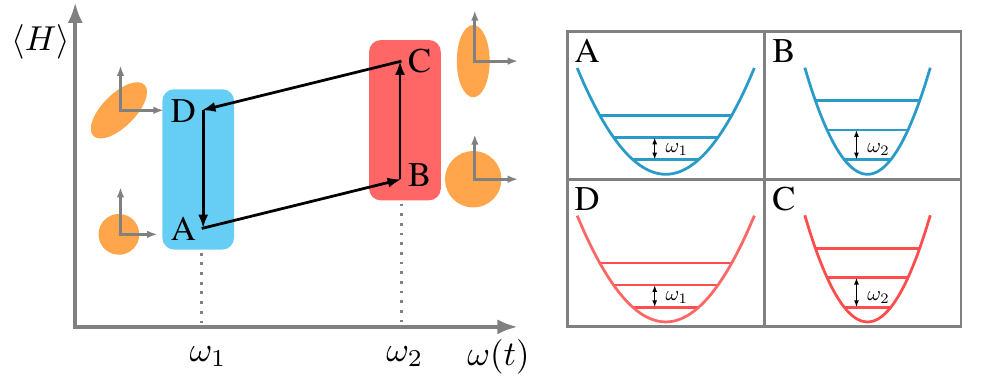}
    	\caption{Sketch of the Otto cycle with a squeezed reservoir and the energy levels of the working medium.~$\text{A}\to\text{B}$ is the isentropic compression stroke, the working medium is decoupled from the thermal reservoir, and the oscillation frequency is increased.~$\text{B}\to\text{C}$ is the isochore stroke; the working medium interacts with the squeezed reservoir and reaches the squeezed thermal state in Eq.~\eqref{eq:nonthermal7}, increasing its energy.~$\text{C}\to\text{D}$ is the isentropic expansion stroke, the working medium is decoupled from the squeezed reservoir, and the oscillation frequency is decreased to the initial value. $\text{D}\to\text{A}$ is the final isochore stroke, the working medium thermalizes, and its energy is released in the cold bath.}
    	\label{neqbathfig3}
    \end{figure}

The idea of a squeezed QHE was first proposed in~\cite{rossnagel14}. An Otto cycle was devised (see also Sec.~\ref{subsec:Enginesintro}), composed of four strokes (ABCD), as depicted in Fig.~\ref{neqbathfig3}.
The working medium is a driven harmonic oscillator described by the Hamiltonian  $H_{S}(t)=\hbar\omega(t)b^{\dagger}b$.
It starts in a thermal state at inverse temperature $\beta_1$.
An isentropic compression stroke ($\text{A}\rightarrow \text{B}$) is then performed, i.e., the frequency of the oscillator is increased from a starting value $\omega_{\rm{1}}$ to $\omega_{\rm{2}}$, while it is decoupled from the reservoir.
Next the oscillator is put in contact with the squeezed bath, while its frequency is held constant ($\text{B}\rightarrow\text{C}$).
Its energy, which is a function of the inverse temperature~$\beta_2$ and the squeezing parameter~$r$, is increased, as is the asymmetry in the variances of position and momentum~\cite{manzano16}. Furthermore, the oscillator is decoupled from the bath and undergoes an isentropic expansion stroke ($\text{C}\rightarrow \text{D}$), where its frequency is brought back to its initial value $\omega_{1}$. Eventually, the cycle is closed by means of another isochoric stroke ($\text{D}\rightarrow \text{A}$), where the oscillator is coupled to the thermal reservoir at inverse temperature $\beta_1$. 
    
It is assumed that the isochoric strokes take place in a much shorter time than the isentropic strokes, and both the time needed to reach the steady-state and the work needed to turn on the interactions with the baths~\cite{Wiedmann_2020} are neglected.
In the weak coupling limit, the final state after thermalization is known, so that the amount of energy exchanged along each stroke can be derived analytically. It follows that the efficiency of the cycle reads~\cite{rossnagel14}
\beq\label{eq:nonthermal8}
\eta^{*}= 1 - \frac{\omega_{1}}{\omega_{2}}\frac{\coth(\hbar\beta_1 \omega_1 /2)-Q^{*}_2 \coth(\hbar\beta_2 \omega_2 /2) \Delta H(r) }{Q^{*}_1 \coth(\hbar\beta_1 \omega_1 /2)-\coth(\hbar\beta_2 \omega_2 /2) \Delta H(r)}.  
\eeq    
%
Here $\Delta H(r)= 1 + (2 + 1/n_{\rm{th}})\sinh^2(r) $ is the ratio between the squeezed and the thermal occupation $n_{\rm{th}}=(\exp(\hbar\beta_2 \omega_2) -1)^{-1}$. Moreover, the factors $Q_i^{*}$ arise from the exact non-equilibrium dynamics of the oscillator during the isoentropic strokes~\cite{husimiexact53,Deffner08}. They explicitly depend on the parametrization of the driven frequency $\omega(t)$ and quantify the degree of adiabaticity of the protocol, i.e., $Q_{i}^{*}=1$ for adiabatic dynamics while $Q^{*}_{i}>1$ for increasingly nonadiabatic protocols. When $r=0$, we have $\Delta H(r)=1$, and the usual thermal equilibrium populations are restored.

\begin{figure}%
    \centering
    \includegraphics[width=8cm]{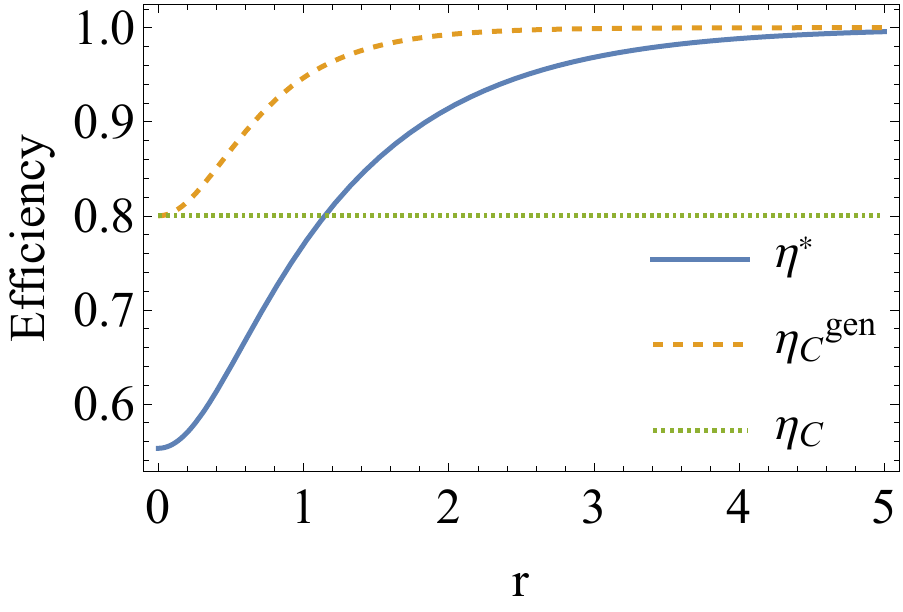} %
    \caption{Efficiency as a function of the squeezing parameter $r$ at the high-temperature limit. $\eta^*$ is the efficiency at maximum power, $\eta_{\text C}$ is the standard Carnot efficiency, and $\eta^{\rm gen}_{\text C}$ is the generalized Carnot bound. Here we chose $\beta_2/\beta_1=0.2$.}%
    \label{fig:effsqu}%
\end{figure}

From Eq.~\eqref{eq:nonthermal8}, under conditions of adiabatic compression/expansion strokes, and in the high-temperature limit, the efficiency at maximum power of the cycle is derived:  
\beq\label{eq:nonthermal9}
\eta^{*}= 1 - \sqrt{\frac{\beta_2}{\beta_1(1 + 2\sinh^2(r))}}.  
\eeq                     
Eq.~\eqref{eq:nonthermal9} shows that the efficiency of the engine equals the Curzon-Ahlborn bound for~$r=0$. 
 However, when the squeezing parameter is increased, the efficiency can surpass the Carnot bound, approaching unity for~$r\gg 1$ as can be seen in Fig.~\ref{fig:effsqu}.   
Yet, this result doesn't violate the generalized Carnot efficiency that accounts for the squeezing as a resource~\cite{huang2012effects,abah2014efficiency} and reads
\beq
\eta^{\rm gen}_{\text{C}}= 1 - \frac{\beta_2}{\beta_1(1 + 2\sinh^2(r))}.
\eeq

The surprising features of squeezed QEs have spurred novel theoretical research efforts aimed at achieving a solid understanding of the thermodynamics involved. The entropy production in squeezed QHEs was first addressed in~\cite{manzano16}.
Here a generalization of the Spohn inequality~\cite{spohn78,breuer,LandiEntProd21} for the entropy production is obtained, in order to deal with the squeezed nature of the bath at inverse temperature $\beta$. 
More specifically, a non-thermal contribution to the \emph{heat} exchange is considered. The time derivative of the entropy production is rewritten as
\beq\label{eq:nonthermal10}
\dot{\Sigma}= -\frac{d}{d t}D(\rho_{s}||\rho_{\rm{st}})= \dot{S}-\dot{\Phi},  
\eeq                     
where $ S=-\tr[\rho_{s}\log \rho_{s}] $ is the Von-Neumann entropy of the working medium, $D(||)$ denotes the quantum relative entropy \cite{Sagawa:2013,parrondo:NatPhys_2015}, and  $\dot{\Phi}=-\tr[\dot{\rho}_{s} \log \rho_{\rm{st}}]$.~From the form of the steady-state in Eq. \eqref{eq:nonthermal6}, it follows that
\beq\label{eq:nonthermal11}
\dot{\Phi}=\beta (\cosh(2r) \dot{Q} -\sinh(2r)\dot{\mathcal{A}}),   
\eeq
where $\dot{Q}=\tr[H_{S}\dot{\rho}_{s}]$ is the heat flow from the reservoir, $H_{S}$ is the Hamiltonian of the quantum system coupled to the squeezed-thermal reservoir, and $\dot{\mathcal{A}}=\tr[A \dot{\rho_{s}}]$ with the asymmetry operator defined as $A=(\hbar \omega/2)( p^2_{\theta/2} -x^2_{\theta/2})$. Notice that Eq.~\eqref{eq:nonthermal10} reduces to the Spohn form for $r=0$.
~The main consequence of Eqs.~\eqref{eq:nonthermal10} and \eqref{eq:nonthermal11} is that, in a dissipative evolution, the total variation $\Delta \Phi$ can become negative, thereby overcoming the total entropy production $\Sigma$ that is nonnegative by definition.~As a consequence, the working medium's  Von-Neumann entropy variation can become negative.
    
If simple two-stroke cycling processes are considered, where the system is put in contact with a single reservoir, it follows from the previous analysis that a positive amount of work $W_{\rm{out}}$ can be extracted from the reservoir that is bounded from above, 
    \beq\label{eq:nonthermal12}
	W_{\rm{out}} \leq \tanh(2r) \Delta \mathcal{A}.   
	\eeq
As a consequence, the amount of work extracted is linked to the increase in the asymmetry of the reduced state.~The maximum work can be extracted when the initial state is thermal, i.e., $\rho_{s}=e^{-\beta H_{S}/Z_{S}}$, and reads
    \beq\label{eq:nonthermal13}
    W_{\rm{max}} = \hbar \omega (2 n_{\rm{th}} + 1)\sinh^2(r).   
    \eeq
Even more striking results appear if a slightly modified Otto cycle is considered~\cite{manzano16}, where an additional unitary transformation is introduced to unsqueeze the state after decoupling from the squeezed bath.
The surplus of work coming from the squeezed reservoir leads to entirely new operating regimes for the Otto cycle.
It follows that the QHE can work as a refrigerator and perform work extraction simultaneously.
The controversial results in the thermodynamic performance of the Otto cycle in Eq.~\eqref{eq:nonthermal7} have been traced back to the additional amount of work extractable from the squeezed reservoir, which is absent in the case of thermal baths. Squeezing, therefore, can be seen as an additional quantum resource to perform thermodynamic work.   
This result implies that the Carnot bound for heat engines operating between two heat baths is not  relevant when squeezing is involved, and  a generalized bound depending on the squeezing parameter $r$ has been shown to hold~\cite{rossnagel14}. 

The derivation of generalized bounds is not limited to driven, four-stroke heat engines. 
In~\cite{Agarwalla:2017}, a generalized bound was derived for a three-terminal continuous heat engine, in which one of the terminals is a photonic squeezed thermal bath. The working medium is a TLS that exchanges energy with a photonic squeezed reservoir with a two-terminal electronic circuit. 
The latter is made of two metal leads acting as thermal reservoirs, with fixed values of inverse temperature and chemical potential, i.e., $\beta,\mu$. In this way, conversion from photonic energy to electrical work is studied through a full counting statistics approach and TPM protocol. A universal bound on the efficiency is derived in terms of effective inverse temperatures, which holds for a generic non-equilibrium reservoir. 
In the instance of the proposed photoelectric device, with leads at the same temperature and within the high-temperature limit, the computed bound reduces to Eq.~\eqref{eq:nonthermal9}.

These controversial theoretical results were later confirmed by an experimental work \cite{Klaers:expsqueeze}, which employs a GaAs nanobeam as a working medium, modeled by a quantum harmonic oscillator. The nanobeam is piezoelectrically coupled to a source of squeezed electronic noise. Both work extraction from a single reservoir and the breaking of the Carnot bound in an Otto cycle have been observed through this method. However, these experimental findings cannot be interpreted as a way to circumvent the second law of thermodynamics.  

Different setups for quantum thermal machines were also considered as sensitive probing devices for temperature estimation~\cite{hofer2017quantum,levy2020single}.
It was shown in~\cite{levy2020single} that small temperature variations between the engine baths can be evaluated by measuring a flywheel that stores the engine's energy output.
In this specific setup, the flywheel is growing  macroscopic oscillations of a trapped ion that can be resolved using fluorescent imaging on a camera.
Squeezing the working medium after the thermalizing strokes significantly amplifies the flywheel energy and the sensitivity in detecting very small temperature differences. Moreover, when the squeezing attains the quantum regime, the sensitivity  is amplified by one order of magnitude.

\subsection{Quantum heat engines with negative temperature baths}\label{subsec:negtemp}

Over the last two decades, reservoir engineering techniques have been employed in several tasks, e.g., quantum state preparation, generation of entangled quantum states, and decoherence reduction.
Negative-temperature quantum states, i.e., inverted-population states~\cite{ramsey56,Hama2018relaxation}, have been studied since the sixties by employing nuclear spins and NMR techniques \cite{Purcell51,Oja97nmr}. More recently, cold atoms in optical lattices have been employed~\cite{rapp2010equilibration,braun2013negative}. Over the years, the quest for a consistent thermodynamic description of these states has spurred many debates~\cite{carr2013negative,dunkel2014consistent}, as they were believed to lead to a reformulation of the second law of thermodynamics.
Quite recently, alternative theoretical schemes have been proposed to describe these states without the need for modification of thermodynamic laws.~Indeed, a negative-temperature state can be recast as a temperature-unstable state \cite{struchtrup2018work}.
In contrast to equilibrium states, which are the focus of the second law, these states can store work, and it can be extracted to perform various thermodynamic tasks, as with a flywheel~\cite{levy16,vonLinden:PRL_2019}.
Adopting this scheme, several experimental proposals of QEs working with engineered negative temperatures have been put forward \cite{de2019efficiency,mendoncca2020reservoir,nettersheim2022power}. An explicit derivation of a Master equation with an effective negative temperature can be found in the appendix of~\cite{levy16}. 

In \cite{de2019efficiency} a QE is described that operates between a thermal reservoir at positive temperatures and a second, engineered reservoir prepared at effective negative temperatures.~The experimental platform consists of $^{13}\text{C}$-labeled $\text{CHCl}_{3}$ liquid and an NMR spectrometer.~In this setting, the nuclear spin $1/2$ of the $^{13}\text{C}$ plays the role of the working medium, while the nuclear spin of the $^{1}H$ atom mimics the reservoir.
The two nuclei mutually interact and they can be driven by RF pulses.
Similar platforms have been employed to engineer Quantum Maxwell's Demon engines (see Sec.~\ref{subsubsec:MDexperiment}).
A pivotal feature of the setup, the application of the RF fields, allows for full control of the spin populations, that is, the state of the working medium can be prepared so as to obey different Boltzmann distributions.
The inverse temperature for a thermal state of the working medium described by the Hamiltonian $H_{i}=\omega_{i}\sigma_{i}$ is related to the populations of the excited states via the relation $\beta_{i}=(1/\hbar \omega_{i})\ln[(1-p^{+}_i)/p^{+}_i]$.
It follows that inverting the population, $0.50<p^{+}\leq 1$, permits simulating a negative-effective temperature state.  

The engine cycle runs through a cooling stroke, where the carbon nucleus is prepared in a thermal state $\rho_{1}=e^{-\beta_{c} H_{c}}/Z_{c}$, with $H_{c}=-(1/2) \omega_{c}\sigma^C_x$.
Then a gap expansion stroke takes place in a finite time $\tau$, which is of the order of $\mu s$, i.e., much smaller than the decoherence time.
In this stroke, the evolution is unitary, the frequency is increased to $\omega_h$, and the Hamiltonian changes to $H_{h}=-(1/2) \omega_{h}\sigma^C_y$, in a rotating frame with the Larmor frequency of the working medium.
Next, the working medium goes through the heating stroke. This thermalization is simulated by employing a combination of RF pulses and free evolution gates under the Hamiltonian $H_{J}=(1/4)\hbar J\sigma^C_z \sigma^H_z$, and the final state is $\rho_{3}=e^{-\beta_{h} H_{h}}/Z_{h}$.
The final, unitary stroke takes the frequency of the working medium back to the initial value. 

The engine's performance shows remarkable properties, namely the efficiency is shown to exceed the Otto efficiency limit.
Note that the Otto efficiency has been fulfilled in similar experimental setups~\cite{peterson2019experimental}, where thermal baths are used. 
Moreover, during the finite-time gap expansion/compression strokes, transitions between the eigenstates of the working medium are shown to occur that increase the average extracted work $\ave{W}$.
Consequently, the engine's performance improves with decreasing running time. 


\subsection{Quantum engines with correlated baths}\label{subsec:corrbath}

\begin{figure}[h]
    	\center
    	\includegraphics[scale=0.80]{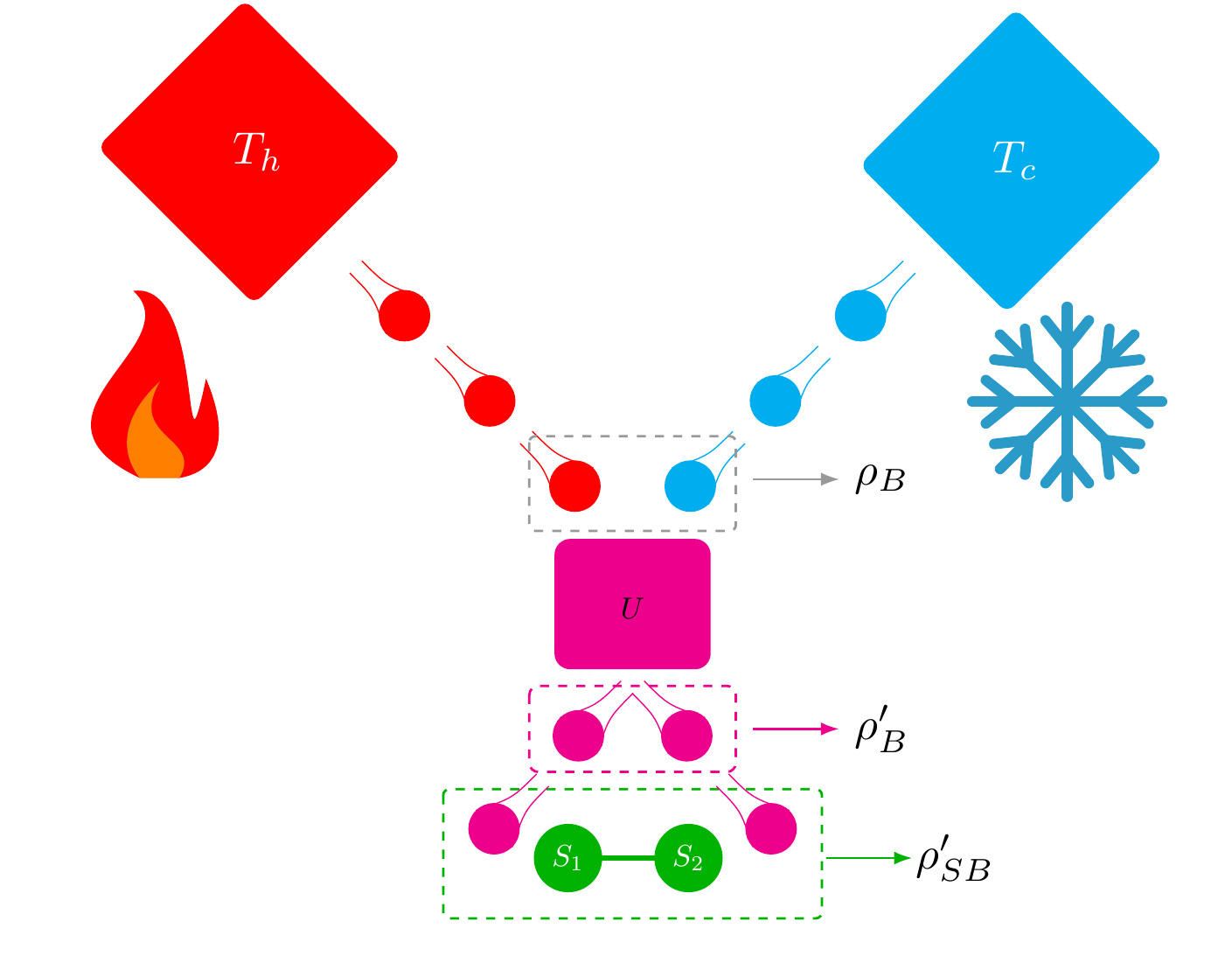}
        \caption{Schematic diagram of the basic mechanism behind the two-qubit correlated engine.
     The reservoirs are modeled with noninteracting qubits.
     They emerge from the reservoirs and correlate after the action of the unitary gate $U$.
     The working medium can thus interact with the correlated reservoir.
     Several operating regimes can be studied by changing the quantum and classical correlations generated by means of $U$. Reproduced from~\cite{DeChiara:PRR_2020}.}
    	\label{neqbathfig4}
\end{figure}

A different type of QEs allows the baths to share correlations, both classical and quantum,  among their constituents \cite{Dille2009}.~A recently devised model of  QE working with correlated baths is based on the theoretical framework of collisional models \cite{barra2015thermodynamic,Strasberg:2017,cattaneo2019local,cattaneo2021collision}.
In~\cite{DeChiara:PRR_2020}, a  QE is proposed whose working medium is made of two XXZ-coupled qubits $S_{1}$ and $S_{2}$, as depicted in Fig.~\ref{neqbathfig4}.
The qubits, described by means of Pauli operators $\sigma_{i},i=1,2$, are also subject to local magnetic fields $B_{1}$ and $B_{2}$.
The reservoirs are modeled by two sets of noninteracting qubits $\tilde{\sigma}_{i},i=1,2$, described by a local Hamiltonian $H_{B}=B_{i}\tilde{\sigma}_{i}$, and they are initially prepared at equilibrium in a product of thermal states  $\rho_{B}=\tilde{\rho}_{\rm th}(\beta_{1})\otimes\tilde{\rho}_{\rm th}(\beta_{2})$. 
In the repeated interactions scheme, $S_{1}$ and $S_{2}$ interact with the environmental qubits by means of collisions, i.e., the interaction takes place over a finite time $\tau$, during which the Hamiltonian is constant in time.
In the continuous limit ($\tau\to 0$), the effect of multiple collisions can be described by means of a Markovian quantum master equation~\cite{cattaneo2021collision}.

In contrast to conventional approaches, before colliding with the working medium, a unitary $U$ is applied to the baths, so that each couple of the bath's qubits is described by the state $\rho^{\prime}_{B}=U\rho_{B}U^{\dagger}$.
As a consequence, classical and quantum correlations develop, which affect the steady-state of the system after a great number of collisions, $\rho^{\rm  steady}(t+\tau)=\rho^{\rm steady}(t)$.
Here the steady state is the state of the working medium and the baths for which the variation of the internal energy amounts to zero,  $\Delta E =\tr[H_{S}(\rho^{\prime}_{SB} -\rho_{SB})]=0$, with $\rho_{SB}=\rho^{\rm steady}_{S}\otimes \rho^{\prime}_{B}\mbox{, }\rho^{\prime}_{SB}=U_{\rm coll}\rho_{ SB}U^{\dagger}_{\rm coll}$, where $U_{\rm coll}=\exp(-i H_{SB} t)$.

In suitable parameter ranges, the system performs thermodynamics tasks, and it can work as a heat engine and refrigerator.
The energetics of this machine is analyzed by studying the work $W$, heat $Q$, and the entropy production $\Sigma$.
However, as with the squeezed reservoir (see \ref{subsec:squeezed},~\ref{subsec:thermotheory}), the cost of maintaining the bath correlations, that is $W_{U}=\tr[H_{B}(\rho^{\prime}_{B}-\rho_{B})]$, directly enters the energetic balance of the machine, such that in the complete case the refrigeration effects are ruled out. A wide range of unitaries are chosen to correlate the two baths. 
In the instance of a swap gate \cite{campisi2015nonequilibrium}, $S(\phi)=\exp[-i\frac{\phi}{2}(\tilde{\sigma}_{x,1}\tilde{\sigma}_{y,2}-\tilde{\sigma}_{y,1}\tilde{\sigma}_{x,2})]$, the populations of the two qubits are partially exchanged, while the reduced states are kept thermal with temperatures  $\beta_{1},\beta_{2}$. 

It is found that the work output is generally amplified with a partial swap and maximized when a complete swap of the bath qubits' populations is achieved for $\phi=m \pi/2$ with $m \in \mathbb{Z}$. 
The engine develops a great amount of both classical and quantum correlations during its operation, and
the mutual information is used to quantify these correlations and the  quantum discord~\cite{Ollivier:PRL_2001,Adesso_2016} to study its quantumness.
Correlations among the baths qubits are absent when the work output is maximal, i.e., the unitary $U$ that maximizes the extracted work performs a complete swap of the state of the two bath's qubits without creating correlations. On the other hand, a considerable amount of quantum correlations is generated and the system-bath qubits correlations play an important role in maximizing the performance of the QE.

\subsection{Thermodynamics of non-thermal quantum engine}\label{subsec:thermotheory}
    
Following the proposals of QEs operating with non-thermal working medium, and with non-thermal heat baths (see~\cite{Gershon16} for useful classification), the quest to understand the thermodynamics of these machines regained momentum.
The apparent violations of thermodynamic laws could be traced to the thermodynamic inconsistency of the adopted definition of heat in these engines.
Competing theoretical pictures have been developed, which rely on different generalizations of fundamental thermodynamical concepts, such as heat, second-law inequality, and entropy production~\cite{LandiEntProd21}.

Quantum correlations induced by non-Gibbsian equilibrium states modify the entropic cost of every quasi-static thermodynamic process, as well as the work exchanged~\cite{Gardas:2015}.
It follows that the thermodynamic cost of correlations, i.e., the \emph{housekeeping heat}, has to be taken into account.
A useful strategy for performing the task was proposed in~\cite{Gardas:2015}, for a rather generic non-thermal state~$\sigma$. The von-Neumann entropy of~$\sigma$ is given by
    \beq\label{eq:nonthermal14}
   S=\beta (\mathcal{E} - \mathcal{F}),  
    \eeq
with $\mathcal{E}=\tr{[\sigma H]}$, $\mathcal{F}=F + \frac{1}{\beta}D(\sigma || \rho_{\rm{th}})$ the information free energy~\cite{Esposito_entropy_2010,Deffnerinfo2013}, and $F$ the thermodynamic free energy of the thermal state $\rho_{\rm{th}}$.
Note that the relation in Eq.~\eqref{eq:nonthermal14} reduces to traditional thermodynamics potential for $\sigma=\rho_{\rm{th}}$.~Using Eq.~\eqref{eq:nonthermal14}, 
the entropic cost of a quasistatic transformation, where $\delta \sigma$ and $\delta H$ denote the slow change of the state and the Hamiltonian of the working medium respectively, can be written in terms of the excess heat $dS=\beta \delta Q_{\rm{ex}}$, with  
    \beq\label{eq:nonthermal15}
    \delta Q_{\rm{ex}}= \delta Q_{\rm{tot}} -\delta Q_{c}.  
    \eeq 
Here $\delta Q_{\rm{tot}}=\tr[\delta \sigma H ]$ is the total heat and $\delta Q_{c}=d\mathcal{F} -\tr[\sigma \delta{H}]$ is the correlation heat, which can be interpreted as the price to be paid to maintain non-Gibbsian correlations.
Analogous reasoning leads to the definition of the excess work as $\delta W_{\rm{ex}}=\delta W + \delta Q_{c}$, so that the first law reads $d \mathcal{E}=\delta W_{\rm{ex}} + \delta Q_{\rm{ex}}$.
The previous definition of cost linked to quantum correlations was included in the evaluation of the QE's efficiency in performing the Carnot cycle. It was found that, when the cost of non-Gibbsian correlation is correctly accounted for in the heat to work conversion balance, the Carnot bound is restored.     
    
Similar information-theoretic approaches have been pursued in studying generalized system-bath correlated systems \cite{Bera:thermocorr}.
The second law can be suitably reformulated if heat is quantified by means of conditional entropy.
Moreover, work extraction from system-bath correlations is possible and, as a result, a generalized form of the Helmholtz free energy can be defined.
As a consequence, for generic correlated systems where non-thermal states are involved, the inclusion of work extractable from correlations permits reconciling the physics of correlated systems with traditional thermodynamics and the Carnot bound.

Owing to the non-equilibrium nature of bath states, \emph{ergotropy} \cite{allahverdyan04,Perarnau-Llobet2015} has been proposed as a key ingredient to explain the energy exchange between the working medium and the bath.
Ergotropy is defined as the maximum amount of work that can be extracted from a given state using only unitary transformation that performs a cyclic variation of the Hamiltonian.
Non-equilibrium bath states, such as the squeezed-thermal state, are non-passive, i.e., by applying a unitary transformation, it is possible to reduce their energy and extract a finite amount of work.
These states are a mainstay in the study of quantum batteries~\cite{alicki13,Binder:2018,Farre:2020}.
Using these concepts of passive states and ergotropy, thermodynamic inequalities have been reformulated to account correctly for the effects linked to non-equilibrium baths~\cite{niedenzu:2018}.

A unified approach to calculating entropy production in open quantum systems applies to any non-thermal environments and dynamics through the analysis of the global unitary evolution of the system and its environment. This involves considering a system $S$ with an initial state $\rho$ interacting with an environment $E$ initially in arbitrary state $\rho_E$,  through a global unitary operation $U$. Assuming the initially the states are uncorrelated, $\rho_{SE}=\rho \otimes \rho_E$, the resultant state of the system-environment is  $\rho_{SE}(t)=U\rho \otimes \rho_E U^{\dagger}$.   
The entropy production in this general case where we don't assume anything about the coupling strength of the system and the environment, their structure, or whether the Hamiltonian is time-dependent, can be cast into the form~\cite{landi2021irreversible}
\beq
\Sigma=\Delta S(\rho(t))+\Phi.
\label{eq:entpro}
\eeq
Here $\Delta S(\rho(t))=S(\rho(t))-S(\rho)$ is the change in the von Neumann entropy from the initial state of the system to  $\rho(t)=\tr_E{[\rho_{SE}(t)]}$, and $\Phi$ is identified as the entropy flux between the system and the environment, expressed as,
\beq
\Phi=\Delta S(\rho_E(t))+S(\rho_E(t)||\rho_E),
\eeq
where $\Delta S(\rho_E(t))$ is the change in the environment's entropy with $\rho_E(t)=\tr_S{[\rho_{SE}(t)]}$, and $S(\rho_E(t)||\rho_E)=k_{\rm{B}} \tr[\rho_E(t) (\log \rho_E(t)-\log \rho_E)]$ is the relative entropy.
The entropy production in Eq.~(\ref{eq:entpro}) is divided into two terms, one that depends solely on the local state of the system and the other that depends on the environment state.

When the environment is in a thermal state $\rho_E=\exp(-\beta H_E)/\tr[\exp(-\beta H_E)]$, with the inverse temperature $\beta$, the entropy flux can be identified as the change in the heat of the bath, i.e.,  $\Phi=\beta Q_E$, and Eq.~(\ref{eq:entpro}) reduces to the Clausius expression for the entropy production. In the weak system-bath environment limit the change in the heat of the bath can be identified as the change in the heat of the system and the result by Spohn for dynamical semigroups can be recovered~\cite{spohn78}. 
Alternative approaches to modeling energetic and entropic costs of non-thermal QEs can utilize studies on the thermodynamics of generalized Gibbs reservoirs~\cite{Lostaglio2017,Manzano18}.

\section{Many-body Quantum  Engines}\label{sec:manybody}

QHEs have been studied extensively with single-particle working medium in both the quasistatic limit as well as in finite time cycles. Recently, many-body systems have been investigated for their influence on the performance of QHEs.
Many-body systems (free or interacting) provide a rich arena for studying and engineering a wide variety of physical phenomena in condensed-matter physics.
Inarguably, the real power of quantum mechanics in fundamental sciences and technology, in general, has been unleashed through the understanding and control of many-body systems. In the case of Quantum Thermodynamics, many-body systems are a natural choice of working medium for exploring quantum effects in meso- and nano-scale thermal devices. 

Many-body cooperative phenomena can already be observed in the two-spin Lipkin-Meshkov-Glick (\acrshort{lmg}) model as the working medium~\cite{anisotropy1}
\begin{equation}
   H= -\frac{J}{4} \Big[ \sigma ^{x}_{1} \sigma ^{x}_{2} + \gamma \sigma ^{y}_{1} \sigma ^{y}_{2} \Big] - \frac{h}{2} \Big[ \sigma ^{z}_{1} + \sigma ^{z} _{2} \Big] -\frac{J(1+\gamma)}{4}
\end{equation}
where $\sigma^i$ are the Pauli matrices. This Hamiltonian exhibits a wide range of behaviors depending on the parameter $\gamma$. When $\gamma=1$, the model is reduced to the isotropic $XY$ model, while, in the range $\gamma \in [-1,1)$, it becomes anisotropic. 
In~\cite{anisotropy1} a detailed analysis of all the Hamiltonian parameters is introduced, showing that anisotropy significantly enhances the work output, even towards high efficiency of the operating regime.

Another approach exploring collective phenomenon was put forward in~\cite{uzdin2016coherence}. In this scheme, coherence was extracted from one single-body engine and injected into another. The collective work extraction from the engines exhibits an enhancement that scales quadratically with the number of engines.
Improvements in efficiency and work extraction are not the only advantages of collective effects. In~\cite{da2022collective} it was further shown that these effects may stabilize the power output fluctuations of the engine, as measured by the constancy~\cite{ptaszynski2018coherence}.   In the following, we discuss some of the many-body QE unique phenomena in detail and present exactly solvable models.   

Collective phenonomena taking place in the working medium can also lead to enhancements in many thermodynamics operations, such as the charging of many-body quantum batteries~\cite{campaiolienhan2017,andolinabatt2019,campaioli2023colloquium}, or the generation of work in a many-body heat engine. In \cite{niedenzu2018cooperative}, a periodically-driven noninteracting $N$-qubits QHE working between two thermal reservoirs was proven to exhibit cooperative enhancement in its output power, computed in the steady-state. This enhancement is due to superradiant behavior \cite{Dicke54} (see also Sec.~\ref{subsec:coherent}), and causes the output power to scale super-linearly with $N$. A quite different form of collective effect has been proven in the instance of $N$ indistinguishable bosonic engines, coupled to two baths and to a distinct quantum system S that can be measured at the start and at the end of the intraction without affecting the state of the engines \cite{watanabestat2020}. Here, the engines are driven, and they are only interacting with S by means of a time-dependent Hamiltonian. A quantum statistical enhancement in the energy transferred to S is proven that occur when the bosonic engines are statistically indistinguishable.

\subsection{Many-body engines and criticality}

 An important caveat in attaining the Carnot bound is the infinite time required to complete a cycle. 
Thus, the power delivered by the engine vanishes at this limit~\cite{levy14a}. Likewise, any realization of finite generation of power in a realistic engine cycle therefore comes at the expense of the engine's efficiency. The engine performance rate, that takes this trade-off into account, can be defined as the ratio between the power and the distance of the efficiency from the Carnot one~\cite{campisi2016}, i.e., $\dot{\Pi} = \frac{\mathcal{P}}{\Delta \eta}$ with the power output $\mathcal P$ and  $\Delta \eta = \eta_{\rm C} - \eta $.

A potential strategy to boost power without compromising efficiency involves enlarging the working substance. Specifically, employing an array of $N$ identical engines in parallel yields a power that is $N$ times greater than a single engine, while maintaining the same efficiency, where the performance rate scales with $N$ as well. Yet, this linear enhancement in performance rate does not translate to an actual advantage, since it necessitates an equivalent linear escalation in resource investment.

\begin{figure}[h]
\centering
\includegraphics[scale=1.2]{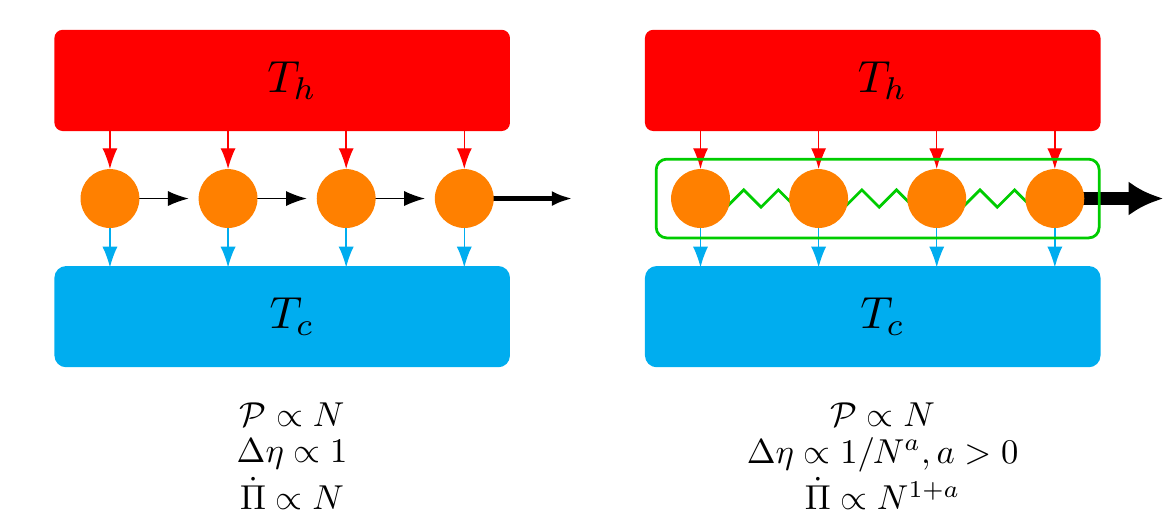}
\caption{Schematic diagram of the many-body heat engine at criticality. Left (Right) panel: scaling of the performance rate $\dot{\Pi}$ of the noninteracting (interacting) many-body engine.~Here $a=(\alpha - z \nu)/d\nu$. Reproduced from \cite{campisi2016}.}
\label{Manybody1}
\end{figure}

The inquiry into whether the performance rate of QHEs can be scaled up beyond linear was positively addressed in several works~\cite{allahverdyanwork2008,allahverdyanopti2010,allahverdyancarnot2013}. Here, a superlinear behavior in the performance rate was found to be a consequence of the occurrence of working mediums with degenerate energy spectra. However, a different mechanism leading to enhancements in the performance rate was found in~\cite{campisi2016}, due to the presence of  many-body interactions in the working medium operating an Otto engine. 

The fundamental concept of this heat engine is illustrated in Fig.~\ref{Manybody1}.
It was found that a universal behavior, with anomalous scaling of the performance rate, is observed
\begin{equation}
    \dot{\Pi} \approx N ^{1 + (\alpha - z \nu)/d\nu},
\end{equation}
when the working medium is on the verge of a second-order phase transition. Here $ \alpha$ is the specific heat, $\nu$ is the correlation length, $z$ is dynamical critical exponent, and $d$ is the working medium dimensions.  The performance rate is influenced by two important parameters: the scaling of the
heat capacity and of the relaxation time. 
The working medium is characterized by a Hamiltonian with a control parameter $\lambda(t)$
\begin{equation}
    \hat{H} = \lambda(t) \mathcal{\hat{K}},
\end{equation}
and the residual part $\mathcal{K}$ makes up the many-body operators for a generic multi-particle system. The heat exchanged with the bath and the work extracted during the different strokes of the QHE, $W$, can be evaluated, leading to an efficiency less than Carnot's efficiency in accordance with heat engine fluctuation relations. The performance of the QHE can be written as
\begin{equation}
\Pi \approx \frac {\partial W }{\partial \Delta \eta} |_{ \Delta \eta=0} \approx N c_{k} (\lambda,T),
\end{equation}
where $c_{k}$ is the specific heat of the working substance. To obtain a super-linear scaling of the performance, an anomalous scaling of the specific heat needs to be achieved, $c_{k} \approx N^{a}$ with $a > 0$,  which can happen on the verge of a phase transition. The underlying physical explanation is that during a phase transition, significant heat exchanges occur alongside negligible temperature variations, resulting in the specific heat capacity becoming infinitely large.

\begin{figure}[h]
\centering
\includegraphics[scale=1.1]{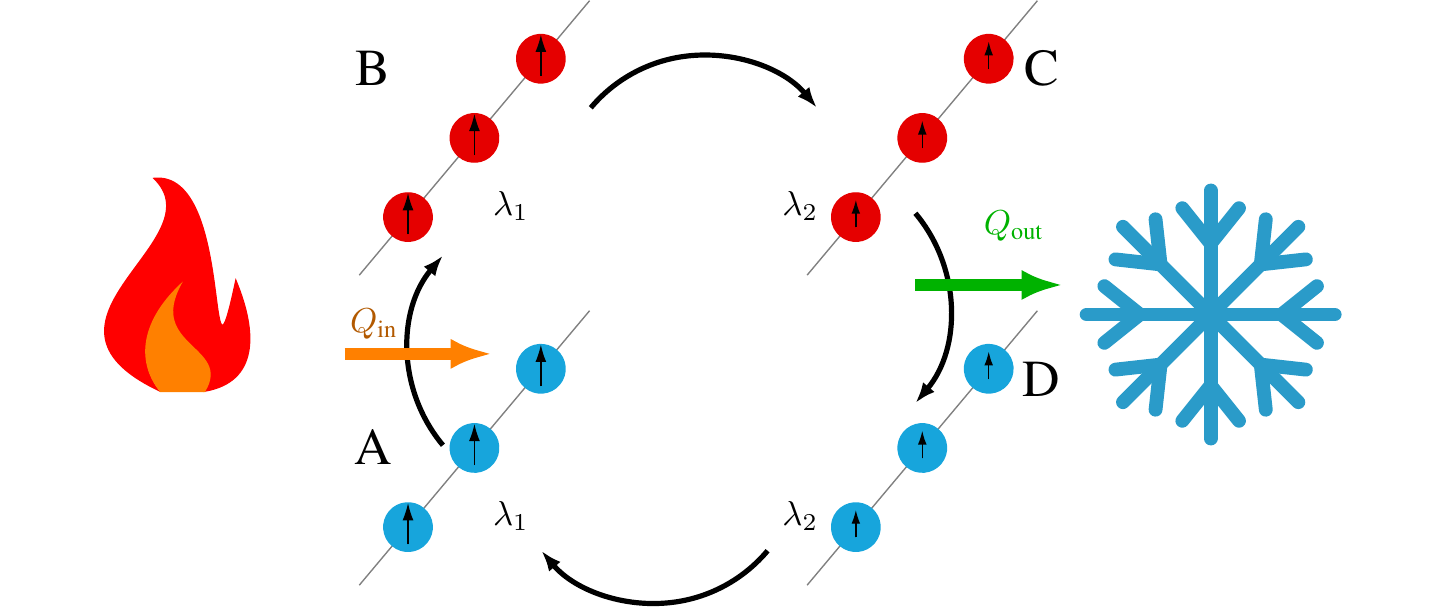}
\caption{Sketch of the many-body Otto cycle. A chain of $N$ spins, interacting through XY interactions, acts as the working medium. It is equivalent to $N/2$ independent thermal machines in the Fourier space. The parameter $\lambda(t)$ modulates the gap in the many-body states.      
$\text{A}\to\text{B}$ ($\text{C}\to\text{D}$) is the energizing (relaxing) isochore stroke, where the working medium is coupled to a hot (cold) thermal reservoir.~$\text{B}\to\text{C}$ ($\text{D}\to\text{A}$) is the isentropic stroke, where the working medium is decoupled from the bath and $\lambda(t)$ is increased (decreased) linearly in time, with finite rates. During the isentropic strokes, the working medium can cross its critical point, depending on $\lambda_1,\lambda_2$. Reproduced from \cite{KZ1}. 
}
\label{Manybody2}
\end{figure}

QHEs dictated by quantum and topological phase transitions and critical points of the many-body working medium have been investigated in various settings~\cite{Dicke1,QT1,QT2, piccitto2022ising}. Transitions between a superfluid and an insulating phase of the quantum matter driving a quantum Otto engine have recently been studied~\cite{fogarty;2020}, showing improved performance in work extraction with a finite-time dynamical protocol for the unitary strokes.

Quantum criticality in many-body QEs has also been studied in analytically solvable models~\cite{KZ1}, accessible in trapped ion experiments. In this study, the Ising and the XY models in a transverse field with a ring geometry were considered in an Otto cycle configuration (see Fig.~\ref{Manybody2}). 
The many-body Hamiltonian reads 
\begin{equation}
H = -  \sum _{j} M_{j} ( c ^{\dagger} _{j} c _{j+1} - c_{j+1} c ^{\dagger} _{j+1}) - \sum _{j} N_{j} ( c ^{\dagger} _{j} c ^{\dagger} _{j+1} - c_{j} c_{j+1})- \sum _{j} R_{j} ( c ^{\dagger} _{j} c _{j} - c_{j} c ^{\dagger} _{j+1}),
\end{equation}
where $c$ and $c^{\dagger}$ are Fermionic annihilation and creation operators and the parameters $M,N,R$ are scalars.
During the two strokes of the quantum Otto engine, the unitary dynamics of the working medium, under the influence of an external parameter (say $\lambda$ with the rate of quench $1/\tau$), might lead to the crossing of the quantum critical point governed by the celebrated KZ scaling law.
The scaling law is characterized by the density of defects (excitations) with respect to the ground state Hamiltonian, $
n_{exc} \sim \tau ^ {\frac{\nu d}{\nu z + 1}}$,
where $d$ is the dimensionality of the system, and $\nu, z$ are the correlation length
and dynamical critical exponent respectively. 
The efficiency of the QHE depends directly on the KZ scaling law and can be diagnosed via its influence on the work distribution statistics in quantum thermodynamical settings \cite{fei,zhang2022}.

\subsection{Long- and short-range interactions}

Many-body solvable Hamiltonians offer a rich array of physics to be explored in the realm of quantum thermodynamics and nonequilibrium dynamics~\cite{mukherjee2021many}. 
A classic example of such a model describes a many-particle (N) dynamics under a time-dependent harmonic trap with frequency $\omega(t)$, along with a two-body interaction potential~\cite{Jaramillo_2016}
\begin{equation}
H(t) = \sum _{i=1}^{N} \left( - \frac{\hbar ^{2}}{2m} \nabla ^{2} + \frac{1}{2}\omega ^{2} (t) r^{2}(t) \right) + \sum _{i \neq j} V(|{r}_{i}-{r}_{j}|). \label{CS}
\end{equation}

The interaction potential is assumed to take the form $V(r/b)=b^{2}V(r)$, which insure a scale-invariant dynamics in the unitary expansion and compression strokes.
This model is then reduced to the 1-d Calogero-Sutherland gas of interacting Bosons in a harmonic trap with a pairwise interaction potential $V(x_{i}-x_{j}) = \frac{\lambda (\lambda -1)}{(x_{i}-x_{j})^{2}}$ .
\begin{figure}[h]
\centering
\includegraphics[scale=1.2]{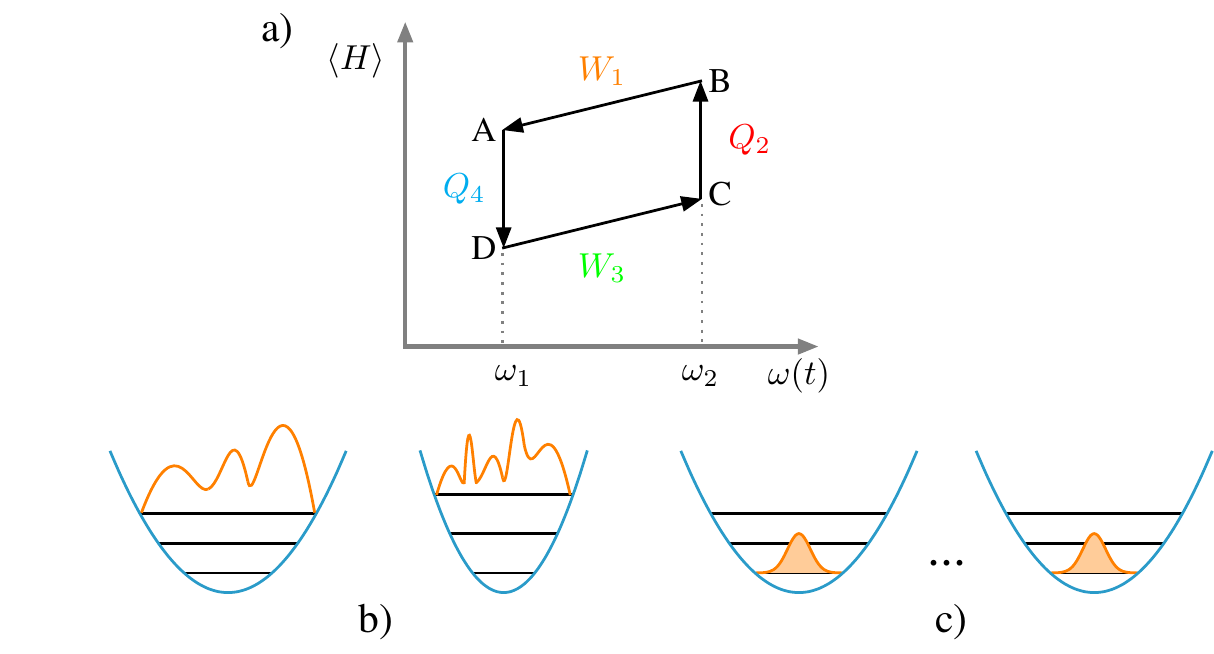}
\caption{a) The thermodynamic cycle of a many-body engine mediated by the time-dependent control parameter $\omega$ that tunes the potential well of the working medium. b) A schematic of the wave-function of a many-body state mediated by the coupling strength $\lambda$ in contrast to c) a single particle trapped in a harmonic bath. Reproduced from \cite{Jaramillo_2016}. }
\label{Manybody}
\end{figure}
Two important parameter values for the interaction strength $\lambda$ are i) $\lambda=0$ where the model becomes a free Bosonic gas in a harmonic trap and ii) $\lambda=1$, a Tonks-Girardeu gas of repulsive Hard-Core Bosons~\cite{girardeau;1971}. For other values of $\lambda$, the model exhibits an inverse square law (long-range) potential and is often reinterpreted as an ideal gas of generalized exclusion statistics~\cite{wilczek;1990}. Thus, the Hamiltonian in Eq.~(\ref{CS}) has a rich  basket of intrinsic many-body physics with an exact spectrum of eigenstates available.

Focussing on the nonequilibrium properties of such a system under external driving, we note that the non-adiabatic average energy of the system turns out to be related to the adaiabtic form through 
\begin{equation}
\langle H(t) \rangle = Q^{*} (t) \langle H(t) \rangle _{ad},
\end{equation}
where $Q^{*}(t)$ is the adiabatic factor that accounts for the non-adiabatic excitations over the adiabatic average.~The adiabatic limit is readily achieved when $ Q^{*}=1 $.
The total work input per cycle -- corresponding to the adiabatic compression phase (AB) and expansion phase (CD) when decoupled from the thermal reservoir -- is $\langle W \rangle = \langle W_{1} \rangle + \langle W_{3} \rangle$, Fig~\ref{Manybody}. Finally, the non-adiabatic efficiency of the QHE is given by
\begin{equation}
\label{eq:efficiency}
\eta = 1 - \frac{\omega _{1}}{\omega _{2}} \left( \frac{ Q^{*} _{CD} \langle H \rangle _{C} -\frac{\omega _{1}}{\omega _{2}}\langle H \rangle _{A} }{ \frac{\omega _{1}}{\omega _{2}}\langle H \rangle _{C} - Q^{*} _{AB} \langle H \rangle _{A}} \right).
\end{equation}
In the adiabatic limit $ Q^{*}=1 $, the efficiency reduces to the Otto efficiency $\eta_O=1-\omega_1/\omega_2$.   
The efficiency Eq.~(\ref{eq:efficiency}) is bound by $\eta\leq \frac{1-\omega_1 Q^*_{CD}}{1-\omega_1}\eta_O$, which is independent of the number of particles and the inter-particle interactions.

To characterize the many-body effects, it is useful to look at the ratio of the optimal power output for the N-Particle engine and that of N single-particle engines, i.e., $P^{N,\lambda}/N P^{1,\lambda}$' for a given coupling $\lambda$. 
Similarly, the ratio between the efficiency at optimal power for these two cases can be explored, i.e., $\eta^{N,\lambda}/\eta^{1,\lambda}$.   
It is concluded in~\cite{Jaramillo_2016} that  many-particle quantum effects and nonadiabtic dynamics can boost both power output and efficiency. Shortcut to adiabaticity methods~\cite{del_Campo_2019,RevModPhys.91.045001} were employed in this model~\cite{e18050168} and others~\cite{Hartmann2020multispincounter,hartmann2020many}, showing further improvement in the performance of the many-particle engines and refrigerators. 

A QHE with a working substance made up of tightly confined potential has also been studied in similar lines~\cite{chen_2019,PhysRevE.98.062119}. The Hamiltonian studied in this case is the  Lieb-Liniger type Hamiltonian
%
%
with a contact interaction portrayed by Dirac $\delta$-distribution. Such models probe the universal nature of Bose gases at low temperature and the role of quantum criticality. 
The influence of many-body effects on quantum thermodynamic cycles has also been explored in Bose-Einstein condensate with nonlinear interactions~\cite{Li_2018}, and in the LMG model mentioned above. In this study, however, their influence is considered in the thermodynamic limit of many particles~\cite{PhysRevE.96.022143}.  

A  variety of many-body condensed matter systems show the presence of long-range inter-particle interactions. Ref.~\cite{PhysRevE.102.012138} considers the implications of such a working medium's on the physics of the devices and quantitatively compares it with the influence of short-ranged interaction on the performance of a QHE. One such Hamiltonian is given by the extension of the famous Kitaev chain to include long-ranged forces:
\begin{equation}
H_{LR} = - J \sum _{j} ^{L} ( c ^{\dagger} _{j} c _{j+1} + c_{j+1} c ^{\dagger} _{j+1}) - \mu \sum _{j} ^{L} ( c ^{\dagger} _{j} c _{j} -\frac{1}{2}) + \frac {\Delta}{2} \sum _{j} ^{L} \sum _{l} ^{L-1} \frac{1}{d_{l}^{\alpha}}( c _{j} c _{j+l} + c^{\dagger} _{j} c ^{\dagger} _{j+l} ).
\end{equation}
Here  $ c$ and $c^{\dagger} $ are the fermionic ladder operators on the $j$th site, and $L$ is the length of the chain. The  parameter $J$ controls the nearest neighbor hopping strength, $\Delta$ controls the fermion pairing interaction, and $\alpha$ characterizes the range of interaction (the short-range limit is achieved for $\alpha \rightarrow \infty$). 
Long-range interaction can play opposing roles in the performance of QHEs, depending on the parameter regime and the thermodynamic cycle. In~\cite{PhysRevE.102.012138}, the long-range effects of both the Otto and Stirling cycles were explored.    
While the Otto cycle exhibits enhancement of the work output and efficiency due to long-range interactions, in the high-temperature regime the Stirling cycle is negatively impacted by them. 
Opposite behavior in the two cycles is also observed near the critical point. The work and efficiency of the Otto cycle decrease with the interaction length, whereas in the Stirling cycle, they increase.

An Otto engine with a many-body localization (\acrshort{MBL}) phase was studied in~\cite{halpern2019quantum}. 
In this setup, the work strokes consist of transitions between thermal states (after thermalization with heat baths) and MBL phases. The QE work per cycle is expressed in terms of the localization length, and it is shown that the MBL spectrum distribution leads to a reduction in the fluctuations and permits a smooth operation of parallel engines compromising the macro-scale engine.       

Optimizing the performance of many-body QHE becomes a difficult task, as the number of possible configurations of the interactions increases tremendously with the number of particles. 
Machine learning~\cite{carleo2019machinelearning_review} provides tools to approach such problems and was recently implemented in optimizing the interaction of many-body nanothermoelectric systems~\cite{machine_learning}.
In particular, the Reinforcement Learning (\acrshort{rl}) approach has been applied to a network topology of interacting thermal systems, utilizing a strategy of optimizing power-efficiency trade-off. A thermal device made up of a many-body working medium is mapped onto a network, i.e.,  a mathematical graph with nodes and vertices.
Each node represents a single-electron level that exchanges electrons with reservoirs and each of the edges between two nodes indicate the presence or absence of interaction between the corresponding level. 
Then a training set for the network is fed into the algorithm by varying the interaction parameters, for the purpose of finding the set of optimized QHEs with respect to power and efficiency.

We note in passing that machine learning is becoming a tool for optimizing the performance of few-body quantum thermal devices as well.  A specific technique in machine learning, called Differential Programming (\acrshort{dp}), has recently been introduced to optimize refrigeration via an Otto cycle in STA conditions~\cite{segal2022optimal_ML}.
This method employs a reward and penalty scheme to a fictitious agent, aiming to minimize the energetic cost of running the QHE  within specific constraints imposed by the \acrshort{STA} scheme. The devised strategies are optimized via deep learning of the training dataset from each engine cycle.

\section{Quantum heat engines beyond the weak coupling and Markovian paradigm}\label{sec:nonMarkov}

\subsection{Quantum Thermodynamics in the strong coupling limit}\label{subsec:strongcoupintro}

In the formulation of thermodynamics, it is often assumed that the interaction between the system of interest and a thermal reservoir is weak. This assumption has several significant thermodynamic implications, one of which is that the reservoir will generally force, in the long time limit, the system to a canonical Gibbs state. This equilibrium state should then grant the possibility of exploring changes in thermodynamic variables defined only at equilibrium. Moreover, the weak system-bath coupling assumption is crucial for consistent energy accounting and for the division of energy into heat and work. Heat can be understood, in this case, as the energy dissipated to the reservoir, and entropy production can be associated with it. 
On the macroscale, weak coupling usually provides a good description, as the system interacts with the reservoir through the boundaries that constitute a small portion of the system’s degrees of freedom and contribute very little to its energy. On the other hand, in the single-particle scale of the system, this assumption doesn’t necessarily hold true. The energy scale of the system becomes comparable to that of the interaction, the boundary between the system and the reservoir is obscured, and the system will generally relax to a non-Gibbisian state. As a consequence, the definition of heat becomes ambiguous. 

Many paths to reconciling the thermodynamics of open quantum systems in the strong coupling regime have been explored. 
Describing thermodynamics with a single reservoir was established using the Hamiltonian of mean force, which overcomes the deviation from a Gibbs state by defining an effective Hamiltonian~\cite{ jarzynski2004nonequilibrium,talkner_hangii,rivas2020strong}  to construct thermodynamic potentials. 
Quantum thermodynamics in the strong coupling regime was also explored using well-established nonequilibrium quantum dynamics theories and methods. This includes the Polaron transformation~\cite{xu2016polaron,gelbwaser;2015}, noninteracting blip approximation~\cite{wang2015nonequilibrium}, Reaction coordinates~\cite{nazir2014_r_c_mapping,strasberg2016_r_c_mapping,nazir2018r_c_mapping}, non-equilibrium Green’s function methods~\cite{esposito2015quantum,whitney2018non,ludovico2014dynamical,bruch2016quantum,ludovico2016dynamics, bergmann2021_nefg,seshadri2021entropy}, hierarchical equations of motion~\cite{schinabeck2016hierarchical,dou2020universal,batge2022nonadiabatically},  the stochastic surrogate Hamiltonian~\cite{katz2016quantum},  and alternative approach of strong coupling via heat exchangers~\cite{uzdin2016quantum_strong} . The list above is limited and the subject accedes the scope of  this review, and in the following, we will focus on a few of these methods in the context of QHE and refrigerators.

\subsubsection{Study of quantum heat engines using the polaron transformation method}

To study the dynamics of an open quantum system, out of equilibrium,  and strongly coupled to a bath at finite temperature, one of the earliest methods developed follows the celebrated polaron transformation technique~\cite{lang1963kinetic}.
The spin-boson model has been investigated via this method~\cite{wurger_1998_strong}, and later was applied to the externally driven variation of the model~\cite{wilson_2002_strong, mccutcheon_2010_strong}, and to nonequilibrium studies that include more than one bath~\cite{hsieh2019nonequilibrium,batge2022nonadiabatically}.   
To give a brief overview of the method, we follow the model in reference~\cite{mccutcheon_2010_strong} of an exciton-phonon system that is driven by an external field. 
 We thus model a quantum dot described by just two energy levels, which are separated by an energy gap $\omega_{0}$. The lower level is the dot's ground state while the excited state describes a single-exciton state. The dot is driven using a laser of frequency $\omega_l$. To account for the effect of phonons, the effective two-level system also interacts with an infinite set of quantum harmonic oscillators. The Hamiltonian in a rotating frame with respect to the laser frequency $\omega_{l}$ can be expressed in terms of the Pauli matrices as follows 
\begin{equation}
    H_{EP} = \delta \sigma_{+}\sigma_{-} + \frac{\Omega(t)}{2}\sigma_x + \sum_{k} \omega_{k} b^{\dagger}_{k}b_{k} + \sigma_+ \sigma_{-}  \sum_{k} (g_{k}b^{\dagger}_{k} + g^{*}_{k}b_{k}).
\end{equation}
Here $\Omega(t)$ is the Rabi frequency, $\delta$ is the detuning, and $g_{k}$ are the couplings to the phononic modes.
The polaron transform is then carried out, $H_{p}= e^{S} H_{EP} e^{-S}$, with
\begin{equation}
    S=\sigma_+ \sigma_{-} \sum_{k}( \alpha_{k}b^{\dagger}_{k} - \alpha^{*}_{k}b_{k}), \;\;\;{\rm and}\;\;\; \alpha_{k} = \frac{g_{k}}{\omega_{k}}. 
\end{equation}
The final transformed Hamiltonian,
\begin{equation}
    H_{p} = \delta^{\prime} \sigma_{z} + \frac{\Omega_{r}(t)}{2}  \sigma_{x} + \sum_{k} \omega_{k} b^{\dagger}_{k}b_{k} + \frac{\Omega(t)}{2}\big(\sigma_{x}B_{x} + \sigma_{y}B_{y}\big),
    \label{polaron1}
\end{equation}
displays several distinct effects: a) renormalization of the exciton parameters
due to the strong coupling, in particular, a shift in the detuning, $\delta^{\prime}=\delta-\sum_k\omega_k |\alpha_k|^2$, and an effective Rabi frequency, $\Omega_{r}(t) = \Omega(t)\langle B \rangle$; 
b) renormalized bath interaction operators $B_{x}=(B_+ +B_- +2\langle B \rangle)/2$, $B_{y}=i(B_+ -B_-)/2$,  and $B_{\pm} =e^{\pm ( \sum_{k} \alpha_{k}b^{\dagger}_{k} - \alpha^{*}_{k}b_{k})}$. 
For a phononic bath in thermal equilibrium at inverse temperature $\beta$, the expectation value of $B_{\pm}$ is simply $\langle B \rangle= \exp \Big[ -\frac{1}{2} \sum |\alpha_{k}|^{2} \coth (\frac{\beta \omega_{k}}{2})\Big]$.
 The clear advantage of the polaron transformation lies in a separation of the exciton-phonon interaction effects into two contributions, namely, the renormalization of the driven two-level Hamiltonian and a residual coupling to the phononic degrees of freedom. In this new reference basis, the dynamical effects brought in by the first contribution are treated exactly, while the influence of the residual interaction is treated perturbatively. Eventually, this leads to a quantum master equation that, for moderately strong coupling, outperforms conventional perturbative expansions. The dynamics of the system generated by the polaron Hamiltonian in Eq.~\eqref{polaron1} can then be solved using different methods and perturbative expansions.
In~\cite{mccutcheon_2010_strong}, a time-local master equation for the reduced system was derived.
Phonon-mediated electronic transport, with nonequilbrium and coupled quantum dots connected to external leads, have also been explored using master equation techniques for the polaron~\cite{maier_2011_strong, walter_2013_transport, krause_2015_thermodynamics}.
More recently, the polaron transformation, together with hierarchical equations of motion, were implemented to study quantum friction and work extraction in an electronic system coupled to two leads and to a phonon mode, with and without damping~\cite{batge2022nonadiabatically}. This work spans both the weak and strong system-bath coupling regimes and the slow- and fast-driving limits.

The polaron transformation was also adapted to studying the light-harvesting systems modeled as quantum heat engines~\cite{qin2017effects,xu2016polaron}.  The light-harvesting energy process is simplified and reduced to a three-level system, which works as the antenna that captures energy from the sun and transfers it to the reaction center.  Transitions between the ground state and the second excited state are provided by light from the sun, the first and second excited states are mediated by phonons, strong coupling is captured by the polaron transformation, and, finally, the reaction center is modeled as a reservoir that couples the ground state with the first excited one to trap the energy. The polaron transformation covers a wide range of system-phonon coupling strengths. This coupling modifies the steady state to include coherence, which significantly affects the efficiency and energy flux in the process.

A different QHE, with a two-level system as a working medium, was investigated in~\cite{gelbwaser;2015}.
The system Hamiltonian including the external driving field in a rotating wave approximation reads $  H_{S} = \frac{\omega_{0}}{2} \sigma_{z} + \frac{\Omega}{2} \big( \sigma_{+} e^{-i \omega_{R} t} + \sigma_{-}e^{i \omega_{R} t} \big)$. 
%
The system is also simultaneously coupled to two Bosonic heat baths at inverse temperatures $\beta_{c}$ and $\beta_{h}$, with frequencies $\omega_{c,k},\omega_{h,k}$.
The interaction Hamiltonian reads,
\begin{equation}
    H_{SB} =  \lambda_{c} \sigma_{z}   \sum_{k}( g_{c,k}a^{\dagger}_{k} + g^{*}_{c,k}a_{k}) +   \lambda_{h} \sigma_{x}  \sum_{k}( g_{h,k}b^{\dagger}_{k} + g^{*}_{h,k}b_{k}).
\end{equation}
Here, the working medium is coupled differently to the hot and the cold baths. Indeed, not only the coupling strengths $\lambda_{c},\lambda_{h}$ are different, but also the coupling factors to each bosonic mode of the two baths are described by two distinct functions $g_{h,c},g_{c,k}$. Moreover, the cold bath induces dephasing while the hot bath can lead to population decay.
The transformation can be carried out in a manner similar to the one presented above, only with $S=\sigma_z\sum_k (\alpha_k a_k^{\dagger} -\alpha^{*}a_k)$ and $\alpha_{k}=\lambda_{c}g_{c,k}/\omega_{c,k}$. In contrast to the single-bath cases, the polaron transformation in the multi-bath model reveals energy exchange between the two baths through their coupling to the system, which is directly observed in the polaron-transformed Hamiltonian.
The compromised system-bath partition, due to the polaronic transformation performed w.r.t. the cold bath only, makes the identification of heat flows from the baths convoluted in a master equation for the reduced-state approach. 
Furthermore, at the ultra-strong coupling to the baths, the power output is suppressed as the coupling increases, indicating that QHE's require some degree of separability to operate (see Fig.~\ref{fig:powerC}).

\subsubsection{Reaction Coordinate Mapping in studying strongly coupled engines}\label{subsubsec:enginestrongcoupl}
   
A method in both classical and quantum dynamics, the Reaction Coordinate (\acrshort{RC}) mapping has opened up an area of study in the physics of strongly coupled open systems~\cite{nazir2014_r_c_mapping,strasberg2016_r_c_mapping,nazir2018r_c_mapping}.
The spirit of the framework lies in finding a suitable coordinate transformation that can eventually lead to a weak coupling description of a dressed open system in the presence of a bath. This outcome helps in applying the usual weak coupling tools to the case of certain strongly coupled open systems. For the purpose of our discussion, we focus on a specific model, the TLS, which also serves as the working medium for investigating the thermodynamics of a strongly coupled QHE \cite{nazir17}. The Hamiltonian in terms of Pauli matrices and the identity reads,
\begin{equation}
 H_{S}(t) = \frac{\mu(t)}{2} I + \frac{\epsilon(t)}{2} \sigma_{z} + \frac{\Delta(t)}{2} \sigma_{x} \label{TLS1},
\end{equation}
where $\epsilon(t) $ is the TLS bias, $\Delta(t) $ the tunneling term coefficient, and the level-splitting of the eigenstates  $\mu(t)= \sqrt{\epsilon ^{2} (t) +  \Delta ^{2}(t)} $.
The TLS couples to two reservoirs, a model invoked to represent the discrete stroke Otto cycle of a heat engine, in a Caldeira-Leggett form
\begin{equation}
    H_{SB} = \sum_{k} \frac{p_{k} ^{2}}{2 m_{k}} + \frac{m_{k} \omega_{k}^{2}}{2} \Big( x_{k} - \frac{d_{k} \sigma_{z}}{m_{k}\omega_{k}^{2} }\Big)^{2} \label{eq:TLS2}.
\end{equation}
Here $(x_{k},p_{k})$ denote the canonical coordinates of each bath's oscillator of mass $m_{k}$ and frequency $\omega_{k}$, while $d_{k}$ quantifies the coupling to each mode $k$. The bath is characterized by a spectrum of Bosonic oscillators encapsulated in the spectral density (in this case both a cold and hot reservoir) $ J(\omega) \equiv \sum_{k} f_{k} ^{2} \delta (\omega - \omega_{k} )$, where the rescaled system-bath coupling is given by $ f_{k} = d_{k}/ \sqrt{m_{k}\omega_{k}}$ and can be arbitrarily strong.
\begin{figure}[h]
\centering
\includegraphics[scale=0.9]{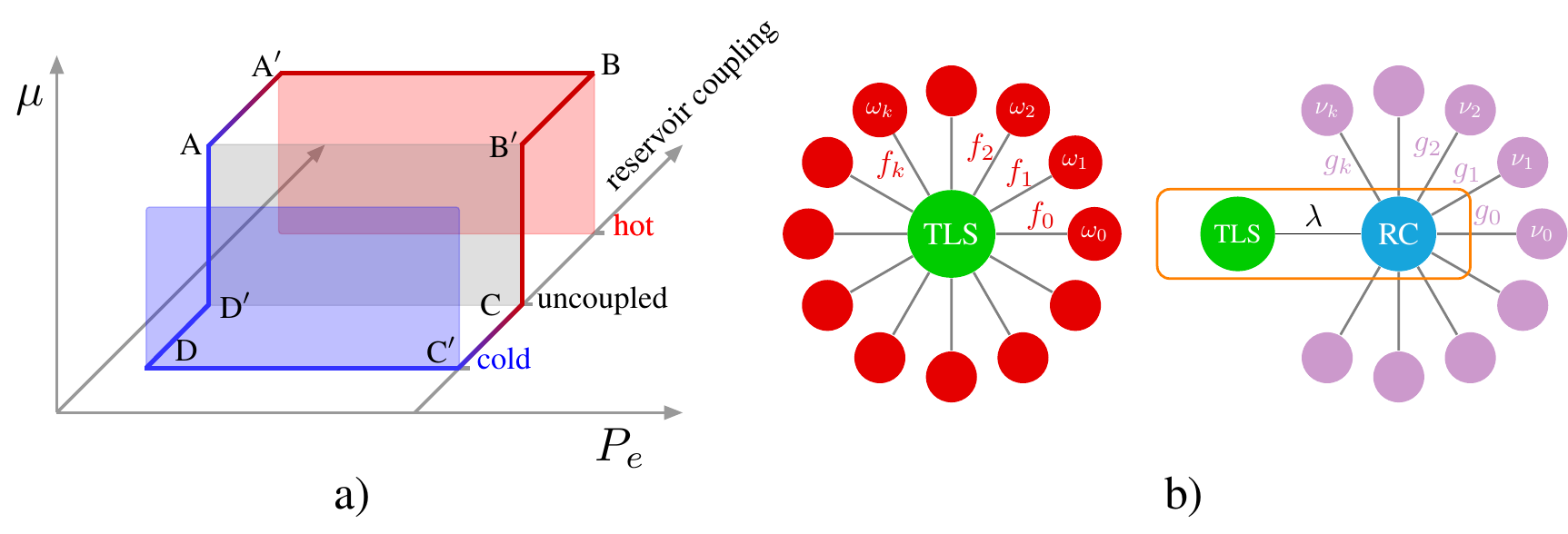}
\caption{Sketch of the Otto cycle in the strong coupling regime. a)  Engine's operating cycle: $A^{\prime}\rightarrow B$ ($C^{\prime}\rightarrow D$) is an isochoric thermalization stroke with the hot (cold) bath, while $B^{\prime}\rightarrow C$($D^{\prime}\rightarrow A$) is an isentropic expansion(compression) stroke. Additional  strokes, i.e., $B\rightarrow B^{\prime},C\rightarrow C^{\prime},D\rightarrow D^{\prime},A\rightarrow A^{\prime}$ are included to model the energetic contributions resulting from the coupling/decoupling operations. b) Schematic diagram of the RC mapping scheme. The star geometry on the left (red circles), where the TLS is coupled to each bath oscillator, is exactly mapped to the scheme on the right (violet). In the new setting, the TLS only interacts with a single oscillator (RC) (natural frequency $\Omega$) with coupling strength $\lambda$, and the RC interacts with a transformed bath of harmonic oscillator modes with frequencies $\nu_{k}$. Reproduced from \cite{nazir17,nazir2018r_c_mapping}.}
\label{NonMarkov1}
\end{figure}

We are now in a position to define the RC mapping scheme. It maps the Hamiltonian in Eqs.~\eqref{TLS1} and \eqref{eq:TLS2} to an enlarged system $S^{\prime}$ in terms of a collective coordinate, which is in turn weakly coupled to a redefined bath $B^{\prime}$ (see Fig.~\ref{NonMarkov1}). This process leads to a transformed system-bath Hamiltonian
\begin{equation}\label{eq:nonmarkov3}
    \tilde{H} = \underbrace{\Big( H_{S}(t) - \lambda \sigma_{z} (a^{\dagger} + a) +  \Omega a^{\dagger}a \Big)}_{\rm{S'}} + \underbrace{\Big( \sum_{k} g_{k} (a^{\dagger} + a)(r_{k}^{\dagger} + r_{k}) + \sum_{k} \nu_{k} r_{k} ^{\dagger} r_{k} \Big)}_{\rm{B'}}. 
\end{equation}
The first parenthesis represents the TLS coupled arbitrarily and strongly (via $\lambda$) to a reaction coordinate, denoted by the annihilation and creation operators, $a$ and $a^{\dagger}$, with frequency $\Omega$. The second parenthesis accounts for the coupling between the collective coordinate and the residual bath modes with couplings $g_{k}$, created and annihilated at natural frequencies $\nu_{k}$ by the operators $r^{\dagger}_{k}$ and $r_{k}$ respectively. 
The analysis of the Otto cycle with the Hamiltonian $H_{S'}(t)$, rather than the reduced system $H_{S}(t)$, is accomplished by virtue of the RC mapping formalism, reduced to the standard weak coupling form with product system-bath states.

\begin{figure}%
    \centering
    \includegraphics[width=8cm]{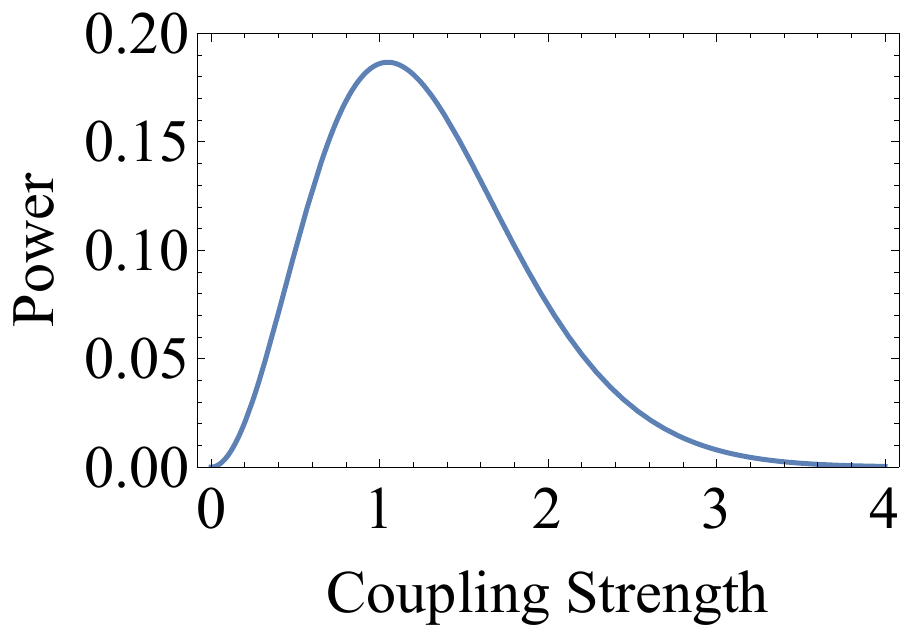} %
    \caption{ The relationship between power output and the system-bath coupling strength for a quantum heat machine studied in Ref.~\cite{gelbwaser;2015}. Noticeable is the turnover point beyond which power output decreases in the ultrastrong coupling regime, attributed to the hindered work extraction due to the breakdown of system-bath separability. Figure courtesy of D. Gelbwaser-Klimovsky.}%
    \label{fig:powerC}%
\end{figure}

The cycle operates as follows: the TLS undergoes an isochoric thermalization stroke $A^{\prime}\rightarrow B$ with the hot bath, followed by an isentropic stroke $B^{\prime}\rightarrow C$, an isochoric thermalization stroke $C^{\prime}\rightarrow D$ with the cold bath and, finally, a second isentropic stroke $D^{\prime}\rightarrow A$. Unlike conventional Otto cycles, here the energetic contributions coming from  coupling (decoupling) the working medium to (from) the baths have been taken into account (see Fig.~\ref{NonMarkov1} a).
Moreover, the usual expressions for work and heat along the isentropic and isochoric strokes are exactly calculated. In the two limits of adiabatic and quenched isentropic strokes, the main source of work costs is the coupling and decoupling of the working medium and the thermal bath, a feature ignored in the weak coupling case. Thus, strongly coupled QHEs compromise the efficiency of work extraction concomitant with the predictions in continuous QHEs~\cite{gelbwaser;2015}.
Figure~\ref{fig:powerC} illustrates the correlation between the power output of the (QHE) and the strength of the interaction between the working medium and the bath. It demonstrates that the engine's power output enhances as the coupling strength intensifies, reaching an optimum level beyond which any additional enhancement in coupling strength negatively impacts the engine's efficiency. 

  
In a three-level QAR (see also Sec.~\ref{subsubsec:absopt}), RC mapping has been applied to figure out the most efficient cooling window and provides a comparison to the weak coupling case~\cite{Ivander2022_rc_strong_coupling}. In this work, the fact (discussed above) that the RC coordinate ultimately couples weakly to the residual bath is utilized to introduce the usual perturbative quantum Master equation approach. 
The merging of these two theoretical techniques leads to remarkable results
in thermal devices with multiple strongly coupled baths: in the case of QAR, the cooling window (see Sec.~\ref{subsubsec:absopt}) is modified as a result of the renormalization of the energy levels of the working medium. Thus, it is possible to achieve refrigeration under different parameter conditions with respect to the weak-coupling case. Moreover,  
a direct heat-flow pathway between the thermal reservoirs takes place, without any change in the population of the
QAR's energy levels. Such inter-bath leakage --which is fundamentally a heat transport phenomenon that was studied earlier via these two techniques in [228]-- leads to a decrease in the current extracted from the cold bath and diminished performance at strong coupling.
Finally, we would like to mention that such a scheme is also useful in characterizing the generation of quantum coherence due to strong system-bath coupling in QHEs fuelled by TLS (as described above in Eqs.~\eqref{TLS1} and~\eqref{eq:TLS2}), leading to bounds on efficiency and power output~\cite{newman_2020_bounds} that cannot be delineated by the Born-Markov master equation. These results are traced back to the generation of coherence in the working medium, which is directly related to the system-bath coupling and not originated from the driving field.

\subsubsection{Perturbative corrections to engine observables due to strong coupling}\label{subsubsec:pertstrongcoup}

A common theme of the above studies is that strong coupling between the system and the bath produces correlations that adversely affect the performance of QHEs. Irreversibility in the dynamics of a nonequilibrium system in contact with a heat bath leads to universal bounds on work extraction of small thermal machines, as well as on heat dissipation and Carnot efficiency. Perturbative methods can be employed to evaluate these corrections due to strong coupling in a systematic and universal manner~\cite{eisert2014quench, eisert_2018_strong_coupling}, without recourse to a specific model per se.

A system S and a bath B are assumed to be coupled through a potential $V$, allowing for strong interactions. Let us consider an N-step process, each step being labeled by $i$ and consisting of three elementary operations: 
(a) S and B are brought into contact, with the total Hamiltonian $H^{(i)}= H_{S} + H_{B} + V$, and the possibility of $V$ being turned on and off on the $i$-th step. The average work gained during such a step is denoted by
$W^{(i)}_{V} = \tr[\rho ^{(i)} V]$; 
(b) a fast quench transformation of the system Hamiltonian $H_{S}$, i.e.,  $H^{(i)}_{S}\rightarrow H^{(i+1)}_{S}$ in which the state $\rho^{(i)}$ remains unchanged and an amount of work, $W^{(i)} = \tr [\rho_{S} ^{(i)} (H_{S} ^{(i)} - H_{S} ^{(i+1)})]$, is obtained.
And; 
(c) a thermalization process at the $(i+1)$-th step, which only includes energy exchange between S and B via the interaction $V$ (which involves zero work cost). At the end of each thermalization step, it is assumed that the composite system SB thermalizes to the Gibbs state computed with respect to the total Hamiltonian $\omega_{\beta}(H^{(i+1)})$, with $ \omega_{\beta} (H) = e^{-\beta H}/ \tr [e^{-\beta H} ] $,  so that the reduced state of the system S reads
$\rho_{S}^{(i+1)} = \tr_{B} [\omega_{\beta}(H^{(i+1)})]$~\cite{Gogolin_2016}. The optimal protocol for maximizing the work extraction can be found for initial uncorrelated state $\rho^{(0)} = \rho_{S} \otimes \omega_{\beta}(H_{B})$. In addition, the initial Hamiltonian is denoted with $H_{S}^{0}$, $H^{1}_{S}(H^{N}_{S})$ is the Hamiltonian at the step 1(N) when $V$ is turned on (off) and, at the end of the protocol, the system Hamiltonian gets back to $H_{S}^{0}$, i.e., $H^{(N+1)} = H^{(0)}$. By considering the established optimal work extraction protocol valid in the weak coupling regime~\cite{eisert2014quench}, the total work at the end of the protocol for arbitrary $V$ can be divided into three contributions,  
\begin{equation} \label{strong_correction}
    W = W^{({\rm weak})} - \Delta F ^{({\rm res})} - \Delta F^{({\rm irrev})}. 
\end{equation}
Here $W^{({\rm weak})}=F(\rho_{S},H_{S})-F(\omega_{\beta}(H_{S}),H_{S})$, with $F(\rho,H)=\tr [\rho H] +(1/\beta)\tr [\rho \log \rho]\geq 0$, is the maximum amount of work that can be extracted in the weak coupling regime.
$\Delta F^{(\rm irrev)}=F(\rho^{(0)},H^{(1)})-F(\omega_{\beta}(H^{1}),H^{(1)})\geq 0$ is a correction term generated due to irreversible energy dissipation through the bath $B$ when the interaction $V$ is turned on, and $\Delta F^{({\rm res})}=F(\omega_{\beta}(H^{N}),H^{0})-F(\omega_{\beta}(H^{0}),H^{0}) \geq 0$ is the extractable work left in the final state.
It is immediately evident that strong coupling diminishes work output over and above the weak coupling term $W^{({\rm weak})}$. 
Taking a perturbative route, i.e., by defining the interaction potential as $gV$ (with $g$ being a perturbative parameter), a rigorous minimum bound on $\Delta F^{(\rm irrev)}$ can be evaluated. Up to the second order in $g$, the term fundamentally depends on the generalized Kubo-Mori covariance matrix of the interaction, i.e., $\Delta F^{(\text{irr})}_{\text{min}}\simeq (\beta g^2/2)\mathrm{cov}_{\omega_{\beta}(\tilde{H}^{0})}(\tilde{V},\tilde{V})$, where $\mathrm{cov}_{\omega_{\beta}(H)}(A,B) = \tr [A_{H,\beta}B\omega_{\beta}(H)]-\tr[A\omega_{\beta}(H)]\tr[B\omega_{\beta}(H)]$, $A_{H,\beta}=\int_{0}^{1}\mathrm{d}\tau e^{\beta \tau H} A e^{-\beta \tau H}$ and the modified operators are defined by means of the following equations: $\tilde{V}=V-\tr_{B}[V\omega_{\beta}(H_{B})]$, $\tilde{H}^{(0)}=\tilde{H}_{S} + H_{B}$, $\rho_{S}=\omega_{\beta}(\tilde{H}_{S})$. Analogous results hold for $\Delta F^{(\text{res})}_{\text{min}}$ (see details in~\cite{ eisert_2018_strong_coupling} for the perturbative calculations)

With the essentials of the nonequilibrium dynamics under strong coupling in place, QHEs employing equilibration with two baths at inverse temperatures $\beta_{c}$ and $\beta_{h}$ can be investigated in the minimal dissipation regime,  where the additional amount of dissipated heat due to the strong coupling is minimal. This leads to a corrected maximum efficiency
\begin{equation}
    \eta = \eta_{\rm{C}} - g^{2}  \frac{\beta_{h}}{\beta_{c}} A + O(g^{3}),
\end{equation}
%
 where $A=(K^{h}_{q}/Q^{\mathrm{weak}}_{h} + K^{c}_{q}/Q^{\mathrm{weak}}_{c})$, $Q^{\mathrm{weak}}_{h/c}=(1/\beta_{h/c})\Delta S^{weak}$ and $K^{h/c}_{q}\geq 0$ are positive coefficients derived from the perturbative expansion \cite{eisert_2018_strong_coupling}. Although dissipation and coupling strengths reduce the efficiency of the QHE, the power output can be enhanced by shortening the thermalization time scale ($\tau$). The correction terms for the maximum power output again follow a series expansion, in the powers of the perturbation parameter $g$, and depend on the characterization of equilibration times for the system S. The latter is a challenging problem in itself for a variety of many-body models.

\subsubsection{Strongly-coupled quantum heat engines with Matrix Product States}\label{subsubsec:engineTN}

The strong system-bath coupling regime entails quantum correlations among the working medium and the reservoirs. In this regime, the same definitions of quantum thermodynamics concepts, such as heat and work, become challenging~\cite{Allah:work,esposito2015nature,strasberg_2019_strong_coupling,talkner_hangii,bergmann2021_nefg,seshadri2021entropy}, and energy exchange statistics require exact approaches~\cite{agarwalla2015full,carrega2015functional,wang2017unifying,funo2018path}. In the absence of external driving fields, strong system-bath interactions lead the system to thermalize to the equilibrium state of the whole system + bath~\cite{weissbook,Talkner2020,rivas2020strong}. On the other hand, when external fields are present, the resulting non-equilibrium steady state leads to nontrivial energy flows, where the system-bath interaction channel drains a relevant amount of power~\cite{Carrega:energyexchange,kato2016quantum} and coherence generation takes place~\cite{francica2019role} at the same time.    
The impact of non-Markovianity due to strong system-bath correlations in driven QHEs has been addressed in a number of recent works (see Sec.~\ref{subsubsec:enginestrongcoupl}). Yet, the conventional approaches are generally valid in limited parameter windows and cannot be easily extended to more complex settings, where the working medium can possibly be composed of more than one quantum degree of freedom.

In the last two decades, Matrix Product States (\acrshort{mps}) ansatz, and Tensor Networks (\acrshort{tn}) generalizations~\cite{schollwock2011density,paeckel2019time} have been among the prominent nonperturbative techniques for studying strongly-correlated, many-body systems.
Recent approaches tried to tackle the problem of nonequilibrium energy exchange in the strong coupling regime by making use of nonperturbative~\cite{tamascelli2018nonperturbative} methods. Finite-temperature transport through impurities and many-body systems has also been studied~\cite{bertini2021finite}, employing many-body techniques such as MPS~\cite{rams2020breaking}.

While conventionally employed to study the ground state properties of many-body quantum systems, MPS have proved useful in simulating the dissipative dynamics of quantum systems, where exact diagonalization of the system Hamiltonian becomes unpractical.
Indeed, pioneering approaches based on the chain mapping~\cite{chin2010exact} showed that an exact linear transformation can be performed. The transformation maps the star geometry in Fig.~\ref{NonMarkov1}b into a linear chain comprising the system and new bosonic degrees of freedom, each coupled to its nearest neighbor.
The resulting model allows for a more efficient treatment of popular dissipative Hamiltonians than the spin-boson model~\cite{prior2010efficient} (see Sec.~\ref{subsubsec:enginestrongcoupl}).
More recent works address polariton systems~\cite{del2018tensor} and multi-terminal settings combining MPS with machine-learning methods~\cite{schroder2019tensor}.

Efficient representations of the density matrix in terms of matrix product operators have been achieved in~\cite{strathearn2018efficient}, and this approach has proven successful in describing heat statistics in the strong coupling regime~\cite{popovic2021quantum}.   

More recent approaches address the problem of dynamics of impurities interacting with multiple reservoirs, e.g., described by means of the Anderson model, and the challenge of devising theoretical approaches based on the peculiar properties of the entanglement in the temporal domain~\cite{thoenniss2022non,thoenniss2022efficient}.

Despite the huge theoretical effort in this direction, finite-temperature simulations of open-system dynamics in the parameter regimes relevant to strong-coupling QHEs remain quite challenging; they tend to become increasingly costly due to the need to adopt purification methods, requiring ancillary degrees of freedom.
Thus far, in order to circumvent such limitations, several theoretical solutions have been proposed~\cite{levy2020modeling}. 

In~\cite{tamascelli2019efficient}, a finite-temperature approach based on chain mapping and orthogonal polynomia is developed for a bath composed of a continuum set of quantum harmonic oscillators, described by a spectral density function $\mathcal{J}(\omega)$.
It is based on the insight that the two-time correlation function of the bath -- ~i.e.,~$\ave{X(t)X(0)}=\int_{0}^{+\infty}\mathrm{d}\omega J(\omega)[e^{-i\omega t}(1+\bar{n}(\omega)) + e^{i\omega t}\bar{n}(\omega)]$, where $\bar{n}(\omega)=(\exp(\beta \omega) -1)^{-1}$ (see also Eq.~\eqref{eq:nonmarkov6}) -- is equivalent to that of an extended harmonic bath, comprising positive and negative frequencies having \emph{zero temperature} and corresponding to an effective spectral density $\mathcal{J}^{\text{eff}}(\omega)=(\mathcal{J}(\omega)/2)(1+\coth(\beta\omega/2))$. 
If the starting system-bath state is factorized, and the initial reduced system state is pure, the dynamics of the system can be efficiently simulated as a global pure state. 
By employing standard chain mapping and the MPS approach, the pure global state of the system is propagated without incurring the use of ancillary degrees of freedom. 
In other words, the dynamics obtained by starting from a product state of the system and the vacuum state of the extended bath is  equivalent to the one achieved by taking an initial product state  of the system and a thermal state of the real bath, for any fixed inverse temperature $\beta$. Inverse linear transformations can in principle be applied to map the above state back to the real bath state space.

\begin{figure}[h]
        \centering
\includegraphics[scale=0.35]{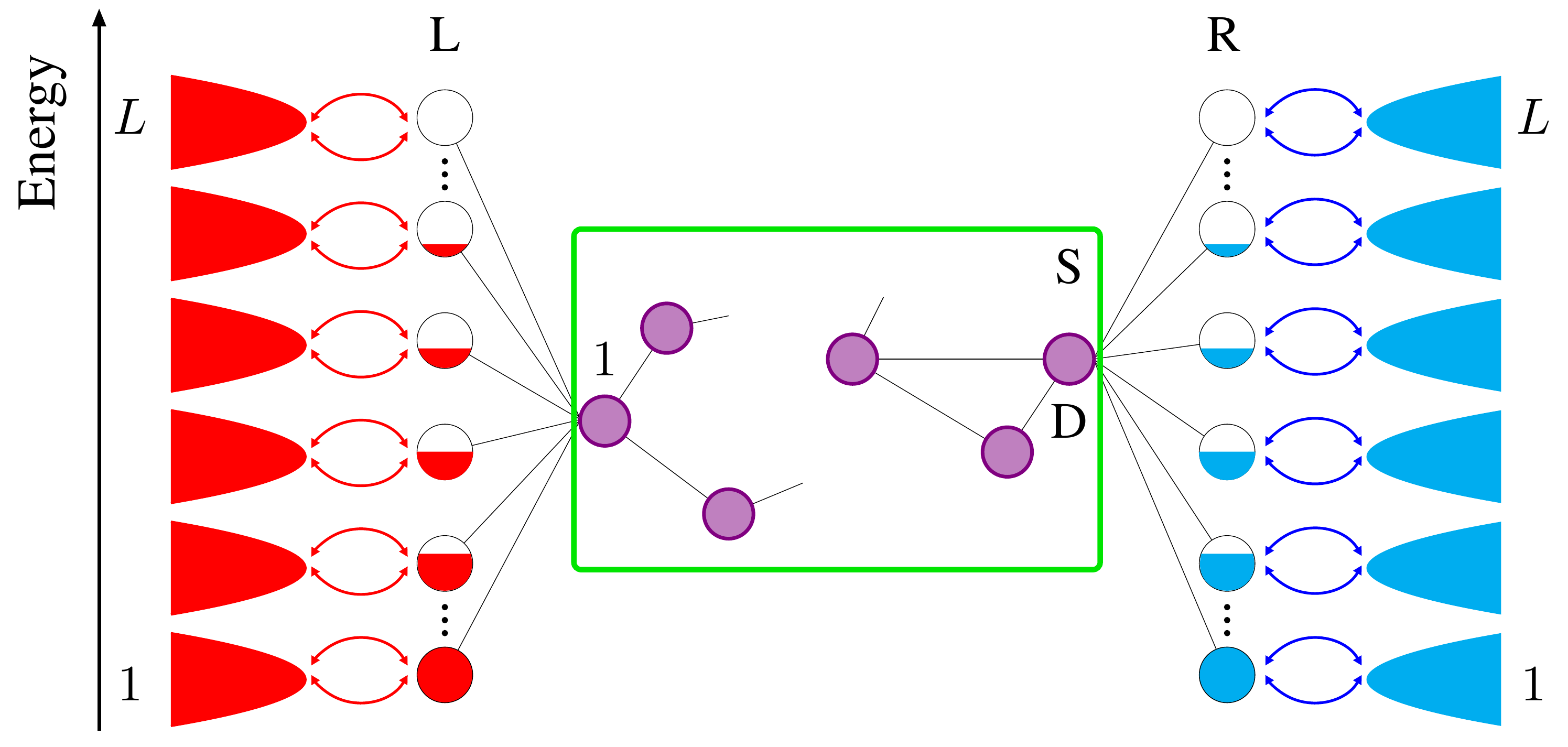}
\caption{Schematic diagram of the TN-based approach to QHE. The system S comprised many strongly-interacting fermionic particles. Through its boundaries, the system is coupled to $L$ effective fermionic levels $\varepsilon_{k}$, and each of the latter is weakly coupled to a fictitious reservoir. The overall effective dynamics is captured by means of a quantum master equation that, after superfermion transformation, can be simulated by means of MPS. Reproduced from \cite{brenes2020tensor}.}
\label{fig:NonMarkovTN}
\end{figure}

In \cite{brenes2020tensor}, a quite different and sophisticated method is presented  to simulate the physics of autonomous quantum thermal devices (see also Sec.~\ref{sec:thermoel}) in the strong coupling regime.
Here two leads at different temperatures are modeled as infinite fermionic systems, with spectral density $\mathcal{J}(\omega)$. The working medium can be either a single fermionic resonant level tunnel coupled to the leads, or an interacting system coupled by means of Coulomb-like interactions. The crucial assumption is that the physical properties of each lead can be reproduced just by making use of a finite number~$L$ of damped fermionic modes, i.e., a mesoscopic approximation for the lead is performed.
In other words, the real fermionic system describing the lead is replaced by a finite number of modes of energy $\varepsilon_{k}$, each coupled to a fictitious infinite reservoir $B_k$ with constant spectral density $J_{k}(\omega)=\gamma_k$.
These additional reservoirs' only role is to guarantee the damping of the effective modes. Indeed, from the Heisenberg equation of motion for each fermionic operator $c_{p}$ belonging to the working medium, it follows that such a scheme is equivalent to assuming for the $L$ modes an effective spectral density $\mathcal{J}^{\text{eff}}(\omega)= \sum_{k=1}^{L} |g_{kp}|^2 \gamma_k/((\omega -\varepsilon_k)^2 + (\gamma_k/2)^2)$, where $g_{kp}$ is the coupling strength of the $p$-th fermion of the working medium with the $k$-th effective mode.

A controlled approximation of the real $\mathcal{J}(\omega)$ is thus achieved using a series of Lorentzian functions, each centered around $\varepsilon_k$.
Although the coupling strengths to each effective mode are assumed to be weak, the average system-bath coupling strength $\Gamma=(1/2 W)\int_{-\infty}^{\infty}\mathrm{d}\omega \mathcal{J}(\omega)$ exceeds the weak-coupling regime.
Therefore, contrary to exact approaches, a master equation for the density matrix of the working medium is employed to compute the long-time nonequilibrium heat and charge currents, 
$ \mathrm{d}\rho/ \mathrm{d}t= i[\rho,H] +\mathcal{D}_{\text{L}}\rho +\mathcal{D}_{\text{R}}\rho$, where $H=H_{S}+H_{\text{L}}+H_{\text{R}}+H_{S\text{R}}+H_{S\text{L}}$ is the full Hamiltonian describing the system S coupling to the mesoscopic modes and the dissipators describe the damping of the modes. 

Another aspect of the proposal is that the whole master equation, comprising the Hamiltonian of the full interacting working medium S, can be recast into a non-Hermitean Schr\"odinger equation for the reduced density matrix operator, combining standard Wigner-Jordan mapping and super-fermion formalism. 
The remarkable advantage of the latter is that the final Hamiltonian operator shows only nearest-neighbor interactions, where each physical fermion interacts with a neighboring fermionic ancilla. The reduced density matrix of the working medium can thus be represented as an MPS, exceeding weak-coupling approximations for what concerns the interacting working medium.        

The steady-state physics of the whole engine, as depicted in Fig.~\ref{fig:NonMarkovTN}, is thus studied focusing on different, strongly-interacting quantum systems chosen as the working medium.
Apart from reproducing and confirming many previous results, the research shows how a strongly-interacting three-fermion engine can achieve higher powers and efficiency with respect to its single-particle counterparts, stressing the relevance of strongly-correlated working mediums to QHEs.     
\subsection {Non-Markovian effects in quantum heat engines}\label{subsec:nonMarkeffects}
  
Exploiting non-Markovianity to boost the performance of thermal devices, especially QHEs, could eventually lead to more efficient miniaturized thermodynamic machines. Non-Markovian effects have already proved beneficial in settings like quantum control~\cite{schmidt2011optimal,koch15b,koch16}, quantum metrology~\cite{chin2012quantum}, and information engines via memory erasure~\cite{bylicka2016_non-markovian} (see also Sec.~\ref{sec:information}), to name a few. 
 
Over the years, several measures of non-Markovianity have been defined, based on the divisibility properties of the dynamical map~\cite{breuer2016colloquium_strcoup, chruscinski2022dynamical}, the distinguishability of pairs of states corresponding to different initial conditions~\cite{rivas2014quantum,  breuer2016colloquium_strcoup}, geometrical properties~\cite{lorenzo2013geometrical}, and negative entropy production~\cite{strasberg2019non,rivas2020strong}. However, the operating cycle of a QHE mainly relies on non-equilibrium dynamics of the working medium in contact with one or more reservoirs.~Therefore, rather than adopting a particular definition of non-Markovianity, following~\cite{de2017dynamics}, we choose to focus on theoretical QHE proposals in which the dynamics of the working medium cannot be successfully described in terms of a Linblad-form quantum master equation. 
  
Open quantum systems beyond standard Lindblad master equations have been the focus of much theoretical research~\cite{weissbook,de2017dynamics}, exploring the deviations from the predictions that occur in memoryless Born-Markov approximations.
Apart from strong system-bath correlations (see Sec.~\ref{subsec:strongcoupintro}), the reasons for these effects in the dynamics of open quantum systems include  structured environmental spectral densities, entanglement~\cite{rivas2010entanglement} 
among the environmental degrees of freedom, quantum correlations in the initial system-environment state, and finite-size bath effects~\cite{de2017dynamics,Pozas-Kerstjens_2018}.
  
For instance,  widely-know quantum bath models, such as the Caldeira-Leggett model~\cite{caldeira:caldeira-leggett}, show non-Markovian features related to quantum correlations and to the bath's spectral density~\cite{groeblacher2015observation}. For sufficiently low temperatures and in the weak-coupling regime, the role of memory effects, due to the presence of colored and multiplicative noise, may alter the horizon of open system dynamics~\cite{weissbook,hangii;1995}. In this limit, the typical separation of timescales pursued in the Markovian setting may no longer be valid.
While analytical approximations have been employed to tackle separate aspects of the problem, the detailed study of these effects often relies on exact methods.
  
One of the issues currently debated in the field of quantum thermal devices is whether non-Markovian effects can lead to real improvements in the engine's performance.
Early results~\cite{Zhang_2014} pointing toward a quantum advantage due to non-Markovianity -- the occurrence of power enhancements and efficiency higher than the Carnot bound -- have frequently relied on invalid approximations of the work needed to switch on and off the system-reservoir coupling~\cite{Wiedmann_2020,shirai2021non}.

Theoretical studies of QHEs beyond the realm of Markovian quantum master equations have followed quite different routes. A more realistic theory of a QHE cycle was pursued employing exact methods \cite{Wiedmann_2020,carrega2022engineering}. Here peculiar physical regimes are addressed, in which the conventional separation of system and bath timescales is not feasible, e.g., due to the interplay between external driving and dissipation or finite-time switching of the system-bath coupling.
Other approaches account for memory effects in the dynamics of the working medium, small deviations from the quasi-static driving regime, correlations among the constituents of each bath, energy backflow~\cite{guarnieri2016energy,shirai2021non}, and short-time function switching due to system-bath interactions~\cite{ishizaki2022switching}, by making use of non-Markovian versions of quantum master equations. Zeno and anti-Zeno quantum advantage due to non-Markovian effects have also been claimed in~\cite{mukherjee2020anti,xu2022minimal}, with models of periodically-driven, single-particle and weakly-coupled heat engines where friction effects can be safely neglected.   
 
Among the wide range of approximations available are second-order Time Convolution-Less (\acrshort{tcl}) master equations~\cite{guarnieri2016energy}, Floquet expansion beyond Markov approximation~\cite{Grifoni1998DrivenQT,mukherjee2020anti}, extended collision models \cite{kretschmer2016collision, mccloskey2014non,uzdin2014multilevel,campbell2018system,Strasberg:2017,ciccarello2022quantum} (see also Sec.~\ref{subsec:corrbath}), and ancillary environmental degrees of freedom correlated with the working medium~\cite{camati2020employing}.
In addition, adopting the framework of resource theories~\cite{lostaglio2019introductory,lostaglio2022continuous}, model-independent theoretical approaches based on  thermal operations have been pursued~\cite{ptaszynski2022non}, in order to find out the conditions under which a QHE could show an advantage in its performance linked to non-Markovian effects.
  
\subsubsection{ Quantum heat engines with system-reservoir coupling control}\label{subsubsection:nonMarkovcoup}
  
An ideal classical heat engine is assumed to work with a perfect architecture, where the working medium's coupling with, or insulation from, the baths does not cause any functioning power loss in the engine. When the engine is miniaturized, quantum effects take over and irreversible thermodynamic effects due to quantum correlations and coherence arise. Any study of these effects must begin by investigating the critical role that the coupling of the working medium to a bath plays in a QHE. The energy cost of such a coupling influences the evaluation of power output and efficiency of QHEs, and the open system dynamics of the working medium can no longer be approximated as a Markovian process. Thermal and quantum fluctuations go hand in hand in such a regime, and a quantitative understanding of their combined effects is also at issue in the study of thermal devices. Not only for purely theoretical interests, but ultimately for the actual fabrication of quantum and mesoscopic thermal devices, such studies deserve closer inspection.

To elucidate the above points, an example involving the theoretical aspect of the problem will be explored in some detail below. 

\begin{figure}[h]
    	\centering
        \includegraphics[scale=1.20]{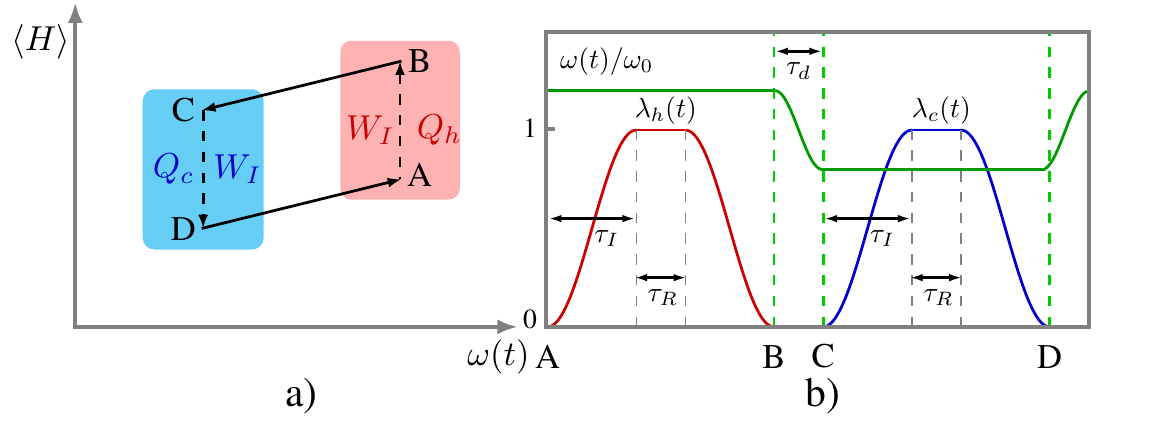}
    	\caption{Schematic diagram of the Otto cycle with time-dependent system-bath coupling switching operation. a) Sketch of the Otto cycle. Additional work input/output contributions $W_{\text{I}}$ are present due to the switching of system-bath interactions. b) Profile of the time dependent coupling functions  $\lambda_{c(h)}(t)$ (red, blue) and level spacing modulation $\omega(t)/\omega_{0}$ (green) along the cycle. The \emph{finite} times $\tau_{I}, \tau_{R},\tau_{d}$ are needed for the evaluation of the engine's performance. Reproduced from \cite{Wiedmann_2020}.}
    	\label{fig:Nonmarkov3}
\end{figure}

%
Consider that the Hamiltonian governing the dynamics of the QHE includes a single particle as a working medium, described by the generic Hamiltonian $H_S(t)=p^{2}/2m + V(q,t)$, where $(q,p)$ are the canonical coordinates. In the framework of the Caldeira-Leggett model ~\cite{weissbook,kato2016quantum}, the particle is interacting with harmonic baths, $H_x$, each described by a continuous spectral density function $\mathcal{J}_{x}(\omega)$, for the $x=c,h$ -bath. The interaction  $H_{I,x}(t)$ couples the position operator $q$ to the bath's position coordinates, and compared to conventional treatments, it is also modulated in time. The total Hamiltonian operator thus reads
\begin{eqnarray}\label{eq:nonmarkov5}
H(t) &=& H_{S}(t) + H_{c} +H_{h} + H_{I,c}(t)+ H_{I,h}(t),\nonumber\\
H_{c(h)} &=& \sum _{k} \omega_{k,c(h)} b^{\dagger}_{k,c(h)} b_{k,c(h)},\nonumber\\
H_{I,c(h)}(t) &=& - \lambda_{c(h)}(t) q\sum _{k} g_{k,c(h)}( b^{\dagger}_{k,c(h)}+ b_{k,c(h)}) + \frac{1}{2} \mu_{c(h)} \lambda_{c(h)} ^{2} (t) q^{2}.
\end{eqnarray}
Here $g_{k,c/h}$ are the usual Caldeira-Leggett coupling coefficients, while $0\leq\lambda_{c(h)}(t)\leq 1$ are functions of time that simulate the coupling/decoupling steps occurring in a finite amount of time. Notice that the additional quadratic contribution (see also Eq.~\eqref{eq:TLS2}), with $\mu_{c(h)}=(2/\pi)\int_{0}^{+\infty} \mathrm{d}\omega\mathcal{J}_{c(h)}(\omega)$ provides a renormalization of the system Hamiltonian and now it explicitly depends on time. Moreover, $b_{k,x}$ and $b^{\dagger}_{k,x}$ are the $x$ bath annihilation and creation operators.
The interaction is depicted in Fig.~\ref{fig:Nonmarkov3}. An initial partitioned system-bath state is assumed, although there have been recent attempts at lifting such approximations in generic open quantum systems~\cite{PhysRevB.95.125124}.
%
The bath quantum correlations in the Caldeira-Leggett paradigm read
\begin{eqnarray}\label{eq:nonmarkov6}
L_{c(h)}(t-t') &=& \langle \xi_{c(h)}(t)\xi_{c
(h)}(t')  \rangle\nonumber\\
&=& \frac{h}{\pi} \int _{0} ^{\infty}\mathrm{d}\omega \mathcal{J}_{c(h)} (\omega) \left[ \coth\left(\frac{\hbar\beta_{c(h)} \omega}{2} \right)\cos (\omega (t-t^{\prime}))-i \sin (\omega (t-t^{\prime})) \right], 
\end{eqnarray}
with $\xi_{c(h)}=\sum _{k} g_{k,c(h)}( b^{\dagger}_{k,c(h)}+ b_{k,c(h)})$. The memory kernel in Eq. \eqref{eq:nonmarkov6} is nonlocal-in-time and it causes the development of memory effects and non-Markovian features in the course of dynamics.
It can be proven that Eq.~\eqref{eq:nonmarkov6} is a result of the flucutuation-dissipation relation for a quantum system-bath dynamics with
initial partitioned condition~\cite{weissbook}.
Such relations for baths with coloured noise, nonlinear system-bath coupling, intrinsically nonlinear and driven baths, etc. have been studied over the years~\cite{weissbook, blokhu1, bhadra;2018, grabert2018} and remain an active field of research.

With a suitable generalized bath spectrum in place, a Stochastic Liouville-von Neumann (\acrshort{sln}) equation
is derived via the Feynman-Vernon Influence Functional Method \cite{SLN_1999,PhysRevLett.88.170407}  

\begin{equation}\label{eq:nonmarkov7}
\dot{\rho}_{\eta} =  - \frac{i}{h}\left[ H_{S}(t), \rho_{\eta} \right] + \mathcal{L}_h\left[\rho_{\eta} \right] + \mathcal{L}_{c} \left[ \rho_{\eta} \right], 
\end{equation}
where $\rho_\eta$ is the stochastic state of a single realization and the Liouvilleans $\mathcal{L}_{\alpha}$ given in~\cite{Wiedmann_2020} that provides a non-perturbative, non-Markovian time-local treatment of the system-bath dynamics.
%
%
 This formalism is applied to evaluate the thermodynamic properties of a four-stroke QHE with a working medium evolving under a time-dependent potential,
\begin{equation}\label{eq:nonmarkov9}\nonumber
    V(q,t) = \frac{1}{2} m \omega ^{2} (t) q^{2} + \frac{1}{4} m \kappa q^{4}.
\end{equation}
The parametric driving is implemented via $\omega (t)$, which varies between $\omega_{0} \pm \frac{\delta \omega}{2}$ within the time-scale $\tau_{d}$ during the isentropic strokes of expansion and compression cycles. 
A given amount of anharmonicity in the spectrum of the working medium, which is quantified by $\kappa >0$, is also introduced to understand how the engine's performance changes with increasing $\kappa$.
During the isochoric strokes of the engine $\omega(t)$ is kept constant (see Fig.~\ref{fig:Nonmarkov3}).

The engine dynamics can be solved numerically to study the combined effect of the system-bath coupling control and the strong dissipation.
The analysis reveals that the additional amount of work $W_{\text I}$ linked to the modulation of the system-bath coupling noticeably impacts the energy balance of the engine. Indeed, two different kinds of contributions arise in determining the latter, $W_{I}=W^{\text{var}}_{I} +W^{\text{corr}}_{I}$. 
It can be shown that $W^{\text{var}}_{I}>0$, which is proportional to $\ave{q^2}$, dominates in the adiabatic and high temperatures limit. On the other hand, $W^{\text{corr}}_{I}$, which is proportional to the correlation $\ave{qp+pq}$, becomes significant at finite-time cycles and in the deep quantum regime, and in certain cases can even counteract the first term, i.e., $W^{\text{corr}}_{I}<0$ in a given parameter range. It follows that the presence of $W_{I}$ is detrimental to the efficiency of the engine, while the $qp$ correlations can be utilized to mitigate its effect and thus improve the engine performance. In addition, strong coupling between the working medium and the baths affects significantly the efficiency while respecting the Curzon-Ahlborn and Carnot bounds, depending on the temporal nature of the isentropic protocol~\cite{Wiedmann_2020}.

Recently, heuristic arguments have been proposed  showing how a slow modification of the system-bath coupling, even in the presence of strong system-bath correlations, enables thermalization speed-ups in isothermal strokes \cite{pancotti2020speed}. 
For a finite time process the work dissipated into the bath is defined as $W_{\text{diss}}=W-\Delta F\geq 0$, with 
\begin{equation}
W=\int_{0}^{\tau_{\text{tot}}}\mathrm{d}t \tr[\rho\dot{H}(t)], \;\;\;\;\; \Delta F=\int_{0}^{\tau_{\text{tot}}}\mathrm{d}t \tr[\rho^{\text{th}}\dot{H}(t)],   
\end{equation}
and with $\rho$ and $\rho^{\text{th}}$ the system-bath state resulting from the finite-time dynamics and the instantaneous thermal state respectively. Notice that, in the limit $\tau_{\text{tot}}\to \infty$, $\Delta F=(1/\beta)\log(Z_{i}/Z_{f})$, where $Z_{i(f)}$ is the partition function of the initial and final system-bath Hamiltonian.

The dissipated work is found to fulfill the scaling  
$W_{\text{diss}}\propto \tau_{\text{tot}}^{-(2\alpha + 1)}$ with $\tau_{\rm tot}$  the total process time, and $\alpha>0$ depends on the specific system-bath control protocol.
Thus, by suitably tailoring the system-bath coupling, it is possible to enhance the performance of an engine going through a Carnot cycle, i.e., reduce the operation time without increasing the dissipated work.
In addition, the efficiency is found to interpolate between the Curzon-Ahlborn and Carnot bound for increasing values of $\alpha$. 

\begin{figure}[h]
    	\centering
    	\includegraphics[scale=0.33]{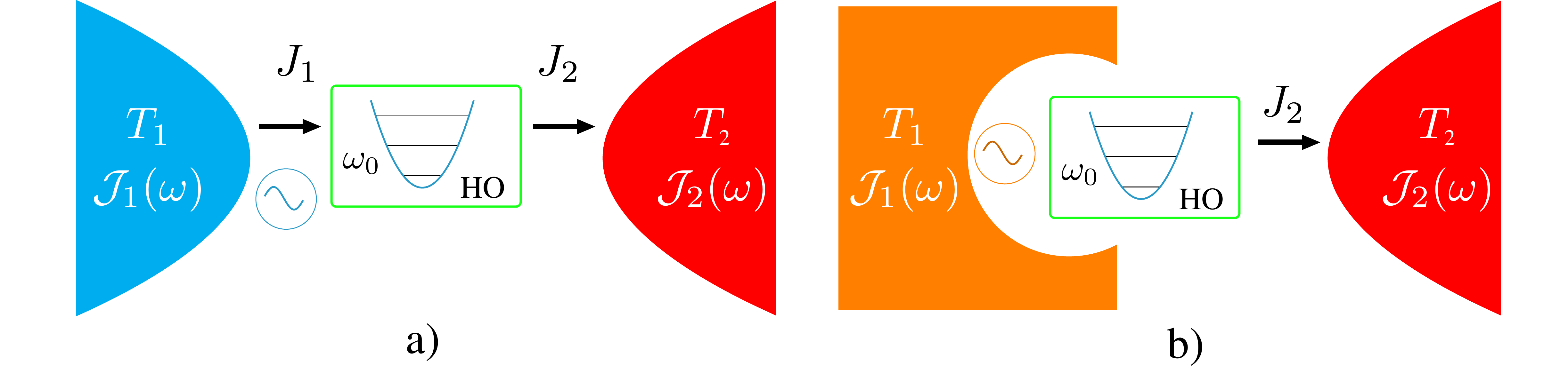}
        \caption{a) Cooling device working with  non-Markovian baths. The coupling with the left reservoir is periodically modulated in time, while the second is hold fixed. The frequency of the working medium (HO) is held constant. Both reservoirs are modeled as low-temperature Ohmic baths. b) Heat engine from similar setting, where on the left a bath with structured spectral distribution is used. Reproduced from \cite{carrega2022engineering,cavaliere2022dynamical}.}
    	\label{fig:Nonmarkov4}
\end{figure}

In~\cite{carrega2022engineering}, it has been shown that engineering time-dependent system-bath couplings in the non-Markovian regime leads to cooling, refrigeration, and heat engine operation~\cite{cavaliere2022dynamical}, without the need for driving the working medium directly. Compared to conventional heat engines described in Sec.~\ref{sec:basic}, these are nontrivial examples of thermodynamic tasks.
Though they share some similarities with heat ratchets~\cite{li2008ratcheting}, they achieve thermodynamic tasks by employing periodical modulation of the system-reservoir couplings.
The principle is demonstrated with a device similar to that sketched in Fig.~\ref{fig:Nonmarkov4}.~A quantum harmonic oscillator of fixed frequency $\omega_{0}$ interacts with two or more reservoirs.
As in~\cite{Wiedmann_2020,pancotti2020speed} and Eq.~\eqref{eq:nonmarkov5}, the baths are described with the Caldeira-Leggett model, with a system-bath coupling,  $H_{I,x}(t)=\sum_{k=1}^{\infty}[-q \lambda_{x}(t)c_{k,x}X_{k,x}+q^2 \lambda^2_{x}(t) c^2_{k,x}/(2 m_{k,x}\omega^2_{k,x})]$, where subscript $x$ labels the reservoir.
Moreover, $q$ is the oscillator position and $X_{k,x},\omega_{k,x}$ and $m_{k,x}$ are the position operators, the frequency, and the mass of each bath oscillator respectively. 
The couplings $\lambda_{x}(t)$ are periodic functions of time, i.e., $\lambda_{x}(t+\mathcal{\tau})=\lambda_{x}(t)$.
Due to the bilinear nature of the coupling Hamiltonian, the non-equilibrium steady-state of the system,~i.e.~the heat currents flowing in the reservoirs, can be analytically computed from the two-time non-equilibrium Green functions~\cite{freitas2018cooling}.

Noticeable results are derived in the instance of only two baths, each described by Ohmic spectral densities,  $\mathcal{J}_{x}(\omega)= m \gamma_{x}\omega e^{-|\omega|/\omega_{c}}$, where $\gamma_{x}$ quantifies the strength of the system-baths interaction and $\omega_{c}$ is the upper cut-off frequency, which corresponds to the highest energy scale of the system. Notice that, in the limit $\omega_{c}\to \infty$, $\mathcal{J}_{x}(\omega)$  gives rise to a local-in-time damping Kernel, $\gamma(t)= 2\gamma_{x}\theta(t)\delta(t)$, where $\theta(t)$ is the step function~\cite{weissbook}.

Assuming the baths are at the same temperature, $T_{1}=T_{2}$, and allowing for asymmetric dissipation strengths $\gamma_{1}\neq\gamma_{2}$ and a simple choice of the time-dependent protocol such that $\lambda_1(t)=\cos \Omega t$ and  $\lambda_{2}(t)=1$, a net heat current can be extracted fromreservoir $1$ and injected into the system.
In other words, by simply modulating the coupling to one bath while keeping the other constant, it is possible to achieve cooling instead of trivial dissipation effects.
This effect takes place in the low-temperature regime, $T \ll \omega_{0}$, and for sufficiently weak dissipation.~It can be thus traced back to memory effects occurring during the engine operation. Indeed, following Eq.~\eqref{eq:nonmarkov6}, non-Markovian memory effects can occur for sufficiently low temperatures, while in the opposite regime, i.e.,  $T\gg\omega_{0}$, Eq.~\eqref{eq:nonmarkov6} reduces to  $\ave{\xi_{x}(t)\xi_{x}(t^{\prime})}\approx 2 m\gamma_{x}\delta(t-t^{\prime})$ ($\hbar=1$, $\omega_{c}\to \infty$), i.e., these memory effects vanish and no cooling effect is present. Moreover, in the presence of a temperature  gradient, $T_{1}\neq T_{2}$, refrigeration effects are also found.

As non-Markovian effects can also result from structured bath spectral distributions~\cite{de2017dynamics}, with the same choice of dynamical system-bath couplings it may also be possible to achieve heat engine operations, i.e., non-Markovianity can be considered a further example of a quantum resource.
Nevertheless, as shown in~\cite{cavaliere2022dynamical}, the presence of non-Markovianity is a necessary but not sufficient condition for obtaining a heat engine.
The heat engine's operations are thus characterized by a Lorentzian spectral distribution $\mathcal{J}_{1}(\omega)=d_{1}m \gamma_{1}\omega/((\omega^2 -\omega^2_{1})^2 +\gamma^2_{1} \omega^2))$ for bath $1$ and an Ohmic distribution for bath $2$.
Furthermore, this  engine is also upper-bounded by the Carnot efficiency, which can be approached in a suitable range of parameters where, of course, the power output vanishes.
  
\subsubsection{Non-Markovian effects in slowly-driven  quantum heat engines}\label{subsubsec:nonMarkslow}

In~\cite{abiuso2019_non_markovian}, the role of non-Markovian effects in the performance of heat engines is investigated by adopting a time-dependent quantum master equation in the slow driving regime~\cite{cavina2017slow}.
The strategy proposed is thus distinct from that reported in Sec.~\ref{subsubsec:enginestrongcoupl} and \ref{subsubsection:nonMarkovcoup}, and it does not address non-Markovian features related to strong-coupling and memory effects.
Here the working medium, a driven two-level system, is coupled to a set of thermal baths in order to perform conventional thermodynamics cycles.~However, it is assumed that the time-dependent generator of the Markovian equation $\mathcal{L}_{t}=-i[H_{t},\cdot] +\mathcal{D}^{j}_{t}$, corresponding to the $j$-th reservoir, is step-continuous, such that the coupling switching occurs over a time-scale that is negligible with respect to the finite operating time.
   
To account for the deviation from the quasistatic limit~\cite{cavina2017slow}, a perturbative expansion is introduced for the reduced system density matrix  $ \rho (t) = \rho^{(0)} (t) + \rho^{(1)} (t) + ...$.~The zeroth-order term describes the standard quasistatic regime, the working medium lies in the instantaneous Gibbs state $\Omega^{j}_{H_t}=\exp(-\beta_{j}H_{t})/\tr[\exp(-\beta_{j}H_{t})]$ related to the $j$-th bath, and it is assumed that the latter is the unique fixed point of the full dissipator, $\mathcal{L}_{t}\Omega^{j}_{H_t}=0$. The perturbation parameter is defined as $\tau_{r}/\tau$, a ratio of the time-scales involved in the dynamics and the relaxation to equilibrium respectively.
Standard thermodynamic quantities (heat and work) pick up these corrections, and an optimization protocol can be implemented to maximize the power and efficiency of the QHE.
  
Some non-Markovian effects can be mimicked by assuming that a few degrees of freedom in the bath, share correlations with the working medium, while all the rest of them give rise to dissipative dynamics.
In other words, system-bath interactions are recast into a local contribution governed by a swap-like Hamiltonian and an effective contribution due to remote baths' degrees of freedom, which is ruled by time-dependent (independent) dissipators, $\mathcal{D}^{j}_{t},(\mathcal{D}^{j})$, acting on the driven working medium and the ancillae states respectively. The latter drive the working medium, as well as the ancillary degrees of freedom, to their instantaneous Gibbs states. By applying optimization protocols, relevant improvements in the maximum power extractable from the engine with respect to the Markovian limit are achieved. A similar approach can lead to improvements in the performance of a quantum Otto refrigerator \cite{camati2020employing}.
\subsubsection{Correlation-induced non-Markovian quantum heat engines}\label{subsubsec:nonMarkcorr}

Signatures of non-Markovianity can also be found in the properties of asymptotic (long-time) states of the reduced system~\cite{de2017dynamics}. Indeed, a steady state may not be invariant under the action of a non-Markovian dynamical map ~\cite{chruscinski2010long}, retaining some memory of the initial state. Many theoretical works have focused on deriving non-Markovian dynamical maps~\cite{breuer2016colloquium_strcoup,chruscinski2022dynamical}, and powerful numerical approaches were developed~\cite{cerrillo2014non}.  
Long-time memory effects linked to non-Markovian maps on the performance of QHEs were investigated in \cite{thomas2018_non-markovian}, showing consistency with the second law and evaluating the minimum work output due to the non-Markovianity of the baths.
  
Extended collisional models~\cite{Strasberg:2017,ciccarello2022quantum} and Markovian embedding techniques~\cite{campbell2018system} have also been developed to describe memory effects, and quite similar techniques have been pursued in the study of engines working with correlated baths~\cite{DeChiara:PRR_2020} (see also Sec.~\ref{subsec:corrbath}).
The main idea behind these methods is modeling the interaction as a set of repeated unitary interactions between the system and each of the individual baths' degrees of freedom, depicted as collisions with ancillae.
To account for memory effects, additional collisions that couple the ancillae together are modeled and described by a suitable interaction Hamiltonian.
By this means, the system correlates with more than one ancillae at a time.    
   
In \cite{Pezzutto_2019}, an extended collisional model was used to study QHEs in the strong nonequilibrium and non-Markovian regime.
Here the working medium is modeled via a harmonic oscillator with a time-dependent frequency $\omega(t)$ (the control parameter), and the work strokes of the Otto cycle are implemented through its variation. During the heat strokes, the working medium is connected to external baths modeled by a collection of $n$ spin-$1/2$ particles, ruled by the Hamiltonian 
  $H^{n}_{e} = \frac{1}{2} \hbar \omega_{e} \sigma ^{z}_{e,n}$ with $e= c,h$.

\begin{figure}[h]
    	\centering
        \includegraphics[scale=1.1]{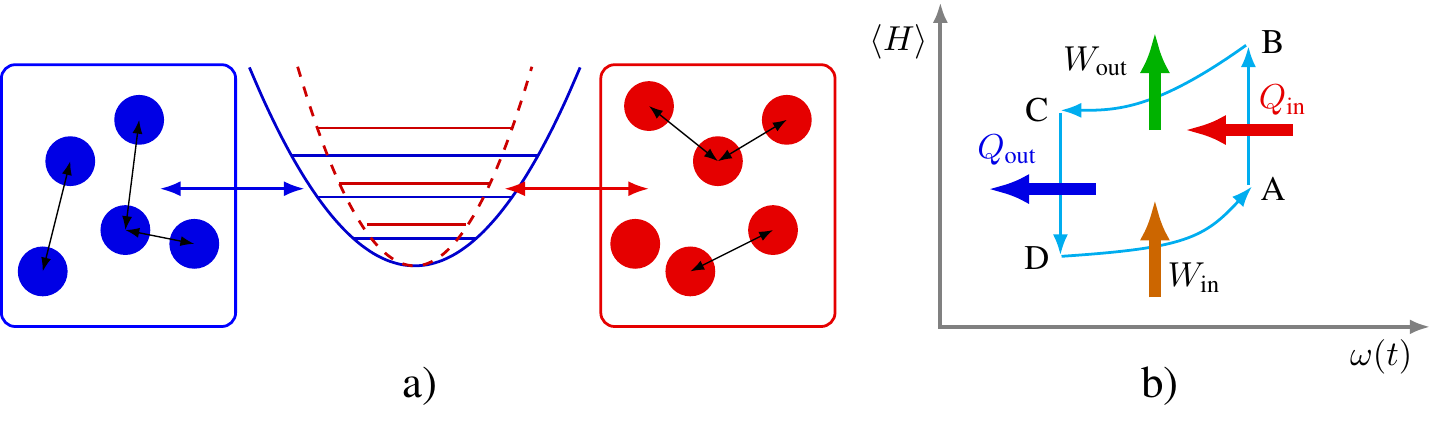}
    	\caption{Diagram of the Otto cycle engine with memory effects. a) The working medium-reservoir interactions are described by means of unitary collisions. Memory effects are simulated by including interactions among the bath constituents, modeled as two-levels ancillae. b) Schematic diagram of the Otto cycle. During each isochoric stroke ($\text{A}\rightarrow \text{B},\text{C}\rightarrow \text{D}$), the working medium gets correlated with bath ancillae.~Reproduced from \cite{Pezzutto_2019}.}
    	\label{fig:Nonmarkov5}
\end{figure}
The collision event between the working medium and the $n$-th spin is modeled with the interaction Hamiltonian as $H_{Ie} = J(a \sigma ^{+} _{e}+ a^{\dagger} \sigma ^{-} _{e})$. Moreover, intra-bath interactions are modeled by a Heisenberg chain, i.e., $ H_{ee} = J_{ee} \Big(\sigma ^{x} _{n} \sigma ^{x} _{n+1} + \sigma ^{y} _{n} \sigma ^{y} _{n+1} + \sigma ^{z} _{n} \sigma ^{z} _{n+1} \Big)$, with $ee=cc,hh$, and $J_{cc} (J_{hh})$ denoting the coupling constant of the cold(hot) reservoirs. 
The analysis of this engine reveals that coherence survives in the stationary state, irrespective of the initial one and that increased non-Markovianity of the system dynamics can slow the relaxation to the stationary state. This slowing down negatively affects the performance of the engine, degrading the efficiency and the extracted power the faster the work strokes are performed.

\section{Thermoelectric devices}\label{sec:thermoel}

Thermoelectric devices are \emph{autonomous}, steady-state heat engines that, unlike those described in Sec.~\ref{subsec:Carnotcycle}, achieve heat to work conversion without an external modulation of the system parameters.~Indeed, such devices have no moving parts, as they work by employing solid-state materials that convert heat flows into microscopic steady-state electronic currents~\cite{AshcroftMermin:1976}.~Thermoelectric effects, such as the generation of voltage from a temperature gradient (Seebeck effect), or the \emph{reversible} generation of heat from current flow in a junction of two materials at equal temperature (Peltier effect), can be quantitatively described in terms of the transport properties of bulk electrons.
In the simplest instance of a macroscopic thermoelectric engine, a thermocouple generates electrical power from two thermal reservoirs at different temperatures.~Thermocouples can be advantageous in many technological applications devoted to achieving more efficient heat to work conversion, e.g., recycling heat from exhaust gases.   
  
Nanoscale thermoelectric devices~\cite{BENENTI20171} have attracted growing attention in the last two decades, following advances in nanosctructure experiments and quantum transport theory~\cite{BEENAKKER19911,BenakkerRandom,nazarov_blanter_2009,Sanchez:2020}.
Understanding heat to work conversion at the nanoscale is crucial for implementing quantum thermal machines, among them quantum refrigerators, on-chip coolers, and caloritronic devices~\cite{Giazotto:2006,Muhonen_2012,Giazotto:caloritronics}. Such machines could achieve better refrigerator performance with respect to conventional cryogenic devices, dragging heat directly from electrons~\cite{BENENTI20171,Whitney:qthermo}.
Moreover, nanoscale thermoelectric devices could be critical to achieving control of energy exchange in upcoming quantum computing platforms~\cite{Buffoni_2020}.       
      
However, the physics of thermoelectrics at the nanoscale poses several challenges to theory.
At sufficiently low temperatures, quantum interference and correlation effects are present and strongly affect electronic transport, due to the size of nanostructures (which are far smaller than the electronic relaxation length).
In this limit, the thermalization process -- via electron-phonon interaction in macroscopic materials -- is governed by electron-electron scattering.
Moreover, since relaxation can occur at a distance longer than the typical lengths of these devices, a fully non-equilibrium quantum theory is required to describe heat to work conversion in such systems.

\subsection{Models of nanoscale thermoelectric transport}\label{subsec:thermodels}
\begin{figure}[h]
    \centering
    \includegraphics[scale=0.5]{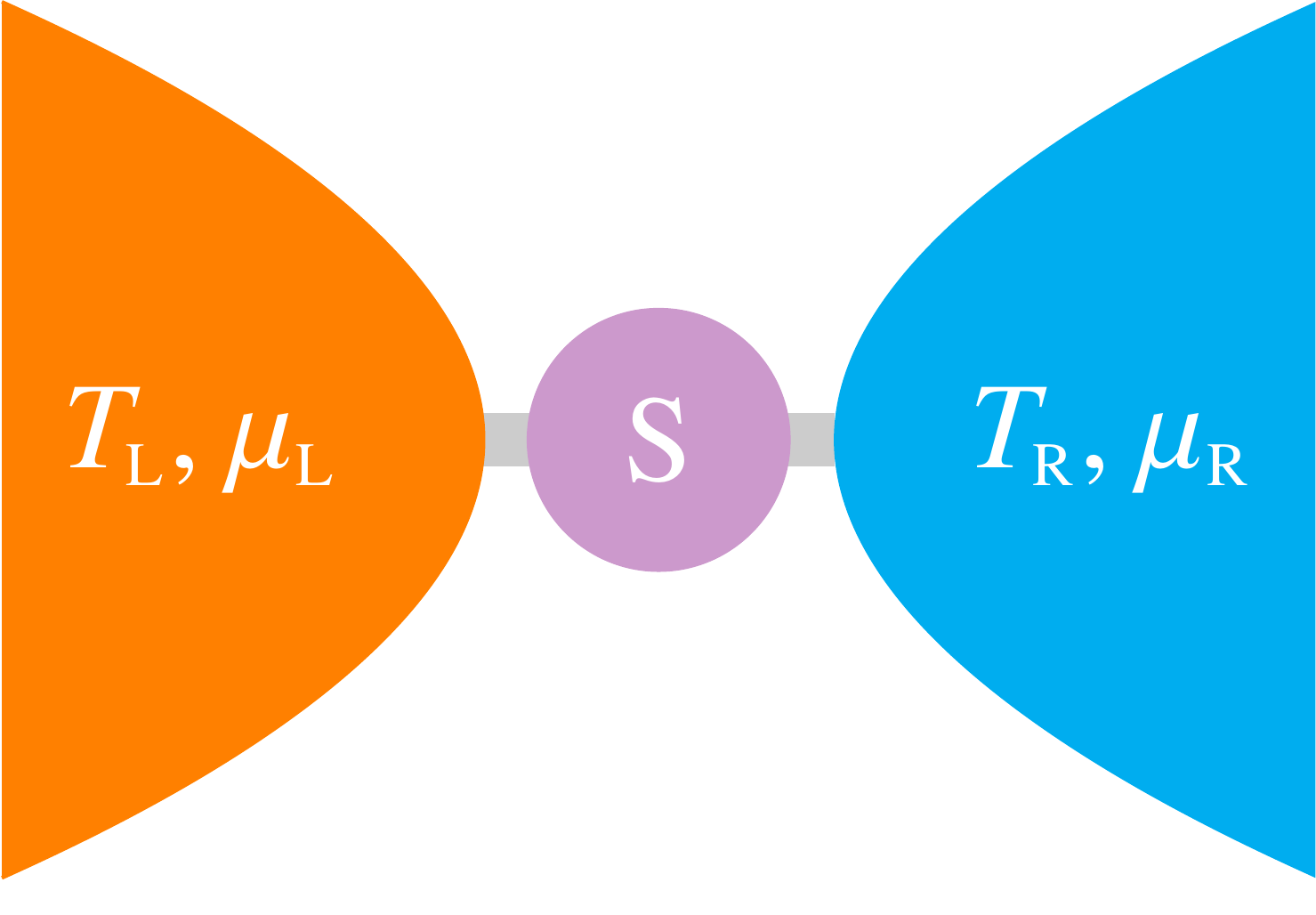}
    \caption{Schematic diagram of a thermoelectric device. A nanoscale system $S$, e.g., a semiconducting heterostructure, is brought into contact with two different reservoirs. Each reservoir can be modeled as an ensemble of noninteracting particles at equilibrium, e.g., an electronic lead, characterized by a temperature $T$ and a chemical potential $\mu$. Charge and energy can flow through the system $S$.}
    \label{fig:thermoel1}
    \end{figure}

 A typical nanoscale thermoelectric device is composed of a nanoscale system interacting with two or more reservoirs, as sketched in Fig.~\ref{fig:thermoel1}.   
Each reservoir can be modeled as an ensemble of free fermionic or bosonic particles at thermodynamic equilibrium, with fixed values of temperature and chemical potentials $(T_{i},\mu_{i})$. Unlike in the bulk of thermoelectric material, here single electrons can move from the filled states of a given reservoir to empty states of another by tunneling through the quantum system.
A simple yet effective explanation of the thermoelectric effect can thus be given by considering the special case of two electronic reservoirs: left ($\rm{L}$) and right ($\rm{R}$), marked by different temperatures $T_{\rm{L}}>T_{\rm{R}}$ and chemical potentials $\mu_{\rm{L}}<\mu_{\rm{R}}$.
In the course of this introductory section, we will therefore focus on the two-reservoir instance.

In order to generate usable power, an electronic current has to take place against a chemical potential gradient, i.e., electrons need to flow from the region L to R, thus overcoming the potential barrier.
It follows that a potential barrier, acting as an energy filter, is needed to block electrons below a given energy threshold from flowing from $\rm{R}$ to $\rm{L}$. In this way, only high-energy electrons occupying states in a window of $k_{B} T_{\rm{L}}$ above the chemical potential $\mu_{\rm{L}}$ can flow through the $\rm{R}$ side. A heat flow due to ``hot'' electrons moving from $\rm{L}$ to $\rm{R}$ reservoir can thus generate an electronic current.
When the device is connected to an external load, a finite power can be produced.
Analogously, refrigeration effects can be obtained by considering $\rm{L}$ ($\rm{R}$) as a cold (hot) reservoir, with $\mu_{\rm{L}}>\mu_{\rm{R}}$.
In this case, an energy filter can be used to block low-energy electrons below (above) the chemical potential of the cold (hot) reservoir from flowing.
This filter generates a current flow from $\rm{L}$ to $\rm{R}$, i.e., a heat flow from the cold to the hot reservoir is set.
As the electron flows from $\rm{L}$ to $\rm{R}$, work has to be spent in order to sustain a finite difference between $\mu_{\rm{L}}$ and $\mu_{\rm{R}}$.
This class of energy filter can be integrated to devise nanoscale thermocouples, in which the electron current is directed toward a load~\cite{BENENTI20171,Whitney:qthermo}.     

Early proposals for nanoscale systems acting as energy filters relied on the idea of Quantum Point Contacts (\acrshort{qpc}).
The point contact was first devised as a constriction in a Two-Dimensional Electron Gas (\acrshort{2deg}) heterostructure, achieved by employing electrostatic gates.~Its typical dimensions are smaller than the electronic mean free path~\cite{Wees88}.
It has been proved to show quantized conductance, in steps of order $2 e^2/h$, for each electronic conduction channel.
The quantized transport through a saddle-point constriction,  i.e., a constriction where the electrostatic potential in the plane $(x,y)$ is well approximated by a $V(x,y)=V_{0}-(1/2)m\omega^2_{x}x^2 +(1/2)m\omega^2_{y}y^2 $, has also been studied~\cite{ButtikerQPC}, showing steps of $e^2/h$ and conductance of $G=(e^2/h) \mathcal{T}$, where $\mathcal{T}$ is the total transmission probability along each channel (see Sec.~\ref{subsubsec:scattering}).
The QPC behaves as a potential barrier with discrete energy levels, depending on the parameters of the constriction.
Electronic wavefunctions flowing through the barrier get transmitted or reflected depending on their energy.
In pioneering studies on thermoelectric properties of QPCs \cite{HoutenThermoQPC_1992}, the Seebeck and Peltier effect were computed and experimentally probed, limiting the linear response regime (see Sec.~\ref{subsubsec:Linresp}).
However, QPCs are still employed as the building blocks of modern proposals for nanoscale thermoelectric devices \cite{Kheradsoud:2019,Pershoguba:CondQPC,Pershoguba:thermoQPC,Yang:thermTrans19}.  
    \begin{figure}[h]
    	\centering
    	\includegraphics[scale=1.0]{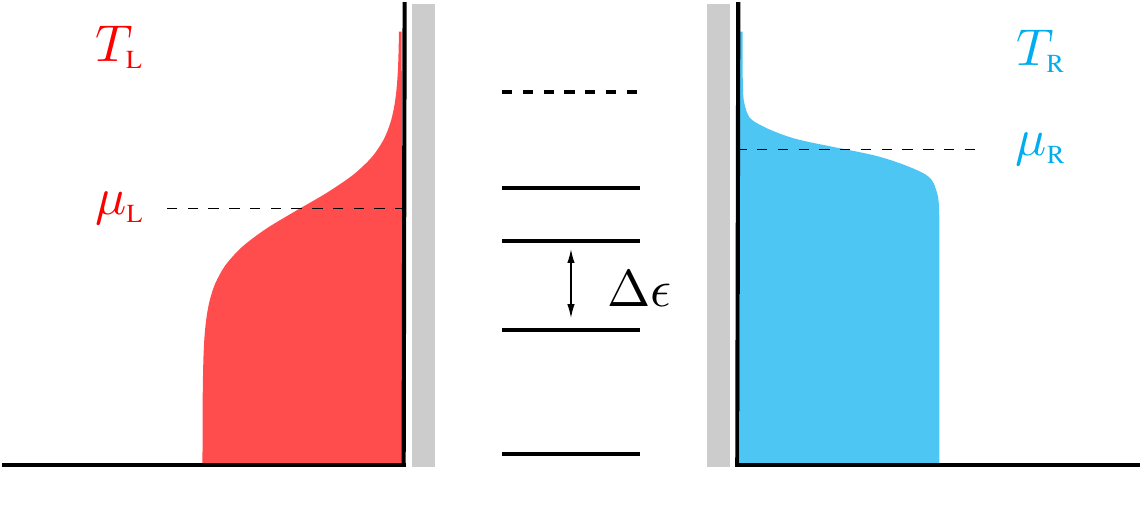}
        \caption{Schematic diagram of a  QD heterostructure. QDs can be described as islands containing a discrete set of electronic levels. The QD is coupled to the L and R reservoirs through insulating barriers. Depending on the level spacing $\Delta \epsilon$ and the reservoir parameters $\mu_{i},T_{i}$, single electrons can tunnel from the reservoirs to the island and vice versa. In the CB regime, an integer number of electrons occupies the QD.}
    	\label{fig:thermoel2}
    \end{figure}
    
Many theoretical and experimental works on nanoscale thermoelectrics have focused on quantum dot (\acrshort{QD}) nanostructures~\cite{Beenakker:QD,Beenakker:QDthermopower,Esposito_2009,Natpathomkun:2010,Erdman:QD17}. A QD can be devised as a semiconducting island, where a number of electrons occupying discretely spaced, single-particle energy levels can be confined. Semiconducting GaAs/AlGaAs heterostructures~\cite{Staring_1993}, and more recently InAs/InAp nanowires~\cite{Svensson_2012,Josefsson:2018,Dominguezadame:2019}, have been used as experimental platforms to investigate QD thermoelectric devices. The simplest thermoelectric device based on a QD is conventionally modeled as in Fig.~\ref{fig:thermoel2}, where the QD works as an energy filter. It is modeled by means of a discrete set of electronic levels and assumed to work in the Coulomb Blockade (\acrshort{CB}) regime~\cite{Beenakker:QD}.

In the CB regime, an integer number $N$ of electrons occupy the QD, and the electrostatic energy of the dot's state in the absence of external bias is $U(N)=(Ne)^2/2C=E_{\rm{C}}N^2$, where $C$ is the finite capacitance of the dot.~Moreover, the dot is weakly coupled to the leads through tunnel barriers, so that the rates of tunneling to the reservoirs are much smaller than the level spacing $\Delta \epsilon$, thermal energy $k_{B}T$, and charging energy $U$ respectively.~In this limit,~i.e.,~the \emph{sequential tunneling regime} (see Sec.~\ref{subsubsec:kinetics}), electronic transport occurs via single tunneling events from the QD to the reservoir. Early works on the theory of thermoelectric effects in QDs have limited themselves to linear response in the bias $V=\mu_{\rm{L}}-\mu_{\rm{R}}$ and temperature $\Delta T=T_{\rm{L}}-T_{\rm{R}}$. These works found that conductance oscillations with the Fermi energy affect the thermoelectric properties, leading to oscillations in the Seebeck coefficient $S$~\cite{Beenakker:QDthermopower}.
Subsequently, the single QD-based platform has been employed to explore various physical configurations.~
Furthermore, intra-dot Coulomb interactions among
electrons and low temperatures can bring in interesting physics linked to the Kondo effect~\cite{Hewson,Dutt:2013,Zimbovsjaya:2015}.
In addition, nonlinear thermoelectric effects in Coulomb Blockaded QDs have also been studied~\cite{Svensson_2013,Sierra:2014,SANCHEZREV2016}. 
    
Molecular junctions~\cite{galperin2007molecular,Dubi:2011,Lee:2013} have also been considered as basic building blocks of thermoelectric devices. The nontrivial interplay of electronic and vibrational degrees of freedom has been elucidated in different works~\cite{Perroni_2016}, and intriguing nonlinear thermoelectric response effects~\cite{SANCHEZREV2016} have been studied.
   
Theoretically describing the prototypical systems depicted in Figs.~\ref{fig:thermoel1} and \ref{fig:thermoel2} requires computing nonequilibrium steady-state electronic heat and energy currents from a given reservoir into the nanoscopic system. Moreover, the efficiency of heat-to-work conversion, as well as the thermoelectric response functions, need to be carefully defined, as they strikingly change moving from linear to nonlinear response~\cite{BENENTI20171}.
 
Before moving to the description of recent proposals for nanoscale thermoelectric devices, we provide a brief overview of the main theoretical tools developed thus far to model these types of devices and refer the reader to the excellent reviews~\cite{BENENTI20171} for further background.

 \subsubsection{Theoretical tools  for studying thermoelectric devices}\label{subsubsec:scattering}    

\paragraph{Scattering theory approach} 

The simple idea of an energy filter is rooted in the scattering theory of quantum transport~\cite{datta_2005,nazarov_blanter_2009}.~In this scheme, the reservoirs are macroscopic leads, connected to the scattering region by means of waveguides, allowing for a given number of electronic modes $N_{\rm{L},\rm{R}}(E)$, for each value of the energy $E$. The scattering region is the space between the leads, and it is occupied by a nanosystem, e.g., a nanostructure or a QD.
On the other hand, the nanosystem in Fig.~\ref{fig:thermoel1} is modeled as a scatterer, described by means of a Hermitian Hamiltonian, which is a time-reversal invariant.
Due to the presence of the scatterer, electrons make transitions from a given mode of a reservoir $\rm{R}$ to another mode of $\rm{L}$, through elastic scattering events occurring at fixed energy $E$.~As with conventional scattering theory, electronic transitions from mode $j$ of reservoir R to mode $k$ of reservoir L are modeled with a unitary scattering matrix $S_{\rm{L},k;\rm{R},j}(E)$, linked to the scatterer Hamiltonian.
Thus, the transition probability can be easily written as $P_{\rm{L},k;\rm{R},j}(E)=|S_{\rm{L},k;\rm{R},j}(E)|^2$ and, for a given value of the energy $E$, the transmission probability between the leads, from $\rm{L}$ to  $\rm{R}$, can be computed from $S_{\rm{L},k;\rm{R},j}(E)$ by summing over all possible paths, i.e., $\mathcal{T}_{\rm{LR}}(E)=\sum_{k,j}P_{\rm{L},k;\rm{R},j}(E)$.

Once the transmission probability is known, the Landauer-B\"uttiker approach can be used to write down the expressions for the electronic and heat currents in the scattering region from a given reservoir. The process can be imagined as counting the electrons going out of the reservoir L through the scattering region, so that the total number of electrons entering the region $S$ equals the total number of electrons going out of L, minus the number of electrons transmitted from R to L. 
The electronic current thus reads
   \beq\label{eq:thermoel1}
    	J_{e,i}=\frac{e}{h}\sum_{j=\rm{L},\rm{R}}\int_{-\infty}^{+\infty}\mathrm{d}E(N_{i}(E)\delta_{ij} - \mathcal{T}_{ij}(E))f_{j}(E),  
    \eeq
    
with $i=\rm{L},\rm{R}$ and $f_{j}(E)=1/(\exp[(E-\mu_{i})/k_{\rm{B}} T_{j}] + 1 )$ the equilibrium Fermi-Dirac distribution, $N_{i}(E)=\sum_{j}\mathcal{T}_{ij}(E)$ is the number of modes available in the reservoir $i$ at the energy $E$ and $\delta_{ij}$ is the Kronecker delta.
Similarly, the energy and heat currents from each reservoir, denoted by $J_{u,i},J_{h,i}$ respectively, can be computed as follows:  
\bea\label{eq:thermoel2}
    J_{u,i}=\frac{1}{h}\sum_{j=\rm{L},\rm{R}}\int_{-\infty}^{+\infty}\mathrm{d}E E (N_{i}(E)\delta_{ij} - \mathcal{T}_{ij}(E))f_{j}(E),\\
    J_{h,i}=\frac{1}{h}\sum_{j=\rm{L},\rm{R}}\int_{-\infty}^{+\infty}\mathrm{d}E (E-\mu_{i}) (N_{i}(E)\delta_{ij} - \mathcal{T}_{ij}(E))f_{j}(E).  
\eea
From Eqs.~\eqref{eq:thermoel2}, it follows that removing an electron below (above) the chemical potential equals to transfer (drag) an amount of heat to (from) the reservoir.
However, the exchanged heat cannot change the reservoir temperature. In addition, employing the relation among
the currents $J_{h,i} = J_{u,i}-(\mu_{i}/e)J_{e,i}$,  the first law can be expressed as
\beq\label{eq:thermoel3}
\sum_{i}J_{h,i}=-\sum_{i}\frac{\mu_{i}}{e}J_{e,i}=P_{\text{gen}},
\eeq
where $P_{\rm{gen}}$ is the total power generated by the system.

In case $P_{\rm{gen}}>0$, that is, the electronic current flows against the biases, the system converts the heat currents into a positive power, while in the opposing limit, it releases heat into the reservoirs.
Notice that the scattering approach is based on the single-electron approximation, i.e., the matrix $S(E)$ can be computed if the scatterer Hamiltonian describes noninteracting electrons.~It follows that correlations effects due to Coulomb interactions can be included by employing ab-initio Density Functional Theory (\acrshort{dft}) and Local-Density Approximation (\acrshort{lda})  schemes.~On the other hand, when Coulomb interactions are the prominent physical mechanism, as in the CB regime, alternative approaches have been developed. 
 
\paragraph{Rate equations}\label{subsubsec:kinetics}
 
\begin{figure}[h]
    	\centering
        \includegraphics[scale=1.0]{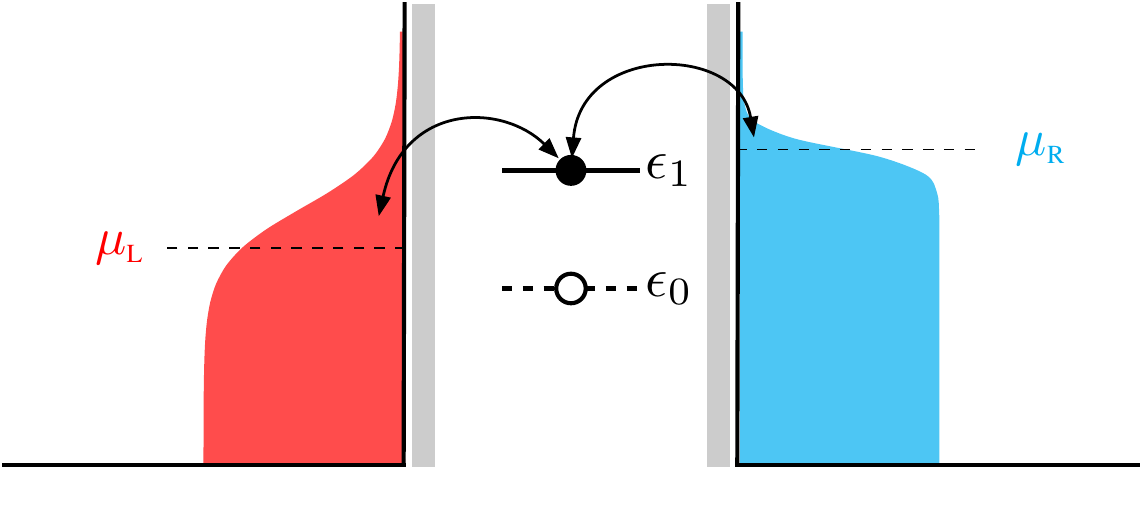}
    	\caption{Schematic diagram of spinless electron tunneling through a single-level QD in the sequential approximation. The level broadening is neglected. The level can be empty or singly-occupied. The probability of occupation of the single level is computed from a rate equation approach.}
    	\label{fig:thermoel3}
\end{figure}

 Thermoelectric effects in the presence of Coulomb interactions among the electrons in the nanoscale system, e.g., a QD in the CB regime, can be described by employing rate equations.~In early works~\cite{Beenakker:QD}, the QD has been modeled as a set of $N$ single-electron levels, where the Coulomb interactions among the electrons are described in the mean-field approach.~For the purpose of this section, it will suffice to consider the prototypical case of a single-level QD, as sketched in Fig.~\ref{fig:thermoel3}.
 Here the electron is assumed to be spinless, such that the level can be empty or singly-occupied.
 In more general settings, spin degeneracy has been considered.
 Adopting this scheme, the QD is modeled by means of two states, $\ket{0}$ and $\ket{1}$ with energies $\epsilon_{0}$ and $\epsilon_{1}$, as depicted in Fig.~\ref{fig:thermoel3}.
 The QD is tunnel-coupled with the reservoirs $\rm{L},\rm{R}$, and it is modeled with the following Hamiltonian:
     \beq\label{eq:thermoel4}
     H=\sum_{k=0,1}\epsilon_{k}d^{\dagger}_{k}d_{k} + \sum_{i\gamma}\epsilon_{i \gamma} c^{\dagger}_{i,\gamma}c_{i,\gamma} +  \sum_{i,\gamma,k}( V^{i}_{k}(\epsilon_{\gamma})d_{k}c^{\dagger}_{i,\gamma} + h.c.), 
    \eeq   
where $k=0,1\mbox{, }i=\rm{L},\rm{R}$, $\epsilon_{i,\gamma}$ denote the energy of the electronic levels in the $i$-th reservoir, $c_{i,\gamma}(c^{\dagger}_{i,\gamma})$ are electronic creation (annihilation) operators in the reservoirs and $V^{i}_{k}(\epsilon_{\gamma})$ is the energy-dependent tunneling element.~However, high tunneling barriers are assumed, so that the rate of tunneling events to the reservoirs $\Gamma^{(i)}$ is small with respect to the level spacings and thermal energy respectively, i.e., $\Delta \epsilon$\mbox{, }$k_{B}T\gg \Gamma^{(i)}$.~In this limit, the sequential tunneling approximation can be employed: it is assumed that transport takes place through single-electron tunneling events from QD to the reservoirs and vice versa.
Therefore, the broadening of the levels of the dot can be safely neglected, so that the state of the dot is described by a set of occupation numbers, one for each electronic of the dot.

It follows that first-order tunneling contributions are the leading terms in the dynamics, while cotunneling effects,  i.e., second-order system-reservoir interaction processes~\cite{tesser2022heat}, are neglected.
The transition rates from one state to another can be computed using the Fermi golden rule.
Moreover, electrons cannot go to linear superposition states, thus physical effects linked to quantum coherence are neglected.
A master equation for the probabilities of occupation of each state $k$ of the dot, $P_{k}(t)$, can be written as 
\beq\label{eq:thermoel5}
    \frac{d}{d t} P_{k}(t) =\sum_{l,i={\rm L,R}} (-\Gamma^{(i)}_{lk}P_{k}(t) +\Gamma^{(i)}_{kl}P_{l}(t)).   
\eeq   
Here $\Gamma^{(i)}_{k,m}$ is the tunneling rates matrix linked to the reservoir $i$.
From the general form of the tunneling Hamiltonian in Eq.~\eqref{eq:thermoel4}, the Golden rule tunneling rates read 
\beq\label{eq:thermoel6}
    \Gamma^{(i)}_{k,m}=\frac{1}{h}\nu_{i}(\epsilon_{k}-\epsilon_{m})f_{i}(\epsilon_{k}-\epsilon_{m})|V^{i}(\epsilon_{k}-\epsilon_{m})|^2,  
\eeq     
where $\nu_{i}(\epsilon)$ denotes the density of states of the reservoir $i$ and $f_{i}(\epsilon)$ is its Fermi distribution.~Once the stationary state of Eq.~\eqref{eq:thermoel5} is known, electronic and heat currents analogous to Eqs.~\ref{eq:thermoel2} can be computed directly from the stationary value of the r.h.s. of Eq.~\eqref{eq:thermoel5}.~Due to its simplicity, this approach has been widely employed to model electronic and heat transport in more complex thermoelectric devices.    
    
\paragraph{Other methods}\label{subsubsec:NEGF}
Capturing effects manifested in the strong system-lead coupling regime -- at low temperatures and with coherence and strong correlations -- requires more involved methods, some of which were discussed in Sec.~\ref{sec:nonMarkov}.         
Over the last three decades, Nonequilibrium Green Functions (\acrshort{negf}s)~\cite{camsari2023nonequilibrium} have been employed to solve a wide range of electronic transport problems. 
In the context of autonomous nanoscale devices, such as rectificators or, more generally, thermoelectric engines, NEGFs have been employed to describe working media strongly coupled to their reservoirs~\cite{galperin2007molecular,hartle2008multimode,agarwalla2015full,erpenbeck2015effect}. 
A quantum thermodynamics formulation in NEGF terms has been also investigated ~\cite{esposito2015quantum,seshadri2021entropy,bhandari2021nonequilibrium}, and inconsistency between the dynamical and thermodynamical of supersystem-superbaths dynamics was put forward in~\cite{bergmann2021_nefg}. 
Another numerically exact method that has been applied to thermoelectric devices is the hierarchical equation of motion approach~\cite{tanimura1989time,jin2008exact,schinabeck2016hierarchical,tanimura2020numerically,batge2021nonequilibrium}, which has also been used to explore thermodynamic properties, such as quantum friction and work, in externally driven systems coupled to several reservoirs~\cite{dou2020universal,batge2022nonadiabatically}.

\subsubsection{Linear response and performance}
\label{subsubsec:Linresp}

Thermoelectric devices are subject to temperature and potential biases
that can induce a heat-to-work conversion.
In the \emph{linear response} regime, classical nonequilibrium thermodynamics provides a general framework for describing steady-state heat and particle currents~\cite{Peliti2011}.
The linear response formalism permits devising simple relations for the figure of merit of generic steady-state heat engines~\cite{Proesmans2016},~e.g., thermoelectric efficiency and more general thermoelectric response functions.
Let us refer once again to the system depicted in Fig.~\ref{fig:thermoel1} and set the reference temperature $T_{\rm{L}}=T$ and the chemical potential $\mu_{\rm{R}}=\mu$, such that the temperature difference is small, $0<T_{\rm L}-T_{\rm R}\equiv \Delta T\ll T$, as is $ |\Delta \mu| \ll k_{B}T$, with $\Delta \mu \equiv \mu_{\rm L}-\mu_{\rm R}<0$. These conditions define the linear response regime we explore next. Notice that, as long as the linear regime holds, the reference temperature could be either equal to $T_{\text{R}}$ or an arbitrary typical temperature of the system \cite{BENENTI20171}.

In this case, the steady-state currents can be linked to potential biases and temperature gradients by means of the Onsager relations \cite{BENENTI20171}, 	   
    \beq\label{eq:thermoel7}
    \begin{pmatrix}
    	J_{e}\\
    	J_{h}
    \end{pmatrix}=
    \begin{pmatrix}
    	L_{ee}&L_{eh}\\
    	L_{he}&L_{hh}
    \end{pmatrix} 
    \begin{pmatrix}
    	F_{e}\\
    	F_{h}
    \end{pmatrix},
\eeq
where  $F_{e}=\Delta V/T$ and $F_{h}=\Delta T/T^2$, with the applied voltage  $\Delta V=\Delta \mu/e$ and the elements $L_{ij}$ (for $i,j=e,h$) of the Onsager matrix $\mathbf{L}$.
Eqs.~\eqref{eq:thermoel7} display the linear relation between the currents and small variations $\Delta V$ and $\Delta T$.

Having a consistent thermodynamic picture of any classical nonequilibrium steady-state requires non-negative entropy production, $\dot{S}=F_{e}J_{e} + F_{h}J_{h}\geq 0$.
Therefore, the Onsager matrix \textbf{L} is positive semidefinite, i.e.,   the elements of the Onsager matrix obey the following relations: 
\beq
L_{ee} \geq 0 \;\;\; {\rm and} \;\;\; L_{hh}\geq \frac{(L_{eh} +L_{he})^2}{4 L_{ee}}\geq 0.
\eeq
For systems that respect time-reversal symmetry, the Onsager reciprocal relations hold, $L_{eh}=L_{he}$, and thus the Onsager matrix is symmetric.

The great advantage of Eqs.~\eqref{eq:thermoel7} is that they allow us to derive useful relations for thermoelectric devices, such as the efficiency at maximum power and general relations in the different operating regimes.
The thermoelectric efficiency, relating the power output and the heat current flowing from the hot reservoir, can be written as 
    \beq\label{eq:thermoel8}
     	\eta= \frac{P_{\rm gen}}{J_{h,\rm{L}}}=-\frac{\Delta V J_{e}}{J_{h}}=-\frac{F_{e}T(L_{ee} F_{e} + L_{eh}F_{h})}{(L_{he} F_{e} + L_{hh}F_{h})}. 
    \eeq
It follows that, for any fixed value of $F_{h}$, taking $F_{e,\rm{max}}=-L_{hh}/L_{he}(1-\sqrt{\det \mathbf{L}/L_{ee}L_{hh}})$, the efficiency in Eq.~\eqref{eq:thermoel8} is maximized.
Moreover,  the transport coefficients of the system -- including the electronic conductance $G$, the heat conductance $K$, the Seebeck and Peltier coefficients $S$, and  $\Pi$ -- can be written in the linear response limit, in terms of the elements of the Onsager matrix $\mathbf{L}$ as follows:  
\begin{eqnarray}
    G &=& \frac{J_{e}}{\Delta V} \Big|_{\Delta T=0}=\frac{L_{ee}}{T},\label{eq:thermoel9}\\
    K &=&  \frac{J_{h}}{\Delta T}\Big|_{J_{e}=0}=\frac{\det \mathbf{L}}{T^2 L_{ee}},\label{eq:thermoel10}\\
    S &=&  - \frac{\Delta V}{\Delta T}\Big|_{J_{e}=0}=\frac{L_{eh}}{T L_{ee}},\label{eq:thermoel11}\\
    \Pi &=&  \frac{J_{h}}{J_{e}}\Big|_{\Delta T=0}= \frac{L_{he}}{L_{ee}}. \label{eq:thermoel12} 
\end{eqnarray}

\begin{figure}%
    \centering
    \includegraphics[width=8.5cm]{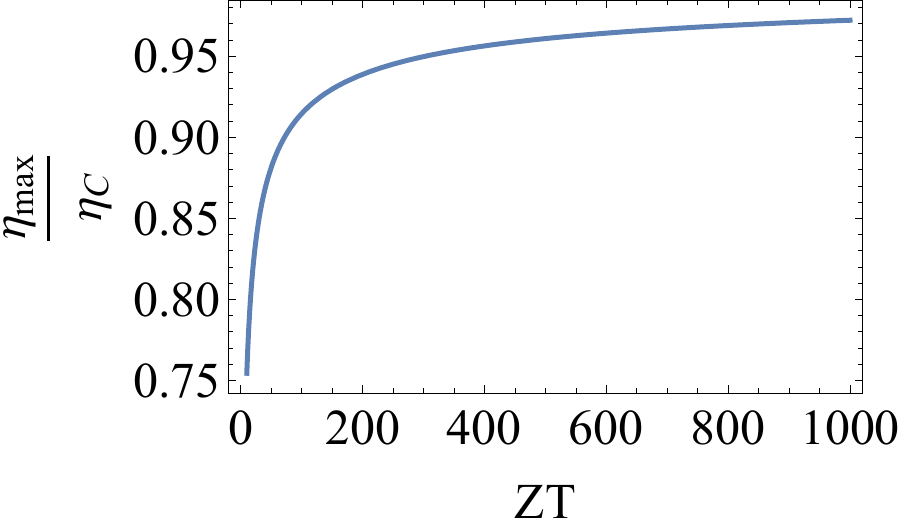} %
    \caption{The maximal thermoelectric efficiency rescaled by the Carnot efficiency as a function of the figure of merit $ZT$. The ratio $\eta_{\rm max}/\eta_{\text{C}}$ approaches 1 as $ZT\rightarrow \infty$. }%
    \label{fig:ZT}%
\end{figure}

The performance of thermoelectric devices can be quantified using the dimensionless figure of merit  $ZT$, which can be written in terms of the coefficients $S,G,K$, and the temperature $T$ as $Z=S^2 G T/K$.
This magnitude is not bounded from above, and in the linear response regime, it can be directly related to the maximum efficiency attainable in heat-to-work conversion.
Inserting the value of $F_{e,\rm{max}}$ in Eq.~\eqref{eq:thermoel8}, it follows that the maximum efficiency can be recast in terms of $ZT$ as follows:
\beq\label{eq:thermoel13}
    \eta_{\rm{max}}=\eta_{\rm{C}}\Big( \frac{\sqrt{ZT + 1} -1}{\sqrt{ZT + 1} + 1}\Big), 
\eeq                        
where $\eta_{\rm{C}}=\Delta T/T$ is the Carnot efficiency.
As an immediate consequence of Eq.~\eqref{eq:thermoel8}, in the linear response regime, for any finite value of $ZT$, thermoelectric efficiency is bounded by the Carnot efficiency, which can only be achieved in the limit $ZT\to \infty$, for the ideal thermoelectric device (see Fig.~\ref{fig:ZT}).
Following a similar route, expressions for efficiency at maximum power can be achieved~\cite{BENENTI20171}.

The previous relations depend neither on the details of the device nor on the approximations employed to compute the currents.
They hold as long as the currents obey Eq.~\eqref{eq:thermoel7}.
In general, due to the reduced dimensions of nanoscale devices, it is experimentally difficult to achieve small temperatures and field gradients \cite{SANCHEZREV2016} over the device length, so the validity of the linear-response picture can be questioned.
On the other hand, linear response expressions for the currents can be obtained starting from the system Hamiltonian, following the different approaches to quantum transport discussed in Sec.~\ref{subsubsec:scattering}.

Below, we briefly introduce the general form of the Onsager coefficients that can be obtained by employing the scattering theory approach.
These expressions are simple enough to provide an intuitive picture of the physical mechanism involved and serve as a starting point for discussing more recent findings.
From Eqs.~\eqref{eq:thermoel1},~\eqref{eq:thermoel2}, expanding the Fermi distributions up to linear order in $k_{\rm{B}}\Delta T,\Delta \mu\ll k_{\rm{B}} T$, the linear-response expression for the currents of the two-terminal device can be obtained. 
The Onsager matrix elements therefore read
\begin{eqnarray}
\label{eq:thermoel14}
    L_{ee}&=&e^2 T I^{(0)},\nonumber\\
    L_{eh}&=&L_{he}=e T I^{(1)},\nonumber\\
    L_{hh}&=&TI^{(2)}, 	
\end{eqnarray} 
with $I^{(n)}=(1/h)\int_{-\infty}^{+\infty}\mathrm{d} E (E-\mu)^{n} \mathcal{T}_{\rm{LR}}(E)(-\partial f(E)/\partial E)$.
Thermoelectric coefficients are calculated straightforwardly from Eq.~(69)-(72): focusing on the Seebeck and Peltier coefficients, we have
\begin{eqnarray}
\label{eq:thermoel15}
    S&=&\frac{1}{e T} \frac{I^{(1)}}{I^{(0)}}=\frac{1}{e T}\frac{\int_{-\infty}^{+\infty}\mathrm{d} E (E-\mu) \mathcal{T}_{\rm{LR}}(E)(-\partial f(E)/\partial E)}{\int_{-\infty}^{+\infty}\mathrm{d}E  \mathcal{T}_{\rm{LR}}(E)(-\partial f(E)/\partial E )},\\
    \Pi&=&TS. 	
\end{eqnarray}
From Eq.~(77), it follows that, if the transmission function $\mathcal{T}_{\rm{LR}}(E)$ is symmetric around the chemical potential $\mu$, the Seebeck coefficient vanishes.

This insight sheds light on a condition sufficient for thermoelectric effects to occur: since electrons and holes bring opposite contributions to $S$, any system breaking the symmetry in the transmission coefficient, below or above the chemical potential, can show thermoelectric features.
In fact, this symmetry can be broken by means of different underlying physical mechanisms, possibly involving nonlinear thermoelectric effects.
In the following sections (Sec.~\ref{subsec:superthermo}), we outline recent proposals of nanoscale systems showing these features.     

\subsection{Multi-terminal devices}\label{subsec:threeterm}
	
\begin{figure}[h]
		\centering
		\includegraphics[scale=0.38]{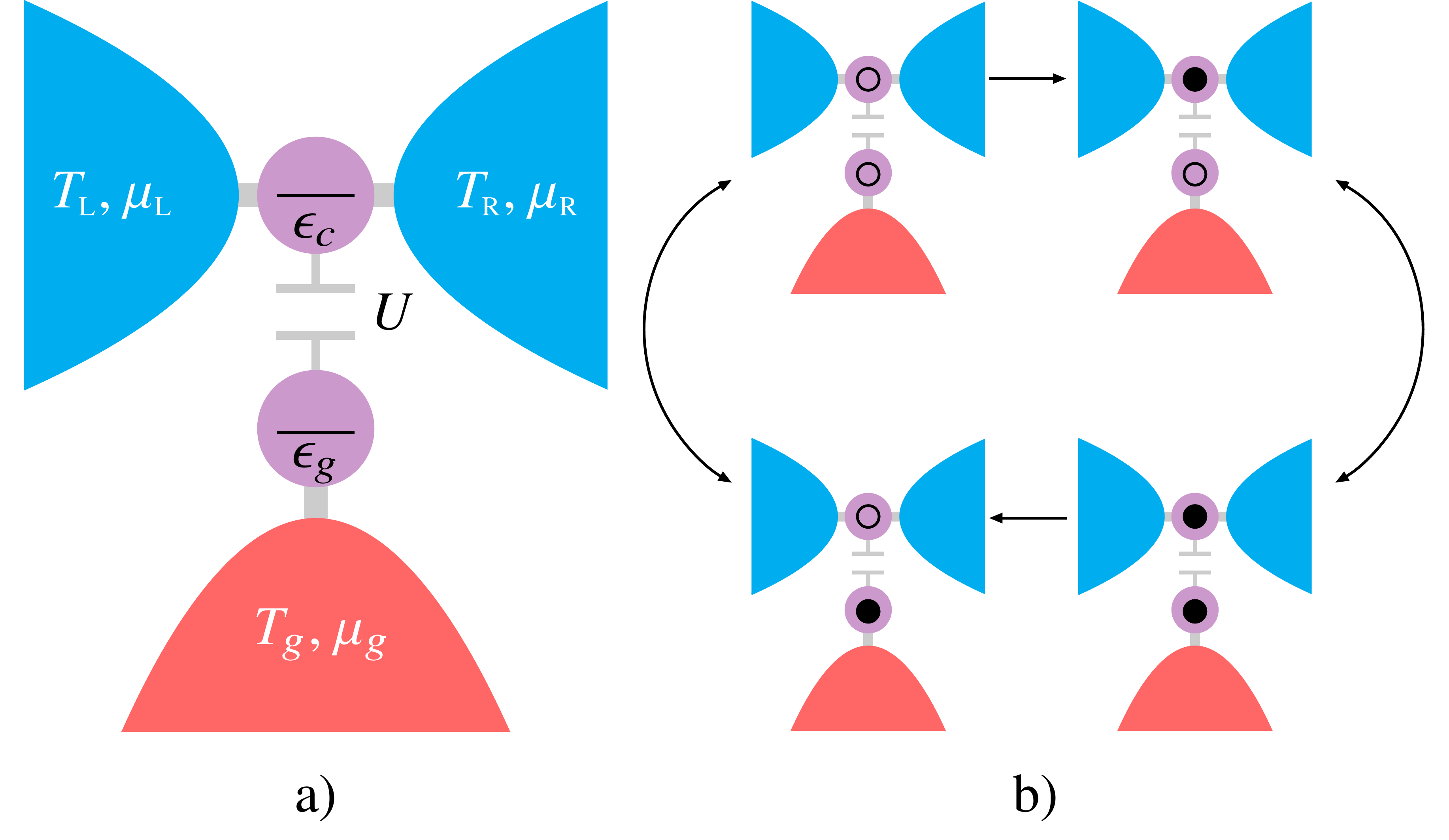}
		\caption{Schematic diagrams of three terminal devices based on Coulomb coupled QDs. a) On the left,  the two QDs are modeled with a single electronic level.~The conductor dot $(\text{C})$ is tunnel coupled to the $\text{L}$ and $\text{R}$ reservoirs, so that charge current can flow among them through $\text{C}$. The gate dot $(\text{G})$ interacts capacitively with the conductor $(\text{C})$, and electrons can tunnel from the gate reservoir to $(\text{G})$. b) On the right, a schematic diagram of the energy harvester's working mechanism. Each cycle comprises single tunneling events from the reservoirs to the QDs. The net result is the absorption of energy $E_{\text{C}}$ from the gate and a single electron tunneling event from $\text{L}$ to $\text{R}$, as shown in Eq.~\eqref{eq:thermoel17}. Reproduced from \cite{Sothmann_2014}. }
		\label{fig:thermoel4}
\end{figure}

A great number of recent studies on nanoscale thermoelectrics have focused on multi-terminal devices.
One common setup is a three-terminal device that makes use of two coupled Quantum Dots (QDs) in the Coulomb blockade
 regime, as sketched in Fig.~\ref{fig:thermoel4}.
 The first QD,~i.e.,~the conductor QD, is tunnel-coupled to the $\rm{L}$ and $\rm{R}$ leads, as in the conventional scheme described by the equilibrium temperatures and chemical potentials $(T_{i},\mu_{i})$.
 It hosts the charge transport between the two leads occurring via electronic tunneling through the dot.
 Furthermore, the conductor is electrostatically coupled to the gate, a second QD that in turn is tunnel coupled to a third lead at temperature $T_{g}$.

Due to the Coulomb coupling between the two dots, no charges can be exchanged between the gate and the conductor.
However, a given amount of energy can be exchanged between the gate reservoir and the conductor, as a consequence of the charging energy $U$ between the dots.~This feature is one of the main advantages of the setup, as it allows a spatial separation between the heat source (the gate), and the working medium (the conductor).

\subsubsection{From the energy harvester to the thermal transistor}\label{subsubsec:harvester}
	  
First proposed in~\cite{SanchezButtiker:2011}, the device in Fig.~\ref{fig:thermoel4} has been proven to work as an \emph{energy harvester}~\cite{Sothmann_2014}, i.e., it converts the heat current coming from the gate into a charge current through the conductor, exploiting the correlations between the charge fluctuations in the gate QD and the electronic tunneling in the conductor. When a load is added, a finite amount of power can be produced with high efficiency.
	
The physics of the device can be understood via the case of the unbiased conductor, i.e., $\mu_{\rm{L}}=\mu_{\rm{R}}$, where the two leads are at equal temperatures $T_{\rm{L}}=T_{\rm{R}}=T_{w}$ and $T_{g}>T_{w}$. 
Heat current is converted to charge current in the conductor when the detailed balance between the tunneling processes in the conductor breaks, due to interdot Coulomb correlations.
This process can be modeled with a rate equation (see Sec.~\ref{subsubsec:kinetics}), describing the probability of occurrence for each of the double-dot system's states that are modeled as $\ket{n_{c},n_{g}}$, where $n_{c},n_{g}=0,1$, depending on whether each dot is empty or occupied.

The correlation between the tunneling and charge fluctuations is modeled by allowing the tunneling rate from the conductor dot to the lead $j$ to depend on the occupation of the dot gate $n_{g}$, i.e., $\Gamma^{\pm}_{j,n_{g}}$, with $j=\rm{L},\rm{R}$ and $\pm$ denoting the probability of leaving and entering the dot from reservoir $j$, respectively.
The conversion of heat to charge current takes place through a cycle, as sketched in Fig.~\ref{fig:thermoel4}.
Each cycle starts when the double dot is empty.
Then an electron tunnels from the lead ${\rm R}$, with rate $\Gamma^{+}_{{\rm R},0}$, while the gate dot is empty.
In the second step, the gate is filled through electron tunneling from the hot reservoir.
In the Coulomb blockade regime, the double occupancy of the QDs requires a finite amount of charging energy, $U=E_{\rm{C}}$, which is extracted from the hot reservoir.
In the third step, the electron in the conductor dot tunnels in the lead ${\rm R}$, and in the last step, the electron in the gate dot tunnels back. 

\begin{figure}%
    \centering
    \includegraphics[width=8.5cm]{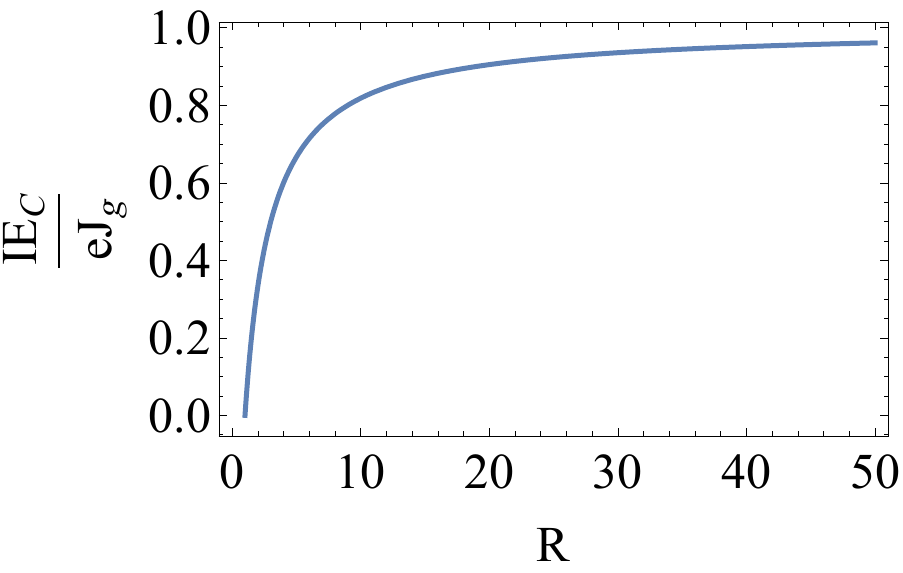} %
    \caption{The current $I$ rescaled by its optimal value $e J_{g}/E_C$ as function of the tunneling rate ratio $R$. Here we chose $R=\frac{\Gamma_{\rm L 0}}{\Gamma_{\rm L 1}} = \frac{\Gamma_{\rm R 1}}{\Gamma_{\rm R 0}} $ to demonstrate the behavior of Eq.~(\ref{eq:thermoel16}) in the limit $\Gamma_{ \rm L0},\Gamma_{\rm R 1}\gg \Gamma_{\rm R 0},\Gamma_{\rm L 1}$ for which the current is optimized as discussed in the text.}%
    \label{fig:current}%
\end{figure}

By solving the corresponding rate equation, it has been proved that a stationary charge current $I$ establishes itself in the conductor, which reads
\beq 
\label{eq:thermoel16} 
	 I=-e \frac{(\Gamma_{\rm{L}1}\Gamma_{\rm{R}0} -\Gamma_{\rm{L}0}\Gamma_{\rm{R}1})}{(\Gamma_{\rm{L}0}+\Gamma_{\rm{R}0})(\Gamma_{\rm{L}1}+\Gamma_{\rm{R}1})} \frac{J_{g}}{E_{\rm{C}}},
\eeq         
where $J_{g}$ is the heat that flows through the gate and $\Gamma_{j,n_g} = \Gamma_{j,n_g}^+ +\Gamma_{j,n_g}^- $. 
Notice that the direction of the current depends on the asymmetry of the tunneling rates.
Moreover, the charge transport can be optimized by imposing that an electron tunnels from the ${\rm L}$ lead to the dot when the gate dot is empty, and from the dot to the right lead ${\rm L}$ only when the gate dot is full, i.e., $\Gamma_{ \rm L0},\Gamma_{\rm R 1}\gg \Gamma_{\rm R 0},\Gamma_{\rm L 1}$ as plotted in Fig.~\ref{fig:current} .
In this limit~\cite{SanchezButtiker:2011}, an amount of energy equal to $E_{\rm{C}}$ is absorbed from the gate, and a single electron tunnels from $\rm{L}$ to $\rm{R}$, so that Eq.~\eqref{eq:thermoel16} reduces to    
	\beq\label{eq:thermoel17}
	\frac{I}{e}=\frac{J_{g}}{E_{\rm{C}}}. 
	\eeq
It is worth noting that the charge current is proportional to the heat current.

When a potential bias $\Delta V$ between the two leads is applied, a steady current develops against it, producing work. The efficiency, $\eta=P_{\rm{gen}}/J_{h}=e\Delta V/E_{\rm C}$, grows linearly with the bias $\Delta V$, until the stopping potential~\cite{BENENTI20171}, $\Delta V_{\rm{stop}}=E_{\rm{C}} \eta_{\rm{C}}/e$, is reached. 
At this point, the device operates with Carnot efficiency.
On the other hand, the efficiency at maximum power, $\eta_{\rm{Pmax}}$, grows as $\eta_{\rm C}/2$ for small temperature bias, $\Delta T=T_{g} - T_{w}$ (linear response regime).
For higher values of $\Delta T$, i.e., in the nonlinear response case, it can grow faster.

Over the last decade, several platforms have experimentally achieved energy harvesting based on QDs~\cite{Roche:2015,Thierschmann:2015,Josefsson:2018,prete2019thermoelectric}.
However, these devices employing Coulomb-coupled QDs  usually have low power production.
Alternative proposals, relying on a couple of resonant tunneling quantum dots~\cite{Jordan:2013}, can deliver higher power and high efficiency at maximum power.
Recently, a similar configuration was experimentally realized~\cite{Jaliel:2019}, with two QDs connecting equal temperatures leads to a higher temperature cavity.
Electrons tunneling into the left QD can only tunnel to the right lead if they gain a prescribed amount of energy, equal to the energy difference between the two-dot levels.
In this way, the thermal energy gained by the electron tunneling into the cavity can be converted into an electrical current.    
	
More recent theoretical studies have produced new insights into a three-terminal device's mechanism of power production.  In~\cite{Jiangthermo2012,Jiangthermo2015}, three-terminal thermoelectric devices composed of two QDs working with phononic baths were devised.
In~\cite{Mayrhofer:2021}, the power generation in three-terminal devices, as in Fig.~\ref{fig:thermoel4}, has been traced back to a purely stochastic process. Thus, the operating mechanism of these devices cannot be related to self-oscillating autonomous heat engines~\cite{Wachtler_2019}.
By employing stochastic unraveling of the master equation describing the three-terminal device, the occurrence of cycles corresponding to power production, like those depicted in Fig.~\ref{fig:thermoel4}, has been characterized as a random process in time. Furthermore, the exact duration of each cycle is not deterministic but rather shows stochastic fluctuations.
The entropy production of the device has been linked to the rate of each cycle, and heat transfer to the conductor has been proven to act as a stochastic piston, which changes the energy of the conductor dot at random in time.
	
Coulomb-coupled, QD-based devices have also been proven effective in gaining control of the heat current flowing between the $\rm{L}$ and $\rm{R}$ leads, when no bias other than the temperature difference $\Delta T=T_{\rm{L}} - T_{\rm{R}}$ is present.
Indeed, the charge fluctuations in the gate, induced by the gate reservoir, can change the thermal currents inside the conductor.
In~\cite{Sanchez_Thermaltrans:2017}, the thermal control of heat currents has been theoretically proposed. 
The possibility to block heat currents along a conductor channel, i.e., the achievement of a thermal transistor, has been investigated.
Moreover, thermal gating has also been proposed as a tool for achieving refrigeration effects, to obtain a QAR~\cite{erdman2018absorption} (see Sec.\ref{subsubsec:absopt}) and nearly-ideal rectification of the heat currents~\cite{Sanchez_thermalgate:2017}.
	
Quite recently, the idea of a thermal transistor was reexamined via a modified version of the three-terminal device, where in place of the conductor QD, a quantum point contact (QPC) is used~\cite{Yang:thermTrans19}.
In this case, the capacitive coupling with the gate QD modifies the transmission probability of the QPC, which is found to depend on the electrostatic potential due to the presence of the dot, and thereby on the occupation state of the dot.
Employing the Landauer-B\"uttiker approach, the differential sensitivity of the average charge current in the dot as a function of the temperature of the gate reservoir, i.e., $\lvert {\mathrm{d}\langle I\rangle_{T_{g}}/\mathrm{d} T_{g}} \rvert$,  has been computed, along with the power gain as a function of $T_{g}$.
Moreover, non-invasive thermometry is attained, 
as only a negligible amount of energy flows back and forth between the dot and the gate, without any energy exchange between them.

\subsubsection{Hybrid thermal machines}\label{subsubsec:hybrid}
 
Theoretical proposals of multiterminal devices performing multiple thermodynamic tasks \cite{entin2015enhanced} at the same time have been considered in~\cite{manzano2020hybrid}.
\begin{figure}[h]
		\centering
		\includegraphics[scale=0.28]{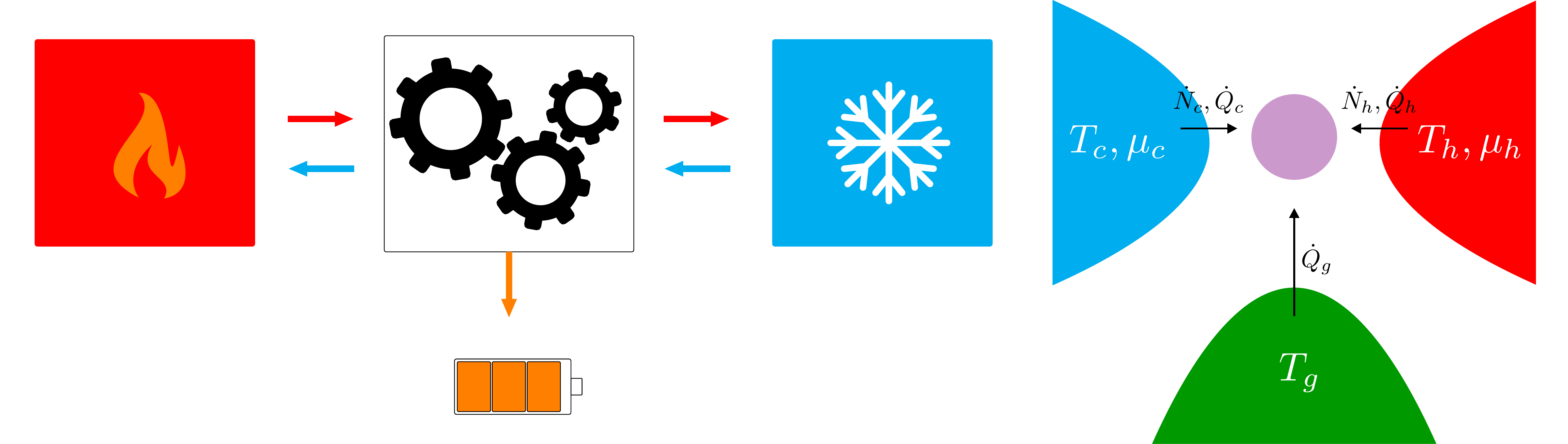}
        \caption{Sketch of a multiterminal device employed as hybrid machine.~By exploiting additional conserved quantities linked to the generalized Gibbs state of the reservoirs, these engines can perform multiple thermodynamic tasks at the same time, expanding the capabilities of the autonomous engines described in Sec.~\ref{subsubsec:harvester}.}
		\label{fig:thermoel5}
\end{figure}
These machines employ additional conserved physical quantities~\cite{guryanova2016thermodynamics,yunger2016microcanonical,Lostaglio2017} that can be exchanged among multiple reservoirs. It follows that unconventional operating regimes can be achieved, in which more than a single task can be performed at the same time, e.g., cooling and power production.
Analogous to QHEs with nonequilibrium baths (Sec.~\ref{subsec:thermotheory}), this class of machines is imagined to work with reservoirs described by generalized Gibbs state,  $\rho^{\rm{gen}}_{\rm G}=\exp(-\beta(H-\sum_{\alpha}\mu^{\alpha}A^{\alpha}))/Z$, with $Z=\tr[ \exp(-\beta(H-\sum_{\alpha}\mu^{\alpha}A^{\alpha}))]$ and $\mu^{\alpha} $ and $ A^{\alpha}$ the thermodynamic potentials and the operators describing the independently conserved quantities respectively.
For example, $A^{\alpha}$ can be the number of particles and angular momentum. With $E$ denoting the amount of energy flowing out of the reservoir, the total heat exchange rate reads $\dot{Q}=\dot{E}-\sum_{\alpha}\mu^{\alpha}\dot{A}^{\alpha}$.
Moreover, the total power output  can be defined as $\dot{W}=-\sum_{\alpha}\mu^{\alpha}\dot{A}^{\alpha}$, such that a positive amount of work output increases the energy of the reservoir, while positive $\dot{Q}$ means a decrease in the reservoir energy.

Let us consider a machine working with multiple reservoirs, each labeled  $i$ and with a temperature $\beta_{i}$.
In the long-time limit, the device approaches a nonequilibrium steady state, where currents flowing among the different reservoirs are established.
While the laws of thermodynamics do not restrict the number of device tasks, they do limit their operation regimes.
The first and second laws of thermodynamics entail energy flow balance $\sum_{\alpha}\dot{W}^{\alpha}=\sum_{i}\dot{Q}_{i}$ and the non-negativity of the  entropy production $\dot{S}=-\sum_{i}\beta_i \dot{Q}_{i}\geq 0$, with the different power output contributions $\dot{W}^{\alpha}=-\mu^{\alpha}_{i}\dot{A}^{\alpha}_{i}$.

Unlike conventional QHEs, conserved quantities imply that the engine can operate with multiple input and output tasks.
As a consequence, a consistent way to classify useful and wasteful processes is needed, as well as to assess whether a given output and input task is more useful or wasteful than the other.
This can be achieved by defining a reference temperature $T_{r}$.
If a reservoir temperature is $T>T_{r}$, then extracting (dumping) heat from it is considered wasteful (useful), and vice-versa for $T<T_{r}$.
As previously discussed in Sec.~\ref{subsec:thermotheory}, an operational definition of the free-energy of a single reservoir is introduced with respect to $T_{r}$. The change in the free-energy of reservoir $i$ is given by $\dot{F}_{i}=-\dot{E}_{i} -k_{\rm{B}}T_{r}\dot{S}_{i}$ and implies that the second law can be recast as~\cite{manzano2020hybrid} 
\beq
\dot{F}_{\rm{tot}}=\sum_i\dot{F}_{i}=\sum_{\alpha} \dot{W}_{\alpha} +\sum_{i} \dot{Q}_{i}(T_{r}/T_{i} - 1) \leq 0.
\eeq

Useful work and heat contributions hold positive signs -- e.g., a current $\dot{A}^{\alpha}$ flowing against a thermodynamic potential difference $\Delta \mu^{\alpha}$, or a finite amount of heat extracted from a reservoir at $T<T_{r}$ -- while, by contrast, wasteful ones bear negative signs, such that the efficiency of these devices can be easily defined as the ratio between the useful terms (the outputs) and the wasteful terms (the inputs),
\beq
\eta_{\rm{hyb}}=-\frac{\sum^{+}_{\alpha} \dot{W}_{\alpha} +\frac{1}{2}\sum_{i} \dot{Q}_{i}(T_{r}/T_{i} - 1)+|\dot{Q}_{i}(T_{r}/T_{i} - 1)|}{\sum^{-}_{\alpha} \dot{W}_{\alpha} +\frac{1}{2}\sum_{i} \dot{Q}_{i}(T_{r}/T_{i} - 1)-|\dot{Q}_{i}(T_{r}/T_{i} - 1)|}.
\eeq
This unconventional definition of efficiency enables an evaluation of multiple, simultaneous tasks that may oppose one another. It is bounded $\eta_{\rm{hyb}}\leq 1$ and can be saturated 
in reversible processes when the entropy generation vanishes.  These cases can be thought of in terms of the Carnot limit in heat
engines, where heat is converted into work at maximum efficiency but infinitely slowly.
The approach is applied to a three-terminal device, as depicted in Fig.~\ref{fig:thermoel5}.  More recent works on hybrid thermal machines address the role of system-bath coupling modulations~\cite{CAVALIERE2023106235} (see also Sec. 6.2.1) and phonon-assisted inelastic processes~\cite{Lumultitask2023}.

\subsection{Superconducting thermoelectric engines and coolers}\label{subsec:superthermo}
\begin{figure}[h]
    	\centering
    	\includegraphics[scale=0.35]{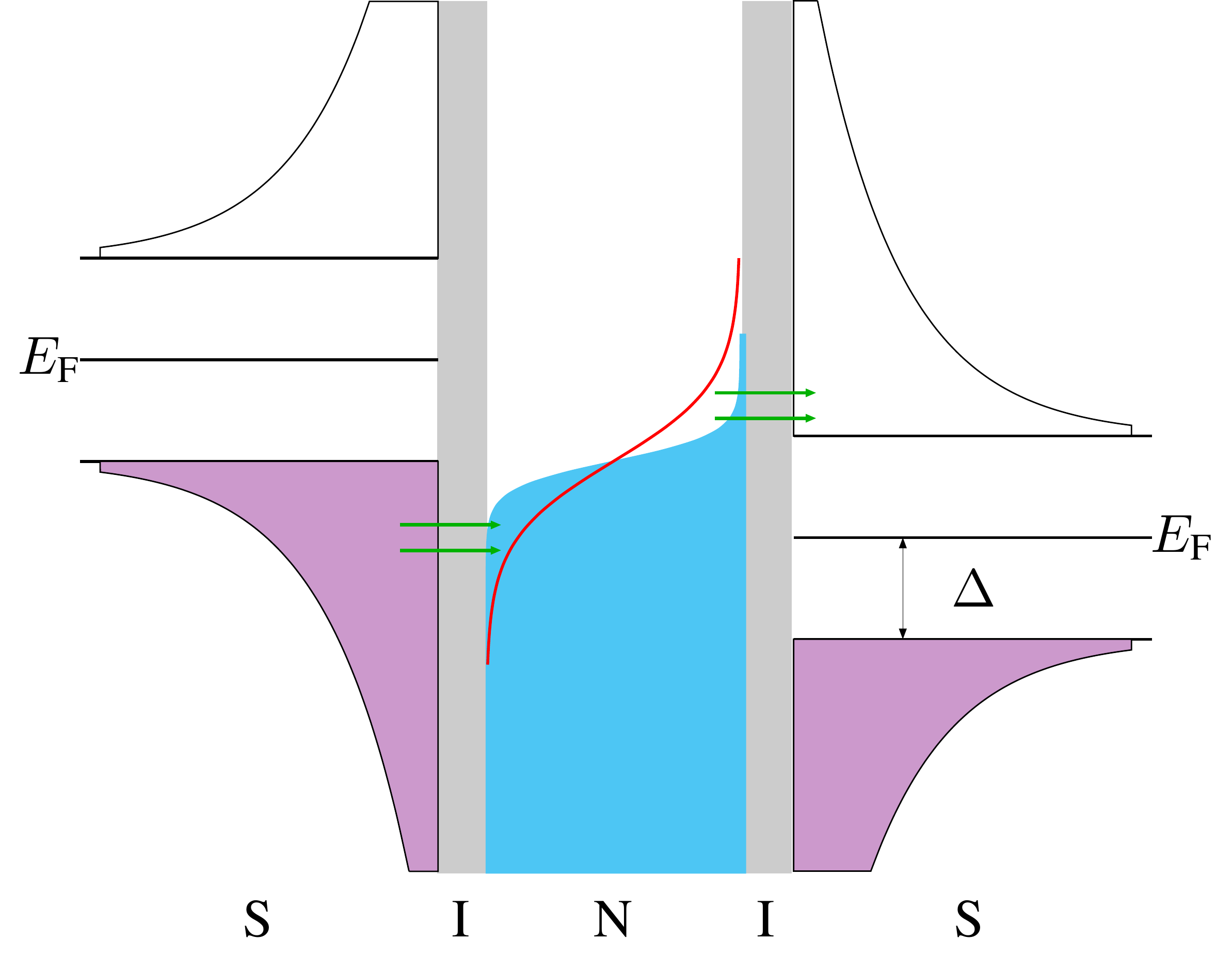}
        \caption{Semiconductor model of a SINIS cooler. For temperatures well below the critical value, in the $\text{S}$ zones the excited quasiparticle states above the superconducting gap are empty. After the application of a bias field, the hot electrons in the $\text{N}$ metal can tunnel to the quasiparticle states of the right $\text{S}$, while cold electrons below Fermi energy can come from the left $\text{S}$.  Reproduced from \cite{Muhonen_2012}.}
    	\label{fig:thermoel6}
\end{figure}

Recent works have brought forward theoretical proposals for nanoscale thermoelectric devices that employ superconducting elements. Traditionally, normal-insulating-superconducting (SINIS) junctions have been studied as suitable experimental platforms for developing on-chip coolers~\cite{Giazotto:2006,Muhonen_2012,Pekola:2021}.
The structure of a SINIS cooler is sketched in Fig.~\ref{fig:thermoel6}. 
Here a normal (N) metal island is tunnel-coupled to a pair of superconducting materials (S), by means of insulating (I) barriers.
When a quasi-equilibrium description is adopted, the S island can be described by the \acrshort{BCS} density of states \cite{Tinkham:1996} with superconducting gap $\Delta$, while the N island is described by the Fermi distribution.
Furthermore, the temperatures of the superconducting elements are assumed to be well below their critical values, $T\ll T_{\text{C}}$.

Cooling of the metallic island N can be achieved by employing the superconductor's gapped density of states as an energy filter.
If a suitable voltage bias is applied, only the high-energy quasiparticles  tunnel from the metallic island to the empty states of the superconductor.
As a result, the average energy of electrons in the N metal decreases, the Fermi distribution tends to become steeper, and a cooling effect is obtained.

Thermoelectric engines and coolers -- based on normal-superconducting-normal (NSN) systems, as well as superconducting-insulator-superconducting (SIS) tunnel junctions -- have been studied extensively. See review paper~\cite{Muhonen_2012} and references therein (see also Sec.~\ref{subsubsec:nonlinear}).
A common feature of these devices is heat transport, which is entirely due to the tunneling of fermionic quasiparticles, while the Cooper pairs condensate, carrying zero entropy, plays no role. On the other hand, over the last few years, several theoretical proposals have focused on thermoeletric devices based on the physical properties of Cooper pairs tunneling, e.g., Josephson junctions~\cite{giazotto2012josephson,marchegiani2016self,Vischi:2019,manikandan2019superconducting,marchegiani2020phase,germanese2022bipolar} and Cooper pairs splitters (\acrshort{cps})~\cite{Tan:2021}.
In addition, non-local thermoelectric effects linked to the quantum spin Hall effect~\cite{Blasi_nonlocal:2020} have been predicted, and topological Josephson engines~\cite{Scharf:2020,scharf2021thermodynamics} have also been proposed.
Ahronov-Bohm  heat engines in the linear response regime have also been studied~\cite{haack2019efficient}. 
	
\subsubsection{Photonic superconducting engines}\label{subsubsec:photosuper}

\begin{figure}[h]
    	\centering
         \includegraphics[scale=0.45]{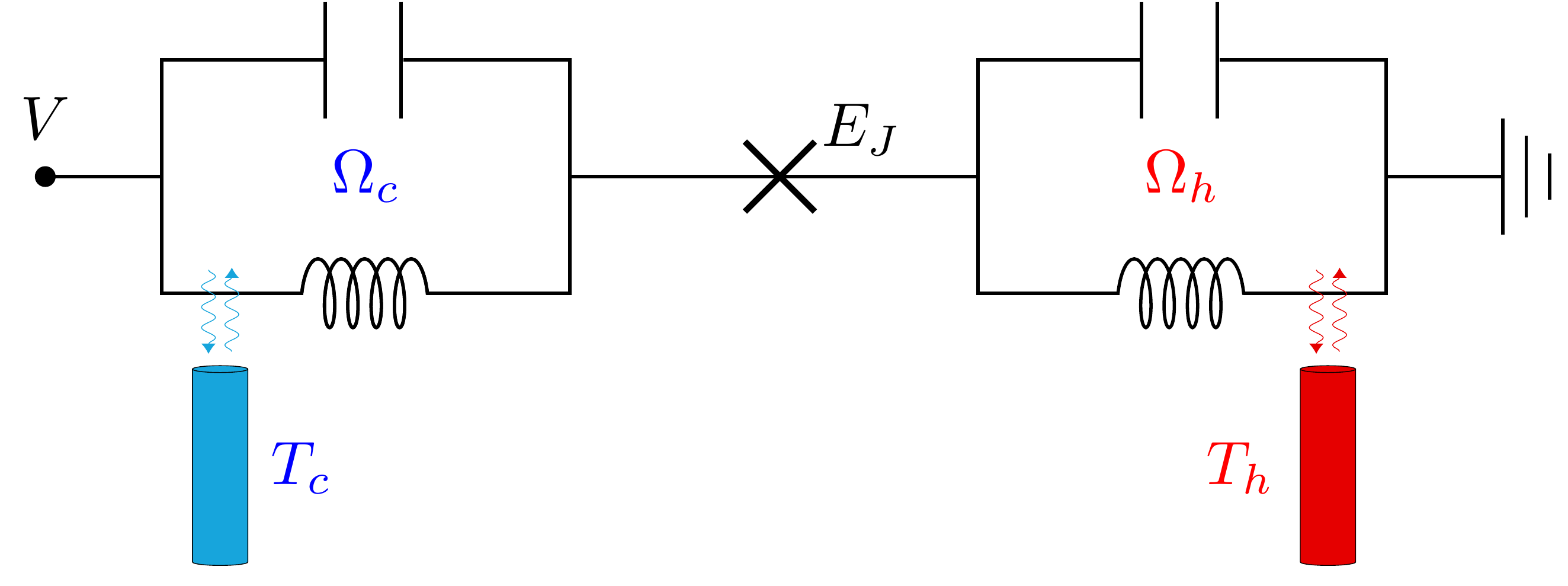}
    	\caption{Schematic diagram of the thermoelectric engine.~A voltage-biased Josephson junction (denoted by $E_{\text{J}}$) is coupled to two microwave cavities at frequencies $\Omega_{h}>\Omega_{c}$.~The coupling is achieved via magnetic fluxes $\varphi_{h}(t),\varphi_{c}(t)$ that modulate the phase of the junction in time.
     Each microwave cavity is coupled to a thermal bath, such that the cavity photons follow a thermal distribution. The setup provides a clear separation between heat and charge currents. Reproduced from \cite{Potts:2016}.}
    	\label{fig:thermoel7}
\end{figure}

The idea of a mesoscopic thermoelectric device based on resonant Cooper pairs tunneling in a Josephson junction has been put forward in~\cite{Potts:2016,Lorch:2018}.
The device is composed of a voltage-biased Josephson junction, coupled to two photonic microwave cavities of frequencies $\Omega_{h}$ and $\Omega_{c}$, as shown in Fig.~\ref{fig:thermoel7}. 
Each cavity is in contact with a hot and cold thermal bath respectively.~It is assumed that $\Omega_{h}>\Omega_{c}$ and the equilibrium occupations of the two cavities fulfill $\ave{n_{h}}>\ave{n_{c}}$. In the case of a Josephson junction irradiated by photons, it is well known that the supercurrent can exhibit finite steps at discrete values of the bias voltage \cite{Tinkham:1996}.~In a rather similar way, here a Cooper pair can tunnel against the bias $V$ of the junction, by exchanging photons with the two cavities.~In particular, When the bias voltage of the junction is
\beq\label{eq:thermoel22}
	2eV= m \Omega_{h} - n \Omega_{c}, 
\eeq       
with $m,n$ integers, the Cooper pair can tunnel against the bias $V$ by annihilating $m$ photons from the hot cavity and creating $n$ photons in the cold one. Thus, the system uses the heat from the two thermal reservoirs to produce a current against the bias, leading to usable power.

An important feature of this engine is that it allows for a full separation between heat and charge currents.
The Cooper pairs carry zero entropy, while the heat arises from energy exchange with the resonator.
The engine's efficiency can be computed for a single tunneling Cooper pair as  
\beq\label{eq:thermoel18}
	\eta= \frac{P_{\rm gen}}{J_{h}}= \frac{2 e V}{m \Omega_{h}} =1-\frac{n \Omega_{c}}{m \Omega_{h}}. 
\eeq
It has been proven that Eq.~\eqref{eq:thermoel12} is bounded by the Carnot efficiency. However, as the heat and charge currents are found to be proportional, the device allows for high power values in the conversion efficiency.
Indeed, at maximum power, the efficiency turns out to be above the Curzon-Ahlborn bound.   
Similarly, an autonomous absorption refrigerator enhanced by quantum coherence~\cite{mitchison15,mitchison2019quantum} can be devised by employing three cavities connected to baths at different temperatures with a phase-biased Josephson junction~\cite{Potts_Refrig:2016,Mitchinson:Qthermo2018}.
In this vein, more recent proposals~\cite{Verteletsky:2020} have focused on thermoelectric engines composed of a pair of qubits, each coupled to a bath at a fixed temperature, coupled together with a Josephson junction.
A complete characterization of the engine performance, i.e., of work and heat fluctuations, has been obtained, along with a study of two-time correlation functions between different energy exchange events in a steady state. 
	
\subsubsection{Nonlocal thermoelectric engines and coolers}\label{subsubsec:nonlocal}

\begin{figure}[h]
    	\centering
    	\includegraphics[scale=0.43]{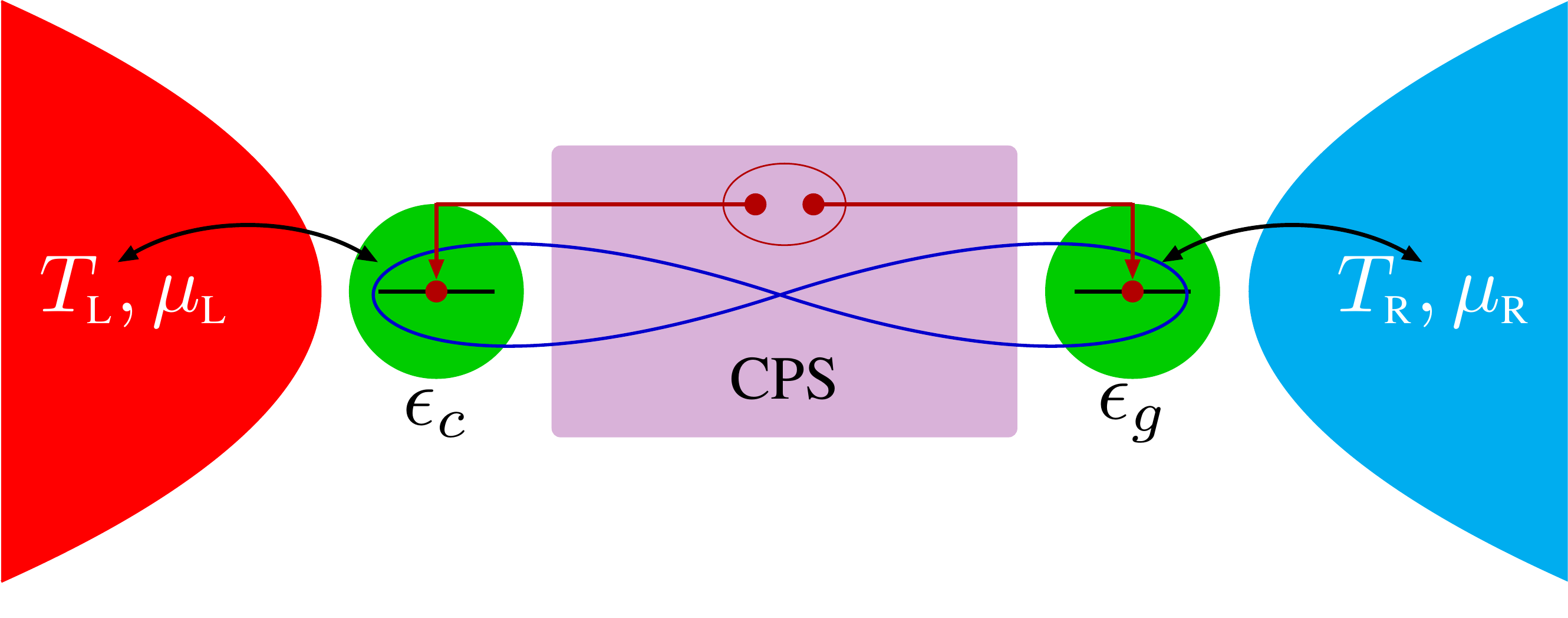}
        \caption{Sketch of the CPS heat engine. The central superconductor is coupled to a pair of QDs.~Due to the Coulomb repulsion, the Cooper pairs coming from the center split into two spin-entangled electrons on each QD. Then they can tunnel into each lead.~The cooling effect takes place as electrons tunnel in the states above (below) the Fermi level of the L (R) lead.~Reproduced from \cite{SanchezCPS:2018}.}
    	\label{fig:thermoel8}
    \end{figure}

Cooper pairs based thermoelectric devices have been the focus of several recent works.
In~\cite{SanchezCPS:2018,Kirsanov:2019,Hussein:2019}, Cooper pairs splitters based heat engines and coolers have been proposed.
A CPS is a device capable of splitting a Cooper pair by producing spin-entangled pairs of electrons in a singlet state~\cite{Hofstetter:2009}. As shown in Fig.~\ref{fig:thermoel8}, correlated electrons, resulting from the splitting of a pair in the central superconducting lead, can tunnel into different QDs.
In this scheme, the superconductor acts as a source of Cooper pairs, and each one of the dots is coupled to a normal electronic lead, denoted by $\text{L}$ and $\text{R}$, such that electrons can tunnel into each of them.

An interesting operating regime is demonstrated when the system shown in Fig.~\ref{fig:thermoel8} works as a cooler. With a parameter regime where $T_{\rm{L}}>T_{\rm{R}}$, the chemical potentials are equal and lower than the superconducting one, $\mu_{\rm{L}}=\mu_{\rm{R}}<\mu_{\rm{S}}$, and the QDs' energy levels obey $\epsilon_{\rm{R}}=-\epsilon_{\rm{L}}$. Electrons below (above) the chemical potential are injected in the R~(L) leads, cooling the R one.~In other words, the heat currents fulfill $J_{\rm{R}}>0\mbox{, }J_{\rm{L}}<0$.~This cooling mechanism is different from previous examples (see Sec.~\ref{subsubsec:absopt} and Fig.~\ref{fig:thermoel6}), as it requires neither a voltage bias between the two leads nor coupling to a third reservoir. Moreover, due to the novel feature of the proposed system, charge is injected into the normal leads through an energy-conserving process.
    
The main mechanism can be understood when the central superconductor is modeled in an effective way, so that the following Hamiltonian can be used to describe the tunneling between the two dots:  
	\beq\label{eq:thermoel19}
	H_{\tau}=-\sum_{\sigma=\uparrow,\downarrow} t_{\rm{EC}} d^{\dagger}_{\text{L}\sigma}d_{\text{R}\sigma} -\frac{t_{\rm{CPS}}}{\sqrt{2}}(d^{\dagger}_{\text{L} \uparrow}d^{\dagger}_{\text{R} \downarrow} -d^{\dagger}_{\text{L} \downarrow}d^{\dagger}_{\text{R} \uparrow}) + \rm{h.c.}, 
	\eeq
where $d^{\dagger}(d)$ are the creation (annihilation) operators of electrons on each of the two QDs, and h.c. is the hermitian conjugate.

The CPS leads to the creation of two spin-correlated electrons in the QDs, while electrons can still move from one side to the other via elastic cotunneling (\acrshort{ec}) processes, without changing the charge state of the superconductor. The peculiar form of Eq.~\eqref{eq:thermoel18} tends to hybridize equal parity states. The robustness of the CPS-induced cooling against EC depends on the parameter ranges of the model.
The nonlocal transport properties of the system can also lead to a heat engine operation. Cooling performance has been studied by computing the Coefficient of Performance (\acrshort{cop}): it approaches the ideal value for decreasing values of the ratio $t_{\rm{EC}}/t_{\rm{CPS}}$, i.e., the stronger the CPS is with respect to the EC.                
Moreover, the non-local Seebeck effect employing a graphene-based CPS device was experimentally observed. Specifically, with an experimental platform conceptually similar to the previous theoretical description, thermal gradients were employed to produce entangled pairs of electrons~\cite{Tan:2021,ranni2022local}.
More recent theoretical proposals of nonlocal heat engines focus on hybrid setups, where a pair of QDs is in contact with superconducting and normal reservoirs~\cite{tabatabaei2022nonlocal}.  
   
   
\subsubsection{Nonlinear thermoelectric engines}\label{subsubsec:nonlinear}
    
\begin{figure}[h]
    	\centering
    	\includegraphics[scale=0.35]{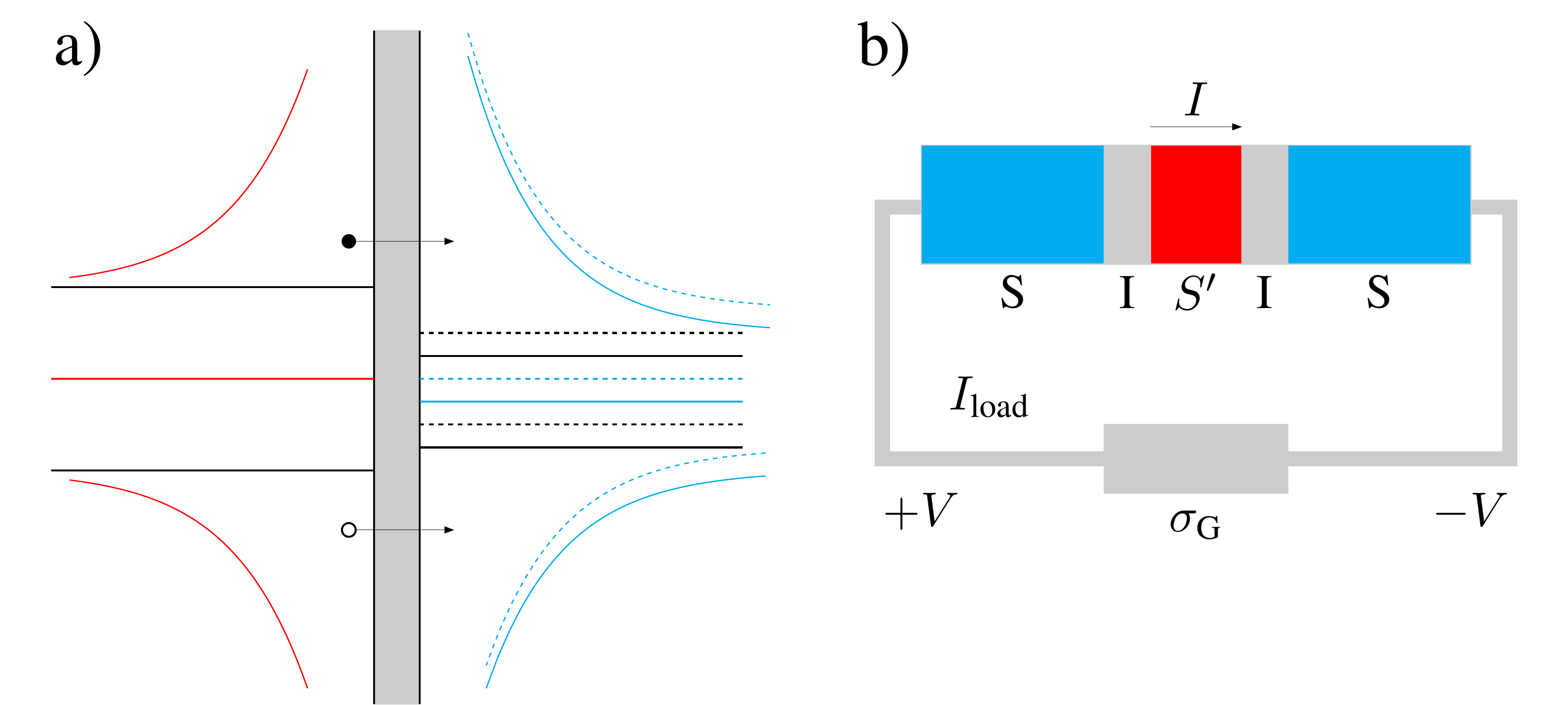}
        \caption{Schematic diagram of an SIS junction working as a nonlinear thermoelectric engine. a) Sketch of the semiconductor model of the SIS. b) Schematic diagram of the biased junction. Reproduced from \cite{Marchegiani:2020,Marchegiani_heateng:2020}.}
    	\label{fig:thermoel9}
    \end{figure}

Besides showing nonlocal thermoelectricity, superconducting elements can also lead to nonlinear thermoelectric response. In~\cite{Marchegiani:2020}, an SIS junction was considered as a platform that could exhibit a peculiar instance of \emph{nonlinear} thermoelectric effect.
The special feature of this platform is the thermoelectric behavior in the presence of Electron-Hole (EH) symmetry, in contrast to the linear-response regime (see Sec.~\ref{subsubsec:Linresp}), where an asymmetry in electron-hole transport is required. The SIS junction is sketched in Fig.~\ref{fig:thermoel9}.        
The two superconductors are modeled with the Bardeen-Cooper-Schrieffer (BCS) theory, and are described by their order parameters $\Delta_{\alpha}(T)$   
and  density of states (DOS) $N_{\alpha}(E)=\theta(|E|-\Delta_{\alpha})|E|/\sqrt{E^2-\Delta^2_{\alpha}}$, where  $\alpha={\rm L,R}$ refers to the left and right superconductors.
It is further assumed that all contributions arising from Josephson effect are neglected.
Thus, when a voltage bias $V$ is applied, $\mu_{\rm L}-\mu_{\rm R}=-eV$ and for $T_{\rm L}\neq T_{\rm R}$ sufficient conditions for a finite thermoelectric power $\dot{W}=-I_{\rm R}V>0$ can be found.
This resource is equivalent to finding a parameter regime for which negative conductance $I(V)/V$ occurs.
When $V>0$ and $T_{\rm{L}}>T_{\rm{R}}$, the expression for the particle and heat currents in the SIS junction is   
\begin{eqnarray}
\label{eq:thermoel20}
I_{\rm L}&=&-\frac{G_{T}}{e}\int_{-\infty}^{+\infty} \mathrm{d} E N_{\rm L}(E_{\rm L})N_{\rm R}(E_{\rm R})[f_{\rm L}(E_{\rm L})-f_{R}(E_{\rm R})], \\ \nonumber
\dot{Q}_{\rm L}&=&\frac{G_{T}}{e^2}\int_{-\infty}^{+\infty} \mathrm{d} E 
 E_{\rm L}N_{\rm L}(E_{\rm L})N_{\rm R}(E_{\rm R})[f_{\rm L}(E_{\rm L})-f_{\rm R}(E_{\rm R})],
\end{eqnarray}
where $E_{\alpha}=E -\mu_{\alpha}$ and $f_{\alpha}(E_\alpha)$ is the Fermi distribution. 

It has been observed that, as a general requirement for achieving a negative current, the hot L \acrshort{DOS} has to be gapped, while the R DOS needs to show behavior that decreases monotonically with $E$. In general, the SIS junction can be designed to fulfill these conditions~\cite{Marchegiani_heateng:2020}.
The thermoelectric effect can be understood in the following manner: for zero bias $V=0$, EH symmetry guarantees that holes and electron currents exactly cancel each other out. Yet with  a finite bias such that $\mu_{\rm L}-\mu_{\rm R}>0$,
R DOS is shifted and the hole transport is increased at the expense of the electron one.
Thus, a net charge transport against the bias is present. 
The nonlinearity of the current-voltage characteristics is then clearly observed within the gap, i.e., $eV< \Delta_{\rm L}(E_{\rm L})+\Delta_{\rm R}(E_{\rm R})$. In the opposite regime, the linearity is restored.

\section{Quantum Entanglement, Information, and Measurement-driven quantum heat engines}\label{sec:information}

In this section, we will provide a brief overview of the fundamental ideas behind Maxwell's Demon (\acrshort{md}) engines. Unlike the conventional machines described in previous sections, these engine models are designed to investigate the fundamental link between information and thermodynamics laws~\cite{goold16}.
After summarizing some of the theoretical approaches developed for this purpose in the classical setting (Sec.~\ref{subsec:ClassicSzilard}), we will review recent theoretical and experimental proposals of MD engines that work in the quantum regime (Sec.~\ref{subsec:MDEng}).
Here new effects can come into play, due to the quantum nature of the working medium as well as the invasive nature of quantum measurement.
As with the QHEs reported in Sec.~\ref{subsec:Enginesintro}, the possible occurrence of a genuine quantum advantage in these machines' performance remains a debated issue.
We will further review engine models fuelled by measurements and entanglement (Sec.~\ref{subsec:MeasFuel}), and,
finally, we will describe QHE models that can generate entanglement (Sec.~\ref{subsec:EntGen}).

\subsection{The classical Maxwell's demon and the Szil\'ard engine}\label{subsec:ClassicSzilard}

Maxwell's demon (MD) is an ideal being capable of measuring the microstate of a physical system, e.g., a gas particle stored in a container that can exchange heat with a thermal reservoir at fixed temperature $T$.
Moreover, it can carry out a feedback protocol that, using the information gained through measurement, allows it to extract an amount of work from the reservoir in a cyclic way, apparently violating the second-law of thermodynamics~\cite{Leff:2002,Maruyama:RevModPhys_2009}.
This theoretical paradox has received a great deal of attention over the last century.
One of the first steps towards the paradox's solution was formalized by Szil\'ard~\cite{szilard29}, who first devised an ideal engine run by the demon.
	
In its simplest realization, the Szil\'ard engine consists of a vessel containing only one gas molecule.
\begin{figure}[h]
        \centering
        \includegraphics[scale=0.75]{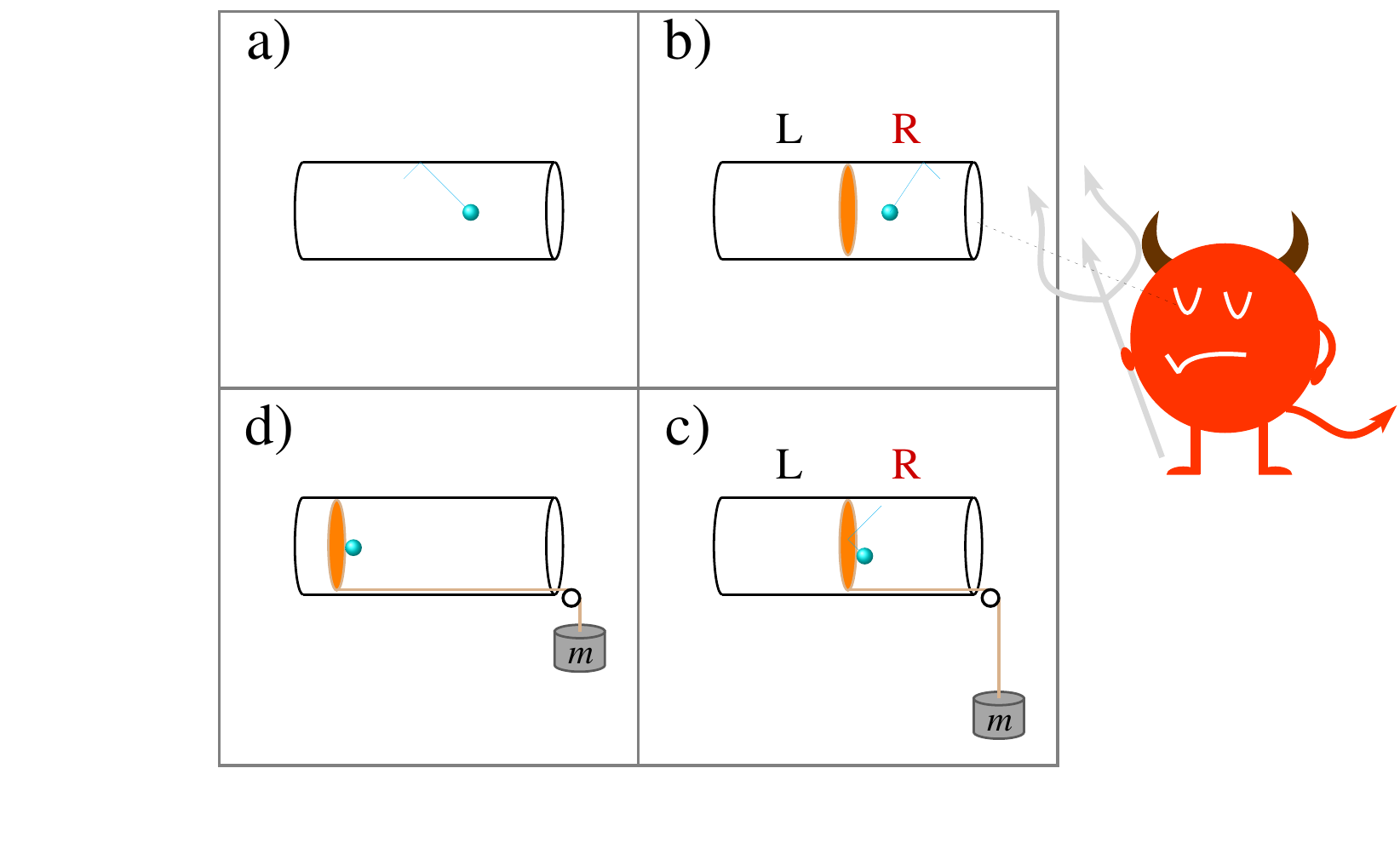}
        \caption{Sketch of the Szil\'ard engine through its four stages. starting at the top left and going clockwise. a) A single gas particle moves in a chamber that is held in contact with a reservoir at fixed temperature $T$. b) The demon inserts a separation in the chamber to divide it equal volumes partitions L and R. c) The demon measures the position of the particle and gets a bit of classical information. Based on this information, it inserts a device to extract work from the chamber. d) The demon lets the gas expand and moving the separation, thus extracting a finite amount of work from the finite-temperature reservoir.}
    \label{fig:Szilard1}
\end{figure}
The vessel can exchange heat with a thermal reservoir held at a fixed temperature~$T$.
The demon inserts a barrier into the vessel, creating a bipartition (L, R) of equal volumes.
The barrier can move along the vessel without friction.
The demon measures the position of the gas particle, either L or R (measurement), and gains an amount of classical information corresponding to one logical bit.
Furthermore, as sketched in Fig.~\ref{fig:Szilard1}, it modifies the apparatus on the side occupied by the particle using a rope tied to a mass, in order to extract work from the system by lifting the weight against gravitational force. 
Given that the gas particle is on the R side,  the demon lets it expand toward the left (feedback protocol) until the barrier reaches the left side of the vessel.
Eventually, the demon removes the barrier and the one-particle gas cycle can start all over again.
The net result of this measurement-feedback protocol is the full conversion of heat extracted from the reservoir into a finite amount of work, which reads          
\begin{equation}\label{eq:erasWork1}
		W= k_{\rm B} T \ln 2. 
\end{equation}

This ideal engine apparently violates the second law of thermodynamics, as all the heat extracted from the bath is converted into work by a device operating in a complete cycle.
This paradox demonstrates how the work extracted is related to the amount of information gained from the demon.
Szil\'ard's theoretical studies link the right hand side of Eq.~\eqref{eq:erasWork1} to a minimum amount of entropy increase~\cite{Maruyama:RevModPhys_2009} due to the measurement.
This insight prompted nearly a century of theoretical and experimental efforts aimed at understanding the interplay between thermodynamics and information~\cite{Leff:2002}.
    
Notable contributions to the solution of the paradox, i.e., the exorcism of MD~\cite{Leff:2002}, were provided by Landauer~\cite{Landauer:IBMJResDev_1961} and Bennett~\cite{Bennett:IJTP_1982}. 
Landauer first determined the minimum energetic cost needed to erase the information gained through measurement~\cite{Deffner:qthermobook_19}.
Indeed, any amount of information resulting from the outcome of a measurement has to be stored in a memory.
The minimum amount of classical information is encoded in one classical bit.
In the context of the Szil\'ard's engine, in order to close the thermodynamic cycle, the demon needs to erase the information stored in the memory.
Landauer recognized that the erasure is a logically irreversible process \cite{Landauer:IBMJResDev_1961}, i.e., it maps many different logical states to a single one.
In the case of a duality relation between thermodynamic and information entropy~\cite{Maruyama:RevModPhys_2009,Lutz:PhysTod_2015,Ciliberto:Inf_2021}, he concluded that the erasure of a logical state entails dissipation of energy into the environment.
This result had profound consequences for the physics of the Szil\'ard engine: the minimum amount of work that has to be dissipated into the reservoir to erase the demon's memory exactly equals the work gained after the feedback protocol, i.e., $W_{\rm{eras}}=k_{\rm B} T \ln 2$.
As a consequence, the entropy growth of the environment due to the erasure process is greater or equal to the entropy reduction linked to work extraction.

\begin{figure}%
    \centering
    \includegraphics[width=8.5cm]{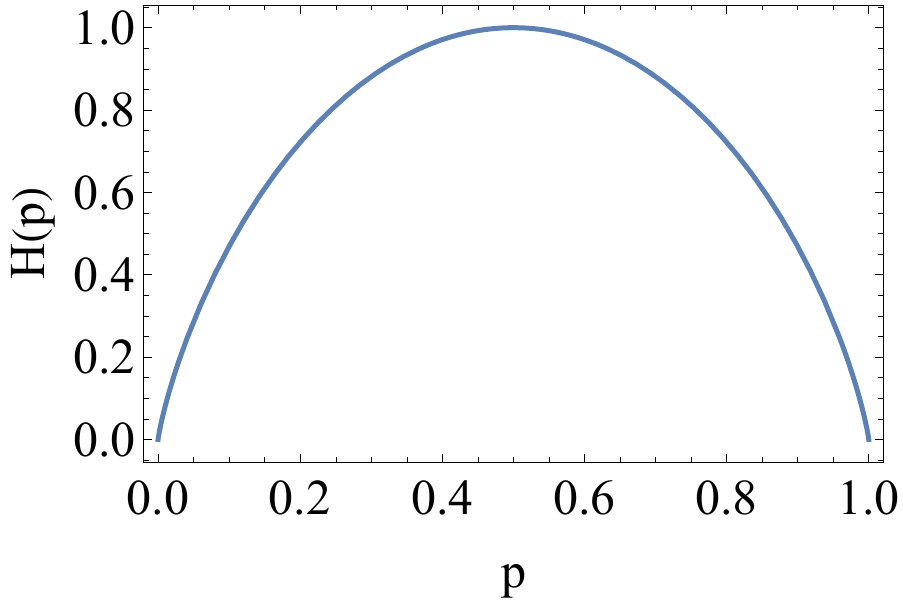} %
    \caption{Binary entropy function $H(p)$. Maximal uncertainty is obtained for $p=1/2$.}%
    \label{fig:Hp}%
\end{figure}

For a generic amount of classical information stored  by the demon, the erasure work is proven to be proportional to the amount of classical information, \cite{Maruyama:RevModPhys_2009} 
	\begin{equation}\label{eq:erasWork2}
	W_{\rm{eras}}= k_{\rm B} T \ln 2 H(p), 
	\end{equation}
where $p$ is the probability of the gas molecule being in a given partition of the vessel, and $H(p)=-p \log_2 p - (1-p)\log_2 (1-p)$ is the binary Shannon entropy,  plotted in Fig.~(\ref{fig:Hp}).
Bennett~\cite{Bennett:IJTP_1982} also pointed out that, given the measurement apparatus is in a blank state (no previous information recorded), the measurement correlates the memory with the system and, at least in principle, it can be carried out reversibly. 

The appearance of the Shannon entropy on the right-hand side of Eq.~\eqref{eq:erasWork2}  hints at a fundamental link between thermodynamics and information.
Over the last two decades, intense theoretical~\cite{Sagawa:PRL_2008,Sagawa:PRL_Minimalcost_2009,Sagawa:PRL_JarzIneq_2009,Esposito_2011,Mandal:PNAS_2012,Sagawa:PRE_2012,Barato:EPL_2013,Strasberg:PRL_2013,Horowitz:PRX_2014,parrondo:NatPhys_2015,faist2015minimal} and experimental~\cite{Berut:NatPhys_2012, Koski:PRL_2014,Koski:PNAS_2014,Roldan:NatPhys_2014,Koski:PRL_2015,Vidrighin:PRL_2016,Ciliberto:PRX_2017,admon2018experimental} efforts have been made to describe information processing tasks -- such as measurement, feedback, and erasure protocols -- as nonequilibrium thermodynamic processes.
The theoretical study~\cite{parrondo:NatPhys_2015} provided generalized definitions of nonequilibrium entropy and free energy, and reformulation of the second law to account for information exchange among different systems, e.g., the demon's memory and the working medium.
Given a classical system described by an Hamiltonian $H_{0}(x)$ and a nonequilibrium state $\rho$, the proposed definition for the nonequilibrium free-energy  reads     
\begin{equation}\label{eq:NeqFree}
		\mathcal{F}=\langle\mathcal{H}_{0}\rangle_{\rho} -TS(\rho),
\end{equation}
where $S(\rho)$ is the thermodynamic entropy related to the nonequilibrium probability distribution $\rho$.
Importantly, $S(\rho)$ is assumed to be proportional to the Shannon entropy of a generic stochastic variable $X$~\cite{Maruyama:RevModPhys_2009,Ciliberto:Inf_2021} described by the probability distribution $\rho$, reading 
\begin{equation}\label{eq:NeqEnt}
    S(\rho)=k_{\rm B}H(X)=-k_{\rm B}\sum_x\rho(x)\ln \rho(x),  
\end{equation}
where $k_{\rm B}$ is the Boltzmann constant and $x$ stands for each of the individual realizations of $X$.
This definition of nonequilibrium entropy permits describing, on the same ground, heat and work exchange with information gain and losses. It has been adopted by many authors in the quantum setting~\cite{vinjanampathy2016_QT_review} (see also Secs.~\ref{subsec:thermotheory},\ref{subsec:MDEng}), where $S(\rho)$ is replaced with the Von-Neumann entropy.

When Eqs.~\eqref{eq:NeqFree} and \eqref{eq:NeqEnt} are adopted, the measurement and feedback control in a Szil\'ard engine can be fully characterized in terms of energy and information balance.
For a classical measurement operation that does not change the energy of the working medium, the change in the nonequilibrium entropy due to the measurement can be written as \cite{parrondo:NatPhys_2015}
    \begin{equation}\label{eq:MeasEntro}
	    \Delta S(\rho)=-k_{\rm B}I(X,M),  
\end{equation}
where $I(X,M)=H(X)-H(X|M)$ is the mutual information between the microstate $X$ and the random outcome of the measurement $M$.~Eq.\eqref{eq:MeasEntro} links the information gain due to measurement to the decrease in the nonequilibrium entropy.
In a generic feedback control process, e.g., one operated by the demon in the Szil\'ard engine, the Hamiltonian is changed by an external parameter over a finite time. Then, at a given point in time, a measurement is performed, and a suitable feedback protocol, $\lambda_{m}(t)$, is operated.
Following Eq.~\eqref{eq:MeasEntro}, a generalized second-law can be written down to describe the whole process~\cite{Sagawa:PRL_2008}
\begin{equation}\label{eq:IIlaw}
	\ave{W} - \ave{\Delta \mathcal{F}}\geq -k_{\rm{B}}T I(X(t_{m}),M).  
\end{equation}
The average work $\ave{W}$ and the average change in the nonequilibrium free energy $\ave{\Delta \mathcal{F}}$ are thus related to the information gained.
From Eq.~\eqref{eq:IIlaw}, it follows that, for a feedback process returning the measured system back to the initial state, there is a minimum amount of work that can be extracted, $\ave{W}\geq -k_{\rm{B}}T I(X(t_{m}),M)$, which, in the case of the Szil\'ard engine, reduces to Eq.~\eqref{eq:erasWork1}. 

In the very first experimental realizations of the Szil\'ard engine, the single molecule gas in the chamber is replaced with a Brownian particle (a polystirene bead in a solution) subject to a spiral-staircase potential~\cite{Toyabe:NatPhys_2010}.~The measurement-dependent feedback protocol consists of switching suitable tilted periodic potentials, using the information to prevent the bead from getting down the staircase.~Moreover, a fully electronic version of the engine has been realized, where an electron evolving in a Single Electron Box (SEB) device plays the role of the working medium~\cite{Koski:PNAS_2014}.
    
More recent proposals involve continuous versions of Szil\'ard engines, that is, refined feedback protocols where the MD performs multiple measurements on the gas to maximize the extracted work~\cite{RibezziCrivellari:NatPhys_2019}.
Moreover, a so-called gambling MD has been proposed, which invests a given amount of work and stops the nonequilibrium dynamics of the medium at stochastic times, in order to earn free energy ~\cite{Manzano:PRL_2021}.

\subsection{Quantum Maxwell's demon heat engines}\label{subsec:MDEng}

The Szil\'ard engine is a basic example of feedback control aimed at converting information into work.~As many advances took place in the classical regime to understand the link between information and thermodynamics, more recent works have focused on the investigation of similar issues in the quantum regime.

The quantum nature of the working medium and the bath can influence the operation of the engine during the measurement-feedback protocol and the information erasure stages.
The former stage clearly involves phenomena unique to the quantum realm.
In this case, the demon can change the quantum state of the working medium by either performing projective measurements or by continuously monitoring it. 
As summarized in Sec.~\ref{subsec:ClassicSzilard}, in the classical setting, nonequilibrium fluctuations occurring in the course of an information-to-work conversion process have been theoretically described via generalized fluctuation relations~\cite{parrondo:NatPhys_2015}.
The quantum versions of the Sagawa-Ueda fluctuation relations~\cite{Morikuni:JStatPhys_2011,campisi2011colloquium,RevModPhys.81.1665,Funo:PRA_2013,Funo:PRE_2013,Albash:PRE_2013,vinjanampathy2016_QT_review} are formally analogous to the classical ones and quantitatively describe the amount of information gained through the feedback control, by means of the quantum mutual information~\cite{Sagawa:2013}.
Following these results, over the last decade, a huge amount of theoretical and experimental effort has been put into describing the measurement-feedback protocol created by a demon acting in the quantum regime, in terms of nonequilibrium quantum fluctuation relations~\cite{solfanelli2021experimental}.
Related experimental works have now reached maturity, and they generally provide robust evidence for quantum fluctuation theorems (see Sec.~\ref{subsubsec:MDexperiment}).

On the other hand, the search for truly quantum effects in the operation of the demon has been pursued via different routes.  In what follows, we will mainly focus on nonautonomous quantum Maxwell Demons, which require external drivings and feedback loops relying on quantum measurements. An excellent overview of the physics of autonomous quantum demons~\cite{sanchez2019nonequilibrium,Ludemon2021}, whose full thermodynamic behavior can be derived by solving the system's Schr\"odinger equation, can be found in~\cite{whitney2023illusory}.
 The issue of quantum effects in demons is rooted in the problem of work extraction from quantum states in the presence of quantum resources~\cite{Perarnau-Llobet2015,korzekwa16,Kammerlander:Scirep_2016,vinjanampathy2016_QT_review} and is therefore linked to the fundamental inquiry of quantum advantage in quantum heat engines~\cite{uzdin15,ghosh2019quantum}.
Thus far, the study of quantum and nonequilibrium effects taking place during the erasure process, possibly occurring in a finite time, has led to several extensions of the Landauer bound~\cite{Peterson:ProcRoySoc_2016,Gaudenzi:NatPhys_2018,Konopik:EPL_2020,Ciampini:arXiv_2021,VanVu:PRL_2021}.
Moreover, quantum coherence and correlations, e.g., entanglement either between the working medium and memory or the working medium and the bath, have been considered as sources of novel quantum effects in the erasure process, also providing advantages to the demon's work extraction (see Sec.~\ref{subsubsec:Trulyquantum}).
In this context, contrasting results have been derived, meaning a clear and conclusive scenario is still in the making.
We give a brief overview of these different research directions in what follows.
	
\subsubsection{Quantum effects in Maxwell's demon engines}\label{subsubsec:Trulyquantum}
     
Early works in the field provided a theoretical analysis of engines run by Maxwell's demons in the quantum regime~\cite{lloyd97}.
One of the main issues that have been investigated is linked to the invasive nature of measurement in the quantum domain.
The influence of measurement and decoherence on a system of two spins coupled to radiation modes at different temperatures and subject to pulses was first analyzed in~\cite{lloyd97}, as an NMR quantum Maxwell's demon.
With one of the spins employed in gaining information on the quantum state of the other, the thermodynamic inefficiency due to the energetic cost of the information gained has been computed and compared with the Carnot efficiency.
Different perspectives on quantum measurement examine its energetic imprint~\cite{Elouard:NPJ_2017}, considering it as an additional quantum resource for QHEs.

Moreover, in~\cite{Kim:PRL_2011}, a quantum version of the theoretical Szil\'ard engine has been proposed, where the working medium is composed of many Bosonic/Fermionic particles.
Here claims are made for the importance of inserting and removing the separation wall between the two sides in energetic accounting, a process that should be considered part of the thermodynamic operation of the engine.
Furthermore, the statistics of the working medium particles determine the amount of work that can be extracted from the engine. Bosonic particles exhibit higher work generation than Fermionic ones, especially in the low-temperature limit. This behavior is also observed in standard QHEs as discussed in Sec.~\ref{subsubsec:Ottocycle}.

In~\cite{DelRio:NatPhys_2011}, the peculiar option of erasing a system S that shares correlations with a quantum memory has been explored. Erasing or resetting a system means bringing it to a known and definite pure state. For a qubit, that state would be the ground state $\ket{0}$.
In a simple setup, the  system S to be erased is maximally entangled with each of the $n$ memory qubits
of the observer Q. 
By accessing the memory qubits, the observer Q erases the system, enacting a process whose work cost doesn't exceed  
    
\begin{equation}\label{eq:QuantumBound}
	       W_{\text{eras}}(\text{S}|\text{Q})= H(\text{S}|\text{Q})kT\ln 2,
\end{equation}
where $H(\text{S}|\text{Q})=H(\rho_\text{SQ})-H(\rho_{\text{Q}})$ is the conditional Von Neumann entropy~\cite{nielsen_chuang_book} and $\rho_{\text{SQ}}$ is the joint quantum state of S and Q.
Contrary to its classical counterpart in Eq.~\eqref{eq:erasWork2}, Eq.~\eqref{eq:QuantumBound} describes genuine quantum features arising from quantum correlations between the system and the memory, i.e., ``side information''~\cite{vinjanampathy2016_QT_review}.
For instance, if the joint state $\rho_{\text{SQ}}$ is maximally entangled, the marginal state $\rho_{\text{Q}}$ is mixed, such that the work in Eq.~\eqref{eq:QuantumBound} is negative.
As a consequence, in the presence of quantum correlations, the erasure can be linked to a net work extraction.
A crucial feature of this erasure process is that the local state $\rho_{S}$ is erased by being set to the reference zero state $\ket{0}$, while the state of the memory Q remains mixed, i.e., the erasure process is locally indistinguishable from the classical one~\cite{vinjanampathy2016_QT_review}.

In the classical setting (see \ref{subsec:ClassicSzilard}), the demon performs measurements directly on the working medium.
However, in the quantum setting, if the working medium S and its environment $\mathcal{E}$ are described through a global bipartite state $\rho_{S\mathcal{E}}$, a demon can be devised to measure the environment without disturbing the state of the working medium. 
In this context, work extraction can be performed by another party with the aid of local operations.
    
The problem of extracting work from a shared bipartite quantum state by employing only local operations, i.e., assisted work distillation~\cite{morris2019assisted}, has been studied in the framework of resource theories~\cite{brandao2013resource}. 
In this case, two parties, Alice and Bob, share many copies of a bipartite quantum state $\rho_{\text{AB}}$. The two parties are allowed to do local operations on their subsystems: Bob can perform only thermal operations, while they can communicate through a classical channel.
In~\cite{Beyer:Steering_2019}, the above theoretical setting has been exploited to devise a fully quantum generalization of the Szil\'ard engine.
Moreover, a theoretical criterion has been proposed to certify the quantum engine's behavior and distinguish it from its classical counterpart.
The latter is based on the derivation of a bound on work that can be locally extracted by Bob, using the notion of quantum steering \cite{wisemansteering2007,cavalcanti2016quantum,Uola:RevModPhys_2020}.  

In this fully-quantum Szil\'ard engine, Alice ($A$) plays the role of the demon.~She can prepare several copies of the system-environment global state $\rho_{S\mathcal{E}}$, and she can perform measurements only on $\mathcal{E}$, while Bob ($B$) can extract work from the local Gibbs state of S, $\tr_{\mathcal{E}}[\rho_{ S \mathcal{E}}]=\rho_{S\text{G}}=(1/Z)\sum_{i}e^{-\beta E_{k}}\ket{k}\bra{k}$.
$A$ performs measurements on $\mathcal{E}$ and communicates the outcome  to $B$ via a classical channel.
Based on this information, $B$ can extract the maximum amount of work, performing only local operations. 
Infinitely many decompositions, $D_{n}=\{p_{i},\rho_{i}\}$, can be found for the local Gibbs state,  which is a linear combination of ensembles, $\rho_{S\text{G}}=\sum_{i}p_{i}\rho_{i}$.
Each of these ensembles can be linked to a different global state by $\rho_i=\tr_\mathcal{E}[(\mathrm{1}\otimes M_i) \rho_{S \mathcal{E}}]$, provided a suitable Positive Operator-Valued Measure (\acrshort{povm}), $M_i$. 

%
\begin{figure}[h]
        \centering
        \includegraphics[scale=0.40]{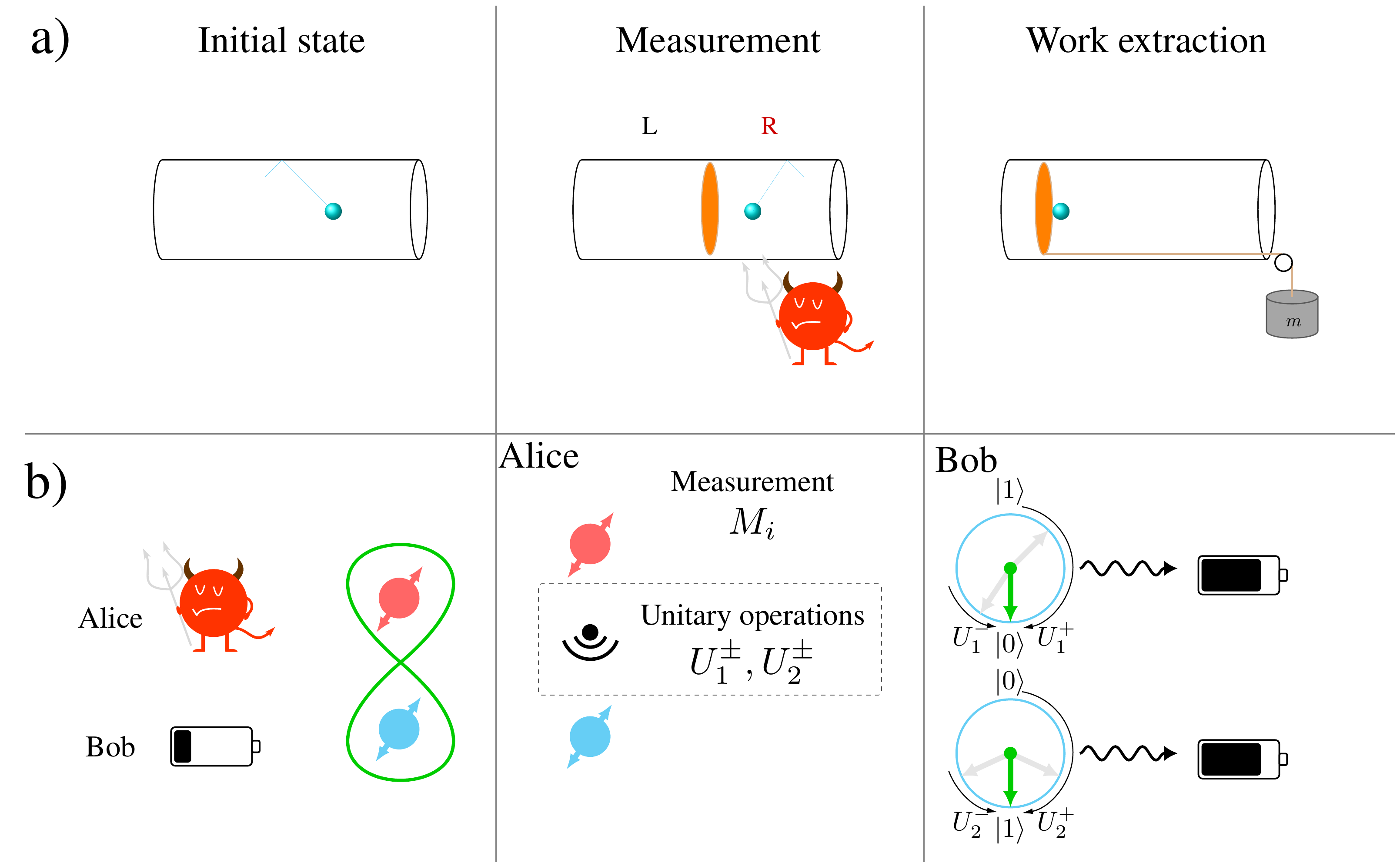}
        \caption{Classical vs. quantum Szil\'ard engine. a) Three stages of the classical engine. b) The demon (Alice) performs measurements on the environment and sends Bob the information on the unitary for use. Once he receives this information, Bob can extract work by applying the right unitary on the working medium qubit and then charging a work storage device. Reproduced from~\cite{Ji:PRL_2022}.}
    \label{fig:QuantumSzilard}
\end{figure}

A set of unitaries $U^{n}=\{U^{n}_{i}\}$ can be used to extract work from each decomposition, $D_{n}$.
The work extraction is performed through the transfer of a nonnegative amount of energy from the system S to a work storage device $\mathcal{W}$, such that the total Hamiltonian of the composite system is $H=H_{S}\otimes\Iop +\Iop\otimes H_{\mathcal{W}}$~\cite{Aberg2014,Perarnau-Llobet2015} and the average work is expressed as $\bar{W}=\sum_{i}p_{i}W_{i}$, with $W_{i}=\tr\{\Iop\otimes H_{\mathcal{W}}(U_{i}(\rho_{i}\otimes\rho_{\mathcal{W}})U^{\dagger}_{i}-\rho_{i}\otimes\rho_{\mathcal{W}})\}$.
To maximize the average work, for a given decomposition, $B$ can pick the unitaries at random and then ask $A$ which particular unitary to use.
On the other hand, $A$ can communicate the unitary to $B$ after reading the outcome of the measurement on $\mathcal{E}$.     

Evidence of quantumness in the engine operation is encoded in the correlations of the state $\rho_{S\mathcal{E}}$ and is based on quantum steering-type arguments.
If the correlations can be described by a Local Hidden State (\acrshort{lhs}) model~\cite{Uola:RevModPhys_2020} with $F=\{p_{\xi},\rho_{\xi}\}$ and $\rho_{\xi}$ the hidden states randomly distributed according to $p_{\xi}$, then there is no assurance of quantumness.
On the other hand, if such a model cannot be found, then $B$ can be sure about the nonlocal character of the correlations.
From a theoretical perspective, a quantum state is steerable if its conditional, or post-measurement states, linked to one party cannot be successfully described by means of an LHS~\cite{Uola:RevModPhys_2020}.
Remarkably, for the Szil\'ard engine described above, if an LHS model is chosen as a decomposition, then the average work locally extractable through the process by $B$ is bounded from above,  $\overline{W}\leq\overline{W}_{\text{cl}}$.
As a consequence, any decomposition leading to a work extraction exceeding the bound $\overline{W}_{\text{cl}}$ is a clear signature of quantum behavior and also provides evidence for quantum advantage over the classical setting. 
This strategy is demonstrated for a two-level system, described by the Hamiltonian $H_{\text{S}}=\ket{1}\bra{1}$, and a local Gibbs state that is parametrized as $\rho_{\text{Gibbs}}=(1+\eta)/2 \ket{1}\bra{1} +(1-\eta)/2 \ket{0}\bra{0}$, with $\eta=(e^{-\beta}-1)/(e^{-\beta}+1)$.
In this case, the average work for any LHS is bounded from above by 
\begin{equation} \label{eq:Workbound}
	   \overline{W}_{\text{cl}} \leq \frac{\eta ( \sqrt{1-\eta^2} + \eta +1) + \sqrt{2 -2\eta^2}}{2(\sqrt{1-\eta^2} +1)}.   
\end{equation}    

If $A$ prepares many copies of the state $\rho_{S\mathcal{E}}$ out of a pure entangled state of the form $\ket{\psi}_{S\mathcal{E}}=\sqrt{(1+\eta)/2}\ket{1}_{S}\otimes \ket{1}_{\mathcal{E}}+\sqrt{(1-\eta)/2}\ket{0}_{S}\otimes \ket{0}_{\mathcal{E}}$ and performs measurements of $\sigma_{x}$ and $\sigma_{y}$, then $B$ can extract on average the optimal amount of work $\overline{W}_{\text{opt}}=(1+\eta)/2$, which exceeds the classical bound in Eq.~\eqref{eq:Workbound} for a wide range of inverse temperatures $\beta$. 
    
Recently, this theoretically proposed Szil\'ard engine was experimentally realized \cite{Ji:PRL_2022}, employing a platform based on NV centers in diamond.
Similar experimental devices have been used to realize MD engines in which the demon performs measurements on the working medium (see Sec.~\ref{subsubsec:MDexperiment}).
In this quantum Szil\'ard engine experiment, the working medium and the bath are implemented with a pair of qubits, making use of the nuclear Nitrogen spin and the vacancy's electronic spin respectively.
The platforms allow for a high degree of spin manipulation and preparation of a wide variety of correlated initial states.
 	
\subsubsection{Experiments on quantum  Maxwell's demon engines}\label{subsubsec:MDexperiment}
	
Over the last decade, experiments with quantum MD engines have been implemented through a wide range of platforms.
Though their perspectives vary, these studies have aimed to provide a solid experimental validation of generalized fluctuation theorems involving energy and information exchange with quantum systems.
They achieve MD feedback control by manipulating quantum systems that exhibit coherence.
However, the various feedback control protocol setups differ in the way the demon gains information, i.e., both the working medium and the detailed experimental technique used to measure the corresponding amount of information.
    
In~\cite{Camati:PRL_2016}, a feedback protocol employing a $\rm{CHCl}_3$ liquid with  $^{13}\rm{C}$ atoms driven using NMR techniques has been performed.
Here the working medium is the spin $1/2$ of the $^{13}\rm{C}$ atom, while the role of the demon memory is played by the nuclear spin of the $^{1}\rm{H}$ atom.
The working medium is initially prepared in a Gibbs state $\rho_G$ at temperature $T$.
Then, employing a resonant RF pulse, a quench is performed to change its Hamiltonian quickly, so that the state at a subsequent time $\tau_1$ is out-of-equilibrium and develops quantum coherences in the energy basis.
The measurement-feedback protocol operated by the demon relies on nonselective projective measurements on the working medium.
For each measurement outcome $m$, ruled by a probability $p(m)$, a feedback protocol is implemented by controlling the evolution of the working medium until a final time $\tau_2$.
The control is based on the engineering of unital operations, each labeled by $k$, where the Hamiltonian of the working medium $H^{k}(\tau)$ can be changed in time.
The feedback control is also affected by errors, i.e., for each measurement outcome $m$, a conditional probability $p(k|m)$ of implementing the $k$-th operation is introduced.

The feedback effect on the entropy production of the working medium can be quantitatively described by the generalized Tasaki identity integral relation, which reads
\begin{equation}\label{eq:integinfo}
	\langle e^{-\beta(W-\Delta F^{k})-I^{(k,m)}}\rangle=1,  
\end{equation}
where $W$ is the stochastic work, $\Delta F^{k}=-(1/\beta)\ln Z_{\tau_2}/Z_{0}$, $Z_{t}=\tr  [e^{-\beta H^{k}(\tau)}]$ is the variation of the free energy for each controlled evolution, and $I^{(k,m)}=\ln \frac{p(k|m)}{p(m)}$ are the elements of the mutual information between the working medium and the memory.
This result is analogous to the classical Sagawa-Ueda integral fluctuation relation with feedback. The proposed measurement-feedback protocol allows for experimental validation of Eq.\eqref{eq:integinfo}.

Moreover, given that the Jensen inequality is applied to \eqref{eq:integinfo}, it follows that, in the presence of feedback, the average entropy production obeys $\langle \Sigma \rangle=\beta\langle W -\Delta F^{k}\rangle \ge -\langle I^{(k,m)} \rangle$
and in principle can become negative. 
This means that the MD can carry out rectification of the average entropy production.
Using quantum state tomography, it is possible to reconstruct all the contributions to the average entropy production of the protocol. As a consequence, the demon rectification can be characterized as a function of the temperature of the state, and an optimal feedback protocol can be found.

Alternative approaches to the experimental investigation of quantum MD engines employ a superconducting circuit platform~\cite{Cottet:PNAS_2017,Masuyama:Natcomm_2018} and NV center in diamond~\cite{Wang:ChinPhysLett_2018}.
In~\cite{Cottet:PNAS_2017}, a quantum MD engine that can work cyclically is constructed.
The working medium is a transmon qubit embedded in a microwave cavity, the latter working as the demon memory.
The transmon and the cavity are coupled by means of the Jaynes-Cummings interaction in the dispersive limit, which reads $H=\omega_q \ket{e}\bra{e} +\omega_{D}b^{\dagger}b -\chi b^{\dagger}b\ket{e}\bra{e}$, where $\hbar=1$,  $\ket{g}$ and $\ket{e}$ are the ground and excited state of the qubit respectively, and $b^{\dagger}$ and $ b$ are the cavity creation and annihilation operators.
A shift in the cavity frequency $-\chi$, of the order of MHz, takes place whenever the qubit is in the excited state.
Reciprocally, if the cavity hosts a finite number $n$ of photons, the qubit energy gap is changed by $-n\chi$.
As a consequence, by suitably driving the cavity with microwave pulses, the qubit and the demon's memory become correlated, and their state can be employed to demonstrate the effect of information on the extracted work.

At the beginning of the cycle, the working medium is prepared in a thermal state at temperature $T_{h}$.
Next, a pulse of frequency $\omega_{D}$ is sent to the cavity and excites the demon's memory only if the qubit is in the ground state $\ket{g}$, i.e, the demon records the state of the qubit.
In the next step, the work extraction is implemented by sending a pulse of frequency $\omega_q$ through a dedicated port.
If the established correlation between the qubit and the demon memory is assumed to be ideal, and if photons are present in the cavity, the qubit cannot make any transition as its energy gap is changed. On the other hand, if the cavity is empty, i.e., the qubit finds itself in the excited state, then stimulated emission takes place, and work is extracted by the working medium.
It follows that the system entropy decreases while the demon memory increases by at least the same amount.
In the final step, the demon memory is reset by thermalizing with a different bath at temperature $T_{c}$, with $T_{c}<T_{h}$.
The extracted power can be deduced from measurements of the photons coming in and out of the port.
The results show that work is extracted over time as a consequence of the information gained by the demon, and it occurs if the cavity is populated enough for the two-qubit states to be distinguished.
However, while signatures of quantumness are present in the reconstructed state $\rho_{D}$ of the demon's memory, no quantum signature linked to quantum coherence in the qubit state is present in the extracted power.

As pointed out in~\cite{Naghiloo:PRL_2018}, many physical features in previous examples of quantum MD engines could be explained without resorting to quantum information.
This outcome is due to the fact that quantum coherence features may be absent, or they could be lost as a result of the projective measurement scheme adopted in the feedback protocol.
In~\cite{Naghiloo:PRL_2018}, a quantum MD experiment is performed in which, differing from previous approaches, the demon implements continuous weak measurements on the working medium, followed by a feedback protocol.
The advantages of this method are twofold: first, experimental verification of fluctuation relations in the presence of information can be achieved at the level of a single quantum trajectory, as it can be done with heat and work~\cite{Naghiloo:PRL_2020}.
Second, the combined effect of quantum coherence and quantum measurement backaction, on the information gained by the demon, can be probed experimentally.
As with~\cite{Cottet:PNAS_2017,Masuyama:Natcomm_2018}, the experimental setup comprises a superconducting transmon qubit, described as a two-level system, dispersively coupled with a cavity.

The noticeable feature in this work is that the qubit is periodically driven by a resonant drive that induces Rabi oscillations. When the phase shift of the cavity is continuously measured, the resulting stochastic dynamics allow for gaining information on the qubit state without resorting to projective measurement.
For each quantum trajectory, 
it is possible to track the amount of information gained by the demon.
Based on this information, a feedback rotation brings the system to the ground state. 
As a result, the average information can be taken over many quantum trajectories, as a function of the initial state of the qubit.
The effect of coherence and quantum backaction is to induce a transition from a regime of information gain, in which the average information is positive, to another marked by information loss, in which the average information turns negative.  

More recent studies provide experimental realizations of autonomous MD engines employing quantum systems.
These engines have been implemented in a cavity QED platform, making use of Rydberg atoms and a microwave resonator~\cite{NajeraSantos:PRR_2020}.
In this case, a system is composed of a qubit (Q), a cavity (C), and a demon (D).
The qubit and the cavity are initially prepared in thermal states at different temperatures $\beta_{\rm{Q}}>\beta_{\rm{C}}$.
The transfer of heat between Q and C is controlled by D, which is engineered utilizing a second qubit.
In this way, the occurrence of heat transfer from the cold qubit to the hot oscillator, through the information gained by  D, is experimentally investigated.
While the idea is closer to the classical formulation of the demon~\cite{Leff:2002}, in contrast to previous works, here the qubit-cavity-demon system is closed and its unitary evolution is addressed.
Furthermore, as the total entropy production is constant, the entropy exchange between Q, C, and D is used to define a generalized second law for the system.

The operations of the demon consist of the readout step, in which D acquires information on the state of Q, and the feedback step: if Q is in the ground state, D stops the interaction between Q and C.
The information gained by the demon is quantified via the change in the mutual information between the system QC and D, $I_{\rm{QC}|\rm{D}}=S_{\rm{Q}} + S_{\rm{C}}-S_{\rm{QCD}}$, during the feedback step. 
This feedback introduces a new contribution to the heat exchange between Q and C,

\begin{equation}\label{eq:infoentro}
	Q_{\rm{C}}(\beta_{\rm{C}}-\beta_{\rm{Q}})=\Delta I_{\rm{QC}|\rm{D}} + D_{\rm{QC}},
\end{equation}
where $D_{\rm{QC}}=D[\rho_{\rm{QC}}||(e^{-\beta_{\rm{Q}}H_{\rm{Q}}}/Z_{\rm{Q}})\otimes (e^{-\beta_{\rm{C}}H_{\rm{C}}}/Z_{\rm{C}})]\ge 0$ is the quantum relative entropy between the joint QC state and the product of initial thermal states, and $\Delta I_{\rm{QC}|\rm{D}}=0$ is the change in the mutual information.
In the absence of feedback, $\Delta I_{\rm{QC}|\rm{D}}=0$, and the conventional form of Clausius inequality is restored.
On the other hand, if the demon gains information on the QC and performs the feedback, it can reduce the entropy of the QC system, and it follows that the r.h.s. of Eq.~\eqref{eq:infoentro} can become negative.
Thus, correlations between the QC system and the demon can be exploited to perform unconventional thermodynamic tasks.

A further implementation of an autonomous MD can be found in~\cite{HGomez:PRXQ_2022}.
Unlike in the previous experimental realizations of MD, here a \emph{dissipative} feedback process is engineered by employing NV centers in diamond and optical pumping techniques.  

\subsection{Measurement-fuelled quantum heat engines and refrigerators }\label{subsec:MeasFuel}

Quantum measurements are crucial to the operation of engines based on feedback protocols, as described in Sec.~\ref{subsubsec:MDexperiment}.
However, contrary to the classical setting, quantum measurements change the state of the physical system~\cite{wisemanbook}, e.g., the engine's working medium, such that the system exchanges a finite amount of energy with the measurement apparatus~\cite{GelbKlim:PRA_2013}.
Repeated projective measurements~\cite{Kurizki:NatPhys_2008,Elouard:NPJ_2017}, as well as generalized quantum measurements, can be interpreted as a source of stochasticity and irreversibility for a quantum system.
Moreover, the stochastic change in the state of the system resulting from quantum measurement induces energy fluctuations with no classical analog, dubbed quantum heat~\cite{Elouard:NPJ_2017,Auff:SciPost_2021}, although there is no consensus on this interpretation in the scientific community. 
Nonetheless, quantum measurement can be considered a quantum resource~\cite{Elouard:PRL_2018}. It can be employed either to gain information to be converted to work via a feedback protocol, as with an MD engine, or as a source of energy in the absence of a hot reservoir, with or without the need for a feedback protocol.  
The first study to suggest the latter was~\cite{levy12}, in which monitoring the interaction of two harmonic oscillators pumps a QAR. Many theoretical and experimental studies~\cite{Elouard:PRL_2017,Yi:PRE_2017,Ding:PRE_2018,Buffoni:PRL_2019,opatrny2021work, misra2022work,Auff:SciPost_2021} followed.

\begin{figure}[h]
        \centering
        \includegraphics[scale=1.35]{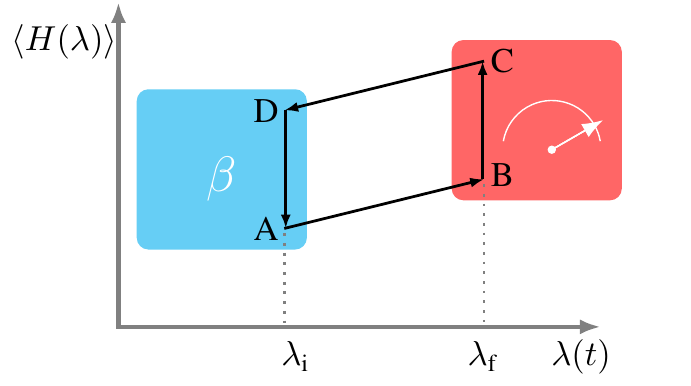}
        \caption{Schematic diagram of the measurement-fuelled engine. The Otto cycle comprises four conventional strokes as detailed in Sec.~\ref{subsubsec:Ottocycle}.~The compression/expansion strokes are governed by the control parameter $\lambda(t)$ that increases/decreases the distance from energy levels. In contrast with the conventional setting, the role of the energizing bath is played by a nonselective measurement channel, modeled using generalized measurement operators.}
    \label{fig:DiagME5}
\end{figure}

In~\cite{Yi:PRE_2017,Ding:PRE_2018}, a single-temperature heat engine fuelled by quantum measurement was theoretically studied.
Here the energy needed to fuel the engine is supplied to the working medium employing nonselective, minimally disturbing quantum measurements.
The engine cycle can be sketched as in Fig.~\ref{fig:DiagME5}.
Analogous to a conventional Otto engine described in Sec.~\ref{subsubsec:Ottocycle}, it is composed of four strokes.
At the beginning of the cycle, the working medium is prepared in a thermal state, with inverse temperature $\beta$.
During the first stroke, the working medium is decoupled from the bath and undergoes a compression, such that its Hamiltonian $H(\lambda)$ changes according to the control parameter $\lambda$ from the initial value $\lambda_{i}$ to the final one $\lambda_{f}$, increasing the gap among the energy levels.
In stark contrast with the conventional Otto engine, during the second stroke, a nonselective measurement of a system observable, which does not commute with the Hamiltonian $H(\lambda)$, is performed, keeping the control parameter fixed at $\lambda_f$.
Due to its minimally-disturbing character, the measurement is modeled with a unital map.
Since the result of the measurement is ignored, the post-measurement state can be written as $\rho_{\rm{pm}}=\mathcal{M}\rho(\lambda_{f})\mathcal{M}$, where $\mathcal{M^{\dagger}}=\mathcal{M}$ is a generalized measurement operator.

After the first measurement, the working medium goes through an expansion stroke, during which the control parameter gets back to $\lambda_{i}$, and the initial energy level spacing is restored.
In the final stroke, the system is connected to a bath at inverse temperature $\beta$, and, depending on the thermalization time, it is brought back to the initial state.
While the compression and expansion strokes can take place either adiabatically or within finite time, it is assumed that energy levels never cross in the course of the strokes.
Provided that the populations of the energy levels after the compression stroke are decreasing for increasing energy levels, the properties of the unital map guarantee that the average energy linked to the measurement stroke is positive, i.e., energy is delivered to the working medium.
Its efficiency reads $\eta=-\langle W\rangle/\langle E_{M}\rangle$, where $\langle W\rangle$ is the average work and $\langle E_{M}\rangle$ is the energy provided by the measurement.

In~\cite{Ding:PRE_2018}, the case of a driven harmonic oscillator subject to Gaussian position measurement was addressed and, importantly, compared with a conventional Otto engine.
The resulting scenario points toward the presence of increasing fluctuations in the energy supplied by the measurement, moving from the adiabatic to the full nonadiabatic regime. However, a measurement-driven engine could perform better than a conventional one concerning the work variance and the delivered power.

\begin{figure}[h]
        \centering
        \includegraphics[scale=1]{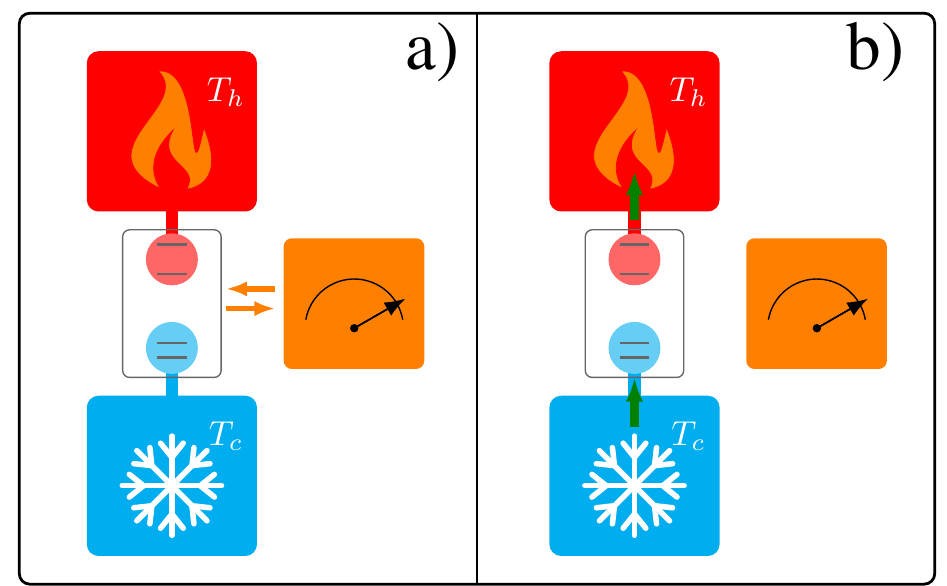}
        \caption{Sketch of the quantum measurement cooling device.
        A two-qubit system is interacting with two thermal baths of different temperatures $T_{h}$ and $T_{c}$. The measurement apparatus is used to perform projective measurements of the two-qubit state. The engine runs through two different strokes. a) The qubits are decoupled from the baths and the measurement is performed (on the right). b) The two qubits are disconnected from the measurement apparatus and are connected back to the thermal reservoirs (on the left). Heat currents from the cold to the hot bath can thus take place.   Reproduced from \cite{Buffoni:PRL_2019}.}
    \label{fig:DiagME6}
\end{figure}

Quantum measurement can also be devised to achieve cooling, as demonstrated in~\cite{levy12,Buffoni:PRL_2019}. In~\cite{Buffoni:PRL_2019}, a two-qubit system working with two thermal baths at different temperatures, $\beta_1 < \beta_2$, is considered.~As sketched in Fig.~\ref{fig:DiagME6}, each of the qubits can be connected with a bath, and they can interact with a measurement apparatus.~The 
two-stroke engine proceeds as follows: the two qubits are initially in contact with the baths so that their state is thermal, $\rho_{\rm{th}}=e^{-\beta_{1}H_{1}}/Z_{1}\otimes e^{-\beta_{2}H_{2}}/Z_{2}$.
In the first stroke, the two qubits are decoupled from the bath and connected with the measurement apparatus, such that their state changes to $\rho^{\prime}=\sum_k \pi_k\rho_{\rm{th}}\pi_k$, where $\pi_k$ are rank-1 projectors on a given measurement basis $\ket{\psi_{k}}$.
In the second stroke, the system is disconnected from the measurement apparatus, and each qubit is brought in contact with one of the baths. In this stage, each of the two qubits (labeled~$i$) thermalizes individually and exchanges an amount of energy $\langle \Delta E_{i}\rangle $ with the baths, which was gained in the measurement stroke. 
Depending on the signs of $\langle \Delta E_{i}\rangle$, the system can operate in a different mode, including as an engine, refrigerator, or heater in which the total energy $\langle \Delta E \rangle=\langle \Delta E_1 \rangle +\langle \Delta E_2 \rangle$ is dumped into the reservoirs.
All these modes of operation obey the second law in the form of $ \beta_1 \langle\Delta E_1\rangle +\beta_2 \langle\Delta E_2\rangle \geq 0. $

The choice of measurement basis plays a crucial role in the mode of operation and its efficiency.
The refrigeration efficiency is maximal if the post-measurement state $\rho^{\prime}$ is diagonal in the energy basis.
Furthermore, through randomly picked measurement basis states, it is shown that the probability of operating in a heater mode is the highest one, i.e., the least useful operation is the most likely one.
Yet the probability of operating in a refrigeration mode is surprisingly higher and more resilient against noise than that of an engine.

Additional examples of quantum measurement-driven engines have been devised to elucidate the mechanism of energy extraction from quantum measurement.
Unlike the previous examples of measurement-fuelled engines, these engines share some similarities with MD engines~\cite{Cottet:PNAS_2017}, as they combine the two aspects of measurement as a thermodynamic resource~\cite{Elouard:PRL_2018}, employing a measurement-feedback protocol to extract work from the whole observation process. On the other hand, they describe the mechanism by which energy is extracted from quantum correlations between the working medium and the measurement apparatus, taking into account the role of the measurement back-action.
In~\cite{levy16}, a two-qubit heat engine pumps energy to a flywheel that stores the energy in a harmonic mode. Monitoring (i.e., weakly and continuously measuring) the two quadratures of a flywheel provides both additional energy transfer from the measurement apparatus (the backaction) and information on the phase and amplitude of the flywheel. This information is then used in the feedback protocol, managing the fluctuation and improving the energy charging efficiency. 

A later proposal~\cite{Seah:PRL:2020} considered a measurement-driven engine composed of a qubit, two thermal reservoirs at inverse temperatures $\beta_h < \beta_c$, and a macroscopic pointer.
The qubit weakly interacts with a hot bath characterized by dissipation rate $k_h$ and is simultaneously coupled to the macroscopic pointer, modeled with a quantum harmonic oscillator.
\begin{figure}[h]
        \centering
        \includegraphics[scale=1.35]{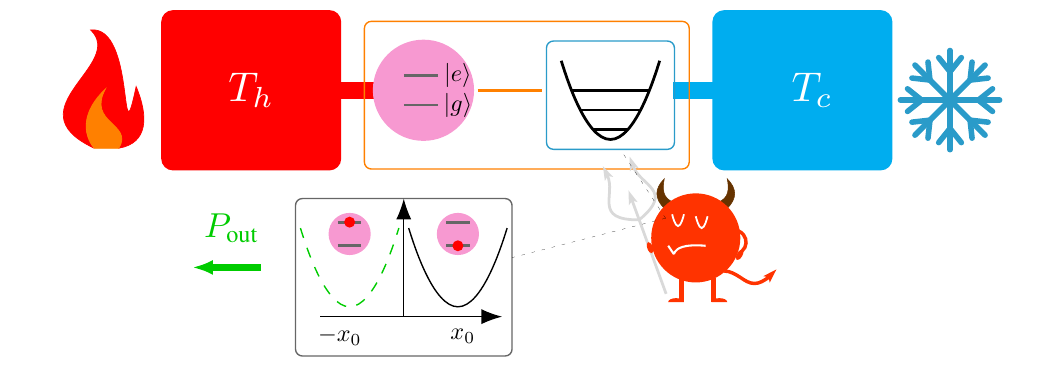}
        \caption{Schematic diagram of the measurement-driven engine. The role of the working medium is played by a two-level system connected to the hot reservoir.~The working medium interacts with a macroscopic pointer, modeled with a quantum harmonic oscillator, which is coupled to a cold thermal reservoir.~The engine works in the strong system-pointer coupling regime,~i.e.,~the working medium and pointer are in a correlated state.~The measurement-feedback mechanism is implemented by a demon that can only probe the displacement of the pointer and, based on the measurement outcome, send a pulse to the qubit to extract work.~The information gained is eventually erased by coupling the pointer to the cold bath.~Reproduced from \cite{Seah:PRL:2020}.}
    \label{fig:DiagME7}
\end{figure}
The pointer is in turn coupled to the cold bath with a dissipation rate $k_c$.
The qubit-oscillator interaction is modeled with a spin-boson Hamiltonian, $H=\hbar \Omega/2 \sigma_z + \hbar \omega (a^{\dagger}a +1/2) +\hbar\omega x_{0}\sigma_{z}(a^{\dagger} + a)/\sqrt{2}$, where the conventional notation of creation and annihilation operators is used and the energy scales obey $\Omega\gg\omega\gg k_{h} \gg k_{c}$, with $x_{0}>1$ (ultrastrong coupling).
As a consequence, the system energy levels both depend on the qubit and the oscillator states, and the pointer can react to any change in the qubit state induced by thermal fluctuations.
In this case, the demon cannot directly measure the qubit state, but it can monitor the pointer.

In the measurement-feedback scheme, the demon probes the pointer's mean displacement at a measurement rate $\gamma$: whenever the pointer's mean displacement is found to be negative, i.e., the qubit lies in its excited state, work extraction is performed by means of a Rabi flip, modeled by the operator $\sigma_x$.
An additional feedback scheme employs a driving field performing a Rabi flip conditioned on the oscillator position.
Remarkably, the information stored in the pointer is erased by dumping the heat into the cold bath, and the expelled heat exceeds the cost of resetting the information stored in the pointer. 
This measurement-driven engine can operate in a temperature regime inaccessible to the conventional Otto cycles and reach both high power and efficiency simultaneously.

A two-qubit engine fuelled by entanglement and local measurement was proposed in~\cite{Bresque:TwoQub_PRL_2021}. Here the two-qubit working medium (A,B) is driven into an entangled state by means of a qubit pairs interaction, $H_{2\rm{qb}} = \sum_{i=A,B} \hbar \omega_{i} \sigma_{i}^{\dagger} \sigma^{-}_{i} + \hbar \frac{g(t)}{2} (\sigma_{A}^{\dagger} \sigma^{-}_{B} + \sigma_{B}^{\dagger} \sigma^{-}_{A})$, where the qubit frequencies are detuned such that $\omega_{B}-\omega_{A}=\delta>0$ and $g(t)$ is a switching function.
Then a local measurement is performed on the qubit B, such that the quantum correlations are erased and the state becomes diagonal in the measurement basis.~In contrast to conventional measurement-fuelled engines, here the quantum state before the measurement is not passive~\cite{Perarnau-Llobet2015,Ding:PRE_2018}.
The energy provided to the system through the measurement is positive and equals the amount of energy required to erase the correlations.
This feature is confirmed by pre-measurement state analysis, in which the quantum origin of the fuelling process is investigated through a single-qubit model of the quantum meter.

In addition, a feedback protocol is adopted to convert the information gained through measurement into work.
The entanglement-measurement cycle takes place in four steps.
In the first, the entanglement step, the qubits are initially prepared in the state $|\psi_{0}\rangle=|10\rangle$, and then the interaction term is turned on so that they coherently evolve into an entangled state $|\psi(t)\rangle=\alpha(\delta,\Omega; t) |10\rangle   + \gamma (\delta,\Omega; t) |01\rangle$, with the Rabi frequency $\Omega = \sqrt{g^{2} + \delta^{2}}$. 
In this stage, their energy remains constant.
In the second step, a local projective measurement of the qubit B is carried out, which turns the qubits' state into a statistical mixture of the form $\rho(\theta) = \cos^{2} {\theta} \ketbra{10}{10}  + \sin^{2} {\theta} \ketbra{01}{01}$, with $\theta=\tan^{-1} {g/\delta}$ such that the quantum correlations between them are erased.

The information gained in this stage is stored in a classical memory M.
Crucially, during this step, the energy and the von-Neumann entropy of the qubits are increased as $E_{\rm{m}}=\hbar \delta \sin^{2} \theta\nonumber \geq 0$ and $S_{\rm{m}} =  - \sin ^{2} \theta  \log_{2} (\sin ^{2} \theta) -\cos ^{2} \theta \log_{2} (\cos ^{2} \theta) $ respectively.~These features fully describe the quantum fuelling of the working medium, as the qubits energy can be increased with vanishingly small entropy increase.
In the case of ideal measurements, the information gained through the process is equal to the mutual information between the system and the memory $I_{m} (S:M)=S_{\rm{m}}$.
The third step involves the feedback process, i.e., a $\pi$ pulse is sent to the qubits whenever the excitation is found in qubit B, such that B emits a photon and A absorbs one. The work extracted is $W=\hbar \delta $, and the system is reset to $\ket{10}$.
Otherwise, if the excitation is
measured in A, no pulse is sent, and the cycle restarts.
Eventually, during the fourth step, a cold bath is used to erase the information stored in M.
The conversion efficiency of the engine is found to be optimal when all the information stored in the memory is consumed.
In the limiting case of no information gain, e.g., in the absence of feedback protocol, the engine can operate as a conventional single-temperature measurement fuelled engine~ 
\cite{Jordan:QSMF_2020,Manikandan:PRE_2022,Son:PRX_Quant_2021,Xiayu:PRL_2022}.  

\subsection{Using quantum heat engines to generate entanglement}\label{subsec:EntGen}

As shown in the previous sections, there is an ongoing debate on whether quantum correlations -- either residing in the reservoirs (see  Sec.~\ref{subsec:corrbath}) or established between the working medium and the bath in an information engine (see Sec.~\ref{subsec:MDEng}) -- can provide a real quantum advantage to the performance of a heat engine operating in the quantum realm, compared to the classical setting. 
The presence of quantum correlations between a cold and hot qubit, when both are locally described by thermal Gibbs states, can lead to intriguing nonclassical physical effects, such as the reversal of heat flow between them~\cite{micadei2019reversing}.
Entanglement in the working medium has been claimed to provide enhancements to the performance of heat engines: quantum refrigerators~\cite{brunner14,brunner15} and Otto engines based on collective effects such as superabsorption~\cite{kamimura2022quantum}, to name a few.  

On the other hand, it has been established that quantum-correlated states can be induced through interaction with baths~\cite{Plenio98,verstraete2009quantum} in a dissipative system.
Indeed, minimal models of autonomous heat engines (see Sec.~\ref{subsec:Enginesintro} and Sec.~\ref{sec:thermoel}) can be devised, leading their working medium to a steady state with markedly nonclassical features, e.g.,  entanglement~\cite{brask2015autonomous,Tavakoli2018heraldedgeneration,tacchino2018steady}.

\begin{figure}[h]
        \centering
        \includegraphics[scale=1.5]{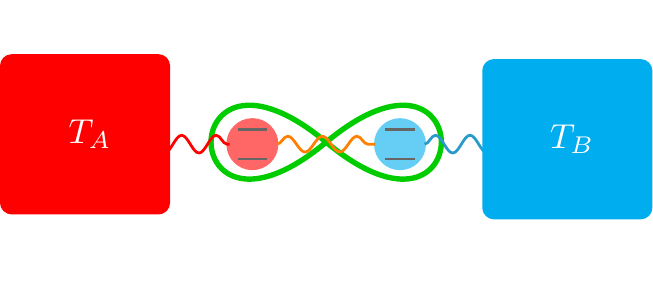}
        \caption{Schematic diagram of the two-qubit autonomous engine to generate entanglement. Reproduced from \cite{BohrBrask2022operational}.}
    \label{fig:Twoqubengine}
    \end{figure}

The emergence of different kinds of operational nonclassicality, namely Bell nonlocality~\cite{horodecki09,brunner2014bell}, steering~\cite{cavalcanti2016quantum,Uola:RevModPhys_2020} and quantum teleportation has been investigated in~\cite{BohrBrask2022operational}, by taking into account prototypical autonomous engines where, as depicted in Fig.~\ref{fig:Twoqubengine}, the working medium is comprised of two interacting qubits.  
The free Hamiltonian of the qubits A, B reads $H_{A}=E\ket{1}\bra{1}_{A}\otimes \Iop_{B}$ and  $H_{B}=\Iop_{A}\otimes E\ket{1}\bra{1}_{B}$, where $E$ is the energy gap. Each qubit interacts with a thermal bath at a fixed temperature, which can have a bosonic and fermionic nature and, when possible, they can be population-inverted (see also Sec.\ref{subsec:negtemp}).
The qubits are coupled through the Hamiltonian $H^{u}_{\text{int}}=g(\ket{01}\bra{10} + \ket{10}\bra{01}) +u\ket{11}\bra{11}$, where $g$ is the magnitude of the term exchanging excitations between the qubits and $u$ is the energy cost of double occupation. Each of the qubits is weakly coupled to its bath, with strengths~$\gamma_{A/B}$. 
In the limit $g<\gamma_{A},\gamma_{B}\ll E$, the dissipative dynamics can be described in terms of a local quantum master equation~\cite{hofer17,de2018reconciliation}.~This form of the master equation allows for the investigation of exceptional points in the dynamics of the engine~\cite{khandelwal2021signatures}.
Furthermore, the analytical form of the working medium's steady state can be studied for the engine parameter.

The steerability of the steady state is explored through the linear programming method of bounding the steering ratio, along with the estimation of the maximum amount of isotropic noise the system can sustain without becoming unsteerable.
Moreover, teleportation has been studied with a standard quantifier, i.e., the fidelity of the state received by Bob with respect to the qubit state $\psi$ to be transmitted, when the shared entangled state~$\rho$ and a classical communication channel are employed.
It reads $f=(1+2F(\rho))/3$, with $F(\rho)=\max_{U}\bra{\psi^{-}}(\Iop\otimes U)\rho (\Iop\otimes U^{\dagger})\ket{\psi^{-}}$, where $\ket{\psi^{-}}=(\ket{01}-\ket{10})/\sqrt{2}$ and the maximization is performed over all the possible single-qubit unitaries $U$.
The state $\rho$ is useful for teleportation if $F>1/2$.
Eventually, Bell nonlocality is studied utilizing the Clauser-Horne-Shimony-Holt (\acrshort{chsh}) inequality.
The results show that the engine generates a sufficient amount of steady-state entanglement to reveal steering in all the bath configurations. On the other hand, Bell's nonlocality and teleportation cannot be achieved without additional resources, such as population inversion. Eventually, by making use of heralding -- that is, the introduction of suitable measurements on the steady state -- the entanglement of the final state is increased so that Bell nonlocality can be enabled.

\section{Concluding Remarks}\label{sec:conclusion}

In this work, we reviewed the active research field of quantum engines and refrigerators, providing a broad overview of the main theoretical proposals and the recent experimental realizations.
While the latter has experienced quite a long period of infancy, over the past several years it has grown tremendously due to the experimental advances in different quantum platforms. These new experimental developments have boosted the field and unfolded it in new directions.  
As the section divisions of this review demonstrate, the study of quantum engines and refrigerators encompasses a wide range of disciplines, from open quantum system dynamics to thermoelectric devices and quantum information theory.
We also wish to note that the study of quantum engines and refrigerators interacts with other topics in quantum mechanics that were not covered in this review. These topics include quantum control theory and optimization strategies that are essential for the efficient operation of the devices, quantum speed limits, quantum heat pumps and nanoscale heat diodes, topological heat engines, algorithmic cooling, and indefinite-causal-order refrigerators.
There is no one answer to the question of what role quantum phenomena play in energy and information conversion on the micro-scale. However, adopting an engine and refrigerator viewpoint when imagining and analyzing these types of processes provides a common ground for defining the efficiency of processes and exploring fundamental aspects of thermodynamics in the quantum regime. Now that energy management is such an urgent concern, as quantum technology evolves in the near future, managing it on the quantum scale will be an essential strategy.

\section*{Acknowledgements}
The authors express their sincere gratitude to Ronnie Kosloff, Robert Whitney, Martin Plenio, Gabriele De Chiara, Karen Hovhannisyan, Jian Hua Jiang, Hiroyasu Tajima, Paolo Abiuso, Paolo Andrea Erdman, and Matteo Carrega for their invaluable discussions and insightful comments on the manuscript. Additionally, we thank Taylor Johnston-Levy for her meticulous assistance throughout the editing process. 
This work was supported by the Israel Science Foundation (Grant No. 1364/21).

 







\end{document}